\pdfoutput=1

\documentclass[12pt]{report}

\usepackage{refcount}
\usepackage{verse}
\usepackage{amsmath}
\usepackage[  
  hidelinks,  
  ]{hyperref}
  \hypersetup{pdfauthor={Dylan Hutchison},pdftitle={ModelWizard: Toward Interactive Model Construction}}
\usepackage{listings}
\usepackage{enumitem}
\setlist[itemize]{leftmargin=*}
\setlist[enumerate]{leftmargin=*}
\usepackage{graphicx}
\graphicspath{ {./img/} }
\DeclareGraphicsExtensions{.pdf,.png,.jpg}
\usepackage{todonotes}
\usepackage{listings,bold-extra}
\usepackage{lstautodedent} 
\usepackage[justification=centering]{caption}
\usepackage{subcaption}
\usepackage{wrapfig}

\usepackage{tikz}
\usetikzlibrary{arrows,automata}
\usetikzlibrary{positioning,calc}

\definecolor{bluekeywords}{rgb}{0,0,0.5} 
\definecolor{greencomments}{rgb}{0,0.5,0}
\definecolor{turqusnumbers}{rgb}{0.17,0.57,0.69}
\definecolor{redstrings}{rgb}{0.5,0,0}

\lstdefinelanguage{FSharp}%
{morekeywords={let, new, match, with, rec, open, module, namespace, type, of, member, %
and, for, while, true, false, in, do, begin, end, fun, function, return, yield, try, %
mutable, if, then, else, cloud, async, static, use, abstract, interface, inherit, finally, val },
otherkeywords={ let!, return!, do!, yield!, use!, var }, 
keywordstyle=\color{bluekeywords}\bfseries,
basicstyle={\onehalfspacing\footnotesize\ttfamily},
breaklines=false,
tabsize=4,
morecomment=[l][\color{greencomments}]{///},
morecomment=[l][\color{greencomments}]{//},
morecomment=[s][\color{greencomments}]{{(*}{*)}},
morestring=[b]",
showstringspaces=false,
stringstyle=\color{redstrings},
}

\usepackage{subcaption}
\usepackage{graphicx}
\graphicspath{ {./img/} }
\DeclareGraphicsExtensions{.pdf,.png,.jpg}

\usepackage {fancyhdr}
\usepackage{amssymb, changepage}
\usepackage{rotating}
\usepackage{setspace}
\usepackage{pgnumchapter_nums}
\usepackage{titlepg}

\def\addsymbol #1: #2#3{$#1$\> \parbox{5in}{#2 \dotfill  \pageref{#3}}\\}

\renewcommand{\headrulewidth}{0 pt}          
\renewcommand{\footrulewidth}{0  pt}         

\renewcommand{\contentsname}{\normalsize{Table of Contents}}
\renewcommand{\listfigurename}{\normalsize{List of Figures}}

\DeclareCaptionType{code}[Code Listing][List of Code Listings] 

\begin{document}

\pagenumbering{roman}

\thesistitlepage

\thesiscopyrightpage              
	
\addcontentsline{toc}{chapter}{Abstract}          		 				
\thesisabstract


\addcontentsline{toc}{chapter}{Acknowledgments}   				
\thesisacknowledgments

\makeatletter \renewcommand{\@dotsep}{10000} \makeatother


\renewcommand{\contentsname}{\normalsize{Table of Contents}}      
\tableofcontents                                                 

\newpage		

\addcontentsline{toc}{chapter}{List of Tables}     					
\listoftables

\newpage     
\addcontentsline{toc}{chapter}{List of Figures}						
\listoffigures

\newpage
\addcontentsline{toc}{chapter}{List of Code Listings}
\listofcodes

%


\newpage
\pagenumbering{arabic}

\chapter{Introduction}

In his 2013 visionary article, Dr. Chris Mattmann issued a call for data scientists: ``To get the most out of big data, funding agencies should develop shared tools for optimizing discovery and train a new breed of researchers'' \cite{MattmannDataScience2013}.
This thesis is an answer to Dr. Mattmann's call, laying groundwork for new tools  to help data scientists gain insight faster.

\section{Data Science as Model Construction}
We address three scientific goals: 
\begin{itemize}
\item \textbf{Prediction} of unseen states.\footnote{\textit{State} is a domain-specific term whose definition is clear in the context of a problem. 
A chemical engineer working on oil refinery design may use \textit{state} to refer to the chemical composition of fluid in a reactor; 
a statistical physicist to refer to a specific configuration of a thermodynamic system; 
a computer architect to refer to the contents of each register on a CPU;
a stockbroker to refer to the price of equities at a given time. 
More examples abound.} \\  
E.g., an apple dropped from a tall building will fall about 122 meters in 5 seconds.
\item \textbf{Understanding} of the structure that naturally generates states. \\
E.g., the force of gravity decreases apple elevation quadratically with time.
Equivalently, gravity linearly decreases a hidden variable, apple velocity, which in turn linearly decreases apple elevation with time.
\item \textbf{Generalization} of common structure to other problems and domains. \\
E.g., quadratic regression generalizes to polynomial regression, a 
relationship we may apply to light dispersion, roller coasters and plenty other datasets.

\end{itemize}

In the context of \textit{data science}, we address these goals by analyzing datasets: predicting unseen data values, understanding the structure that generates data, and generalizing that structure to other datasets. 
We apply these principles to the apple freefall example 
by imagining a dataset of apple elevations over time
in the form of Table~\ref{tAppleFall}. 
Each row consisting of a (time, elevation) pair is a data point. We aim to predict unseen rows with missing time or elevation, to understand the structure that generates these pairs, and to generalize that structure to other contexts.

\begin{table}
\centering
\begin{tabular}{ll}
Time (s) & Elevation (m) \\
\hline
0 & 200 \\
1 & 196 \\
2 & 179 \\
3 & 155.1 \\
6.5 & 0
\end{tabular}
\caption{Apple in Freefall dataset}
\label{tAppleFall}
\end{table}

By \textit{structure} we refer to some mechanism that generates data points.
Assuming the standpoint of epistemic uncertainty,\footnote{in contrast to aleatory uncertainty, which holds that no true structure may ever be known due to intrinsic randomness. Distinguishing the two types of uncertainty may be useful for determining what uncertainty is reducible by further data collection \cite{Kiureghian2009105}, but for our purposes we consider true structure completely discoverable in the limit of infinite data.}
there exists a true mechanism that generates data points 
in reality that we do not fully know.  
We call the unknown true mechanism \textit{true} structure, and 
we call a guess at the true structure a \textit{candidate} structure, 
shortened to structure for brevity. 
We also refer to structure as a \textit{model}.
We use the language of probability to describe uncertain knowledge of true structure.

True structure includes everything up to recording a data point.
Including how we collect data points in models can therefore be helpful,
in addition to how ``nature'' generates data points.
In the Apple Freefall data for example, 
if we know our method of recording elevation 
is particularly inaccurate or imprecise, 
we ought to reflect that knowledge in our model.\footnote{See how the Apple Freefall model noise term reflects data imprecision in Figure~\ref{fAppleLinRegModel} and~\ref{fAppleQuadRegModel}.}
We may also introduce noise to account for model imprecision:
that a model may partly account for the true structure mechanism.
The magnitude of noise reflects how well that model accounts.

Models represent a (joint) \textit{probability distribution} over 
data points.  
Impossible data points that a model cannot generate have zero probability mass.  
Possible data points receive probability mass proportional to likelihood that the model will generate them.
We create candidate models using the subset of possible data points
present in a dataset's non-missing values.

We are specifically interested in structure taking the form of \textit{machine learning models}. 
These are models with a computable \textit{sampling} procedure to output a data point.
Sampling algorithms such as rejection sampling compute samples that satisfy conditions. 
Inference algorithms solve the reverse problem: computing the probability that the model will generate a data point. 

Some models are able to sample and infer data point probabilities for every column in their dataset.  We call those models \textit{generative}, since they specify a distribution that can be used to generate samples over any column.
We may also consider \textit{discriminative} models that do not define distributions over every column.\footnote{See \cite{bernardogenerative} for an overview of generative and discriminative models.} 
We cannot sample or infer data point probabilities of ``unmodeled'' columns in a discriminative model,
instead using them as known inputs to sample or infer other columns.
Data scientists may find discriminative models relieving 
because they do not have to specify distributions behind unmodeled columns,
saving work required to justify why a certain distribution makes sense for a column.

Define \textit{model construction} as the process of creating machine learning models for a dataset.  
A well-built model solves all three goals: predicting unseen states by (conditional) sampling, understanding structure by examining model components, and generalizing structure by applying components to machine learning models for other datasets. 
Prediction is a key link, because high predictive power on unseen data indicates that a machine learning model is close to the true structure generating data.
Simplicity is another guide that increases understanding and generalizability;
see \cite{Freitas2014}
for further reading on scoring machine learning models for comprehensibility.

In constructing our final model we may consider alternative models, comparing them as we imagine them and choosing models that are simpler and fit our dataset better.  
We call this process \textit{model exploration}.  

The goal of the ModelWizard project is to develop a tool 
that supports the model exploration workflow.  
We aim to facilitate creativity, increase interactivity, 
introduce safety and support abstraction
when writing a ModelWizard script 
that ultimately constructs a Tabular machine learning model.

\section{Prelude Example: Freefall in ModelWizard}
\label{sAppleFreefallExample}
We now introduce ModelWizard's exploration workflow through a small example,
focusing on how a user guides exploration through 
questions and fitness metrics.
Subsections address issues to consider before basing judgements
on ModelWizard's output, 
and then on the underlying Bayesian methematics.

Assume the perspective of a data scientist, unaware of gravity, presented with the 
apple-in-freefall data of Table~\ref{tAppleFall},
collected on a windy day or by an imprecise instrument to add non-determinism.
We may ask several questions: where is the apple at 4 or 5 seconds?  What happens past 6 seconds when we observe the apple at elevation zero?  At what time does the apple reach an elevation of 50m?  
We incorporate our questions into the dataset in Table~\ref{tAppleFallPlus}.

\begin{table}
\centering
\begin{tabular}{ll}
Time (s) & Elevation (m) \\
\hline
0 & 200 \\
1 & 196 \\
2 & 179 \\
3 & 155.1 \\
4 & \\
5 & \\
6.5 & 0 \\
7 & \\
 & 50
\end{tabular}
\caption{Apple in Freefall dataset, with ``missing value'' rows}
\label{tAppleFallPlus}
\end{table}

An initial model we may posit for this dataset is a linear relationship between time and apple elevation.
To create this model, we use the ModelWizard operations 
\begin{align}
\label{opLinReg} &\texttt{LinReg "Elevation" "Time"}\\
\label{opModelTime} &\texttt{Model "tmain" "Time"}
\end{align}
Operation~\ref{opLinReg} places a noisy linear regression model on elevation with explanatory variable time.
Operation~\ref{opModelTime} places an uninformative Gaussian distribution on time.

\begin{figure}
\centering
\begin{subfigure}[b]{.495\linewidth}
\includegraphics[width=\linewidth,clip=true,trim=0 0 238pt 0]{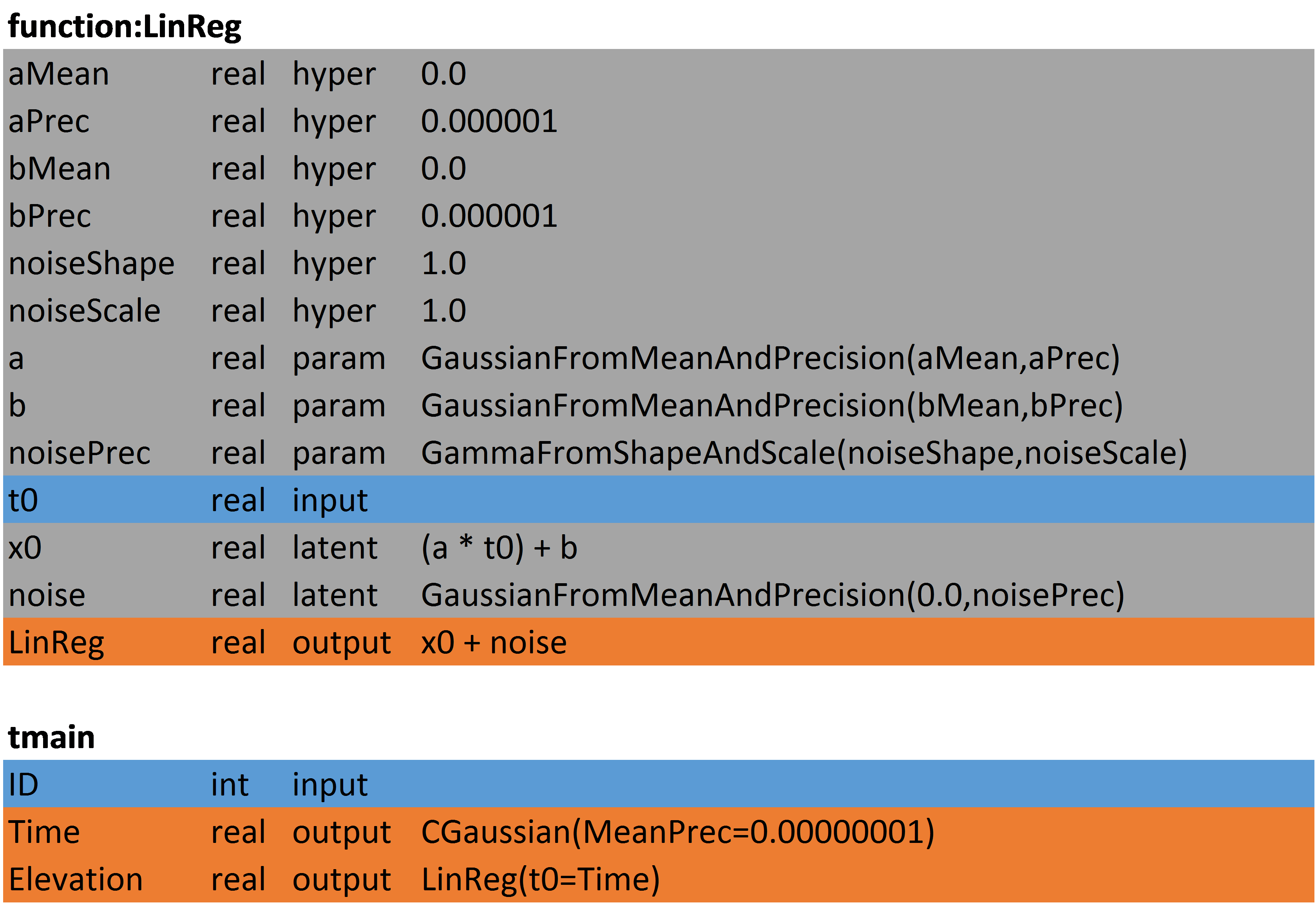}
\caption{LinReg Model}
\label{fAppleLinRegModel}
\end{subfigure}
\begin{subfigure}[b]{.495\linewidth}
\includegraphics[width=\linewidth,clip=true,trim=0 0 238pt 0]{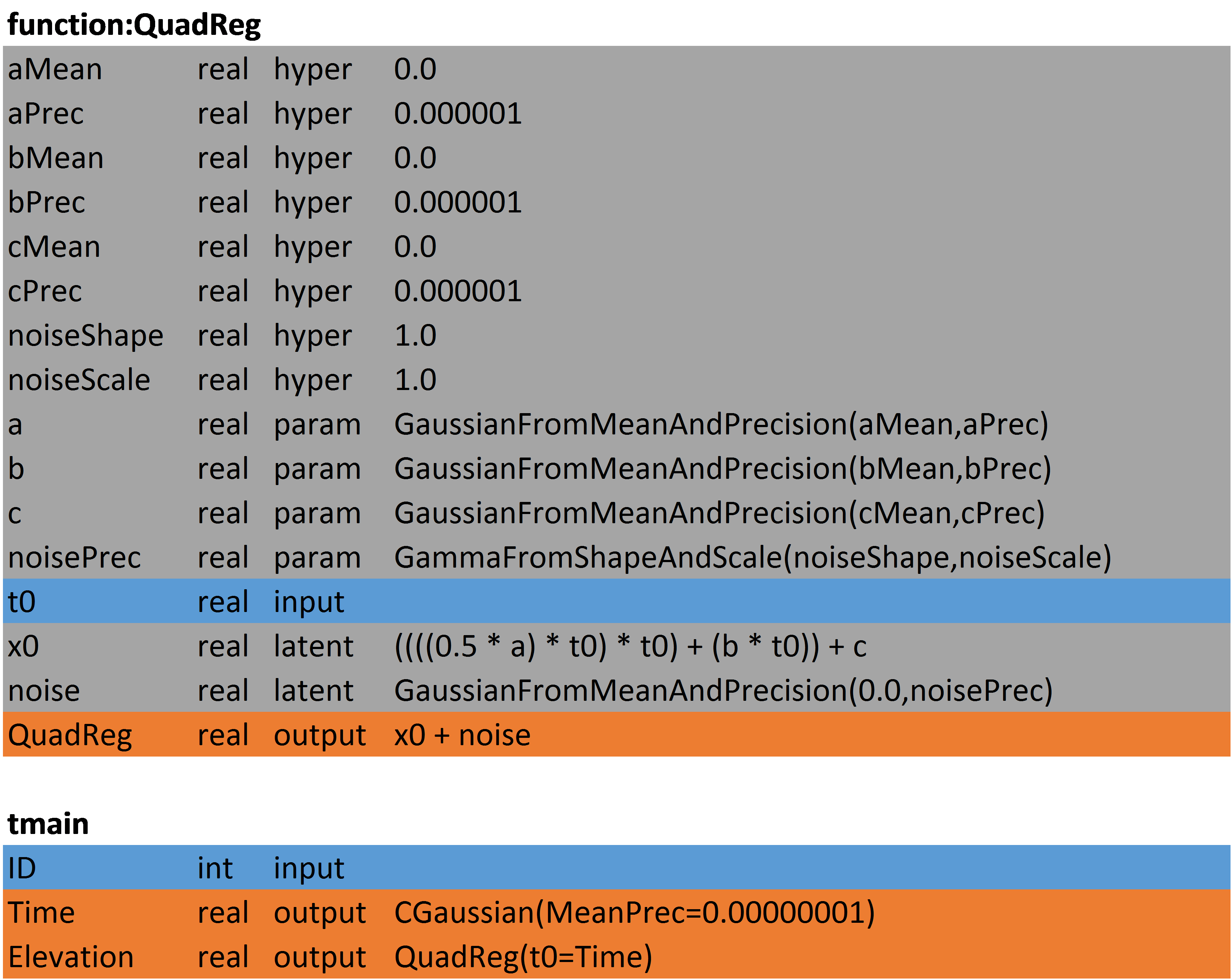}
\caption{QuadReg Model}
\label{fAppleQuadRegModel}
\end{subfigure}
\vspace{15pt}

\begin{subfigure}{.99\linewidth}
\includegraphics[width=\linewidth]{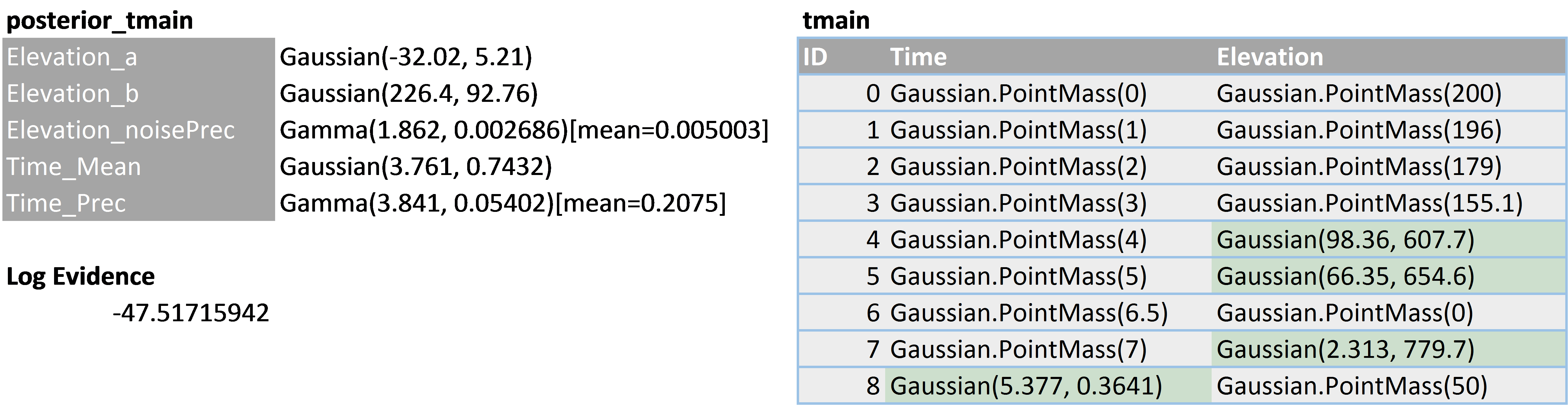}
\caption{LinReg Inference Results}
\label{fAppleLinRegResults}
\end{subfigure}
%
\vspace{12pt}

\begin{subfigure}{.99\linewidth}
\includegraphics[width=\linewidth]{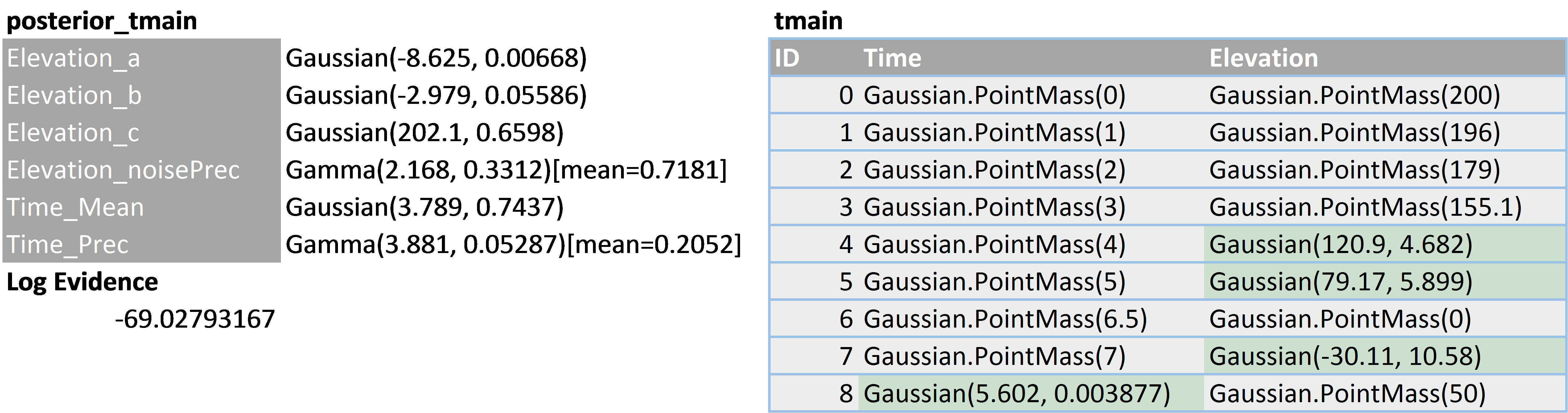}
\caption{QuadReg Inference Results}
\label{fAppleQuadRegResults}
\end{subfigure}
\caption[Apple Freefall Linear and Quadratic Regression]{Apple Freefall Linear and Quadratic Regression\\Green cells are posterior distributions on Table~\ref{tAppleFallPlus}'s missing values.}
\end{figure}

Figure~\ref{fAppleLinRegModel} shows the resulting linear model.
The rows in the bottom table labelled `tmain' 
list the name, type and model of corresponding columns in the original 
dataset of Table~\ref{tAppleFallPlus}.
The giant \texttt{LinReg} function block implements linear regression in Tabular syntax such that the Elevation column model may call 
linear regression as a macro.
Detail-oriented readers may skip to Chapter~\ref{cLanguage} for deeper Tabular syntax and semantics, 
but for now we recommend accepting linear regression as is, focusing instead on ModelWizard's workflow.
The linear function block's one addition to usual regression is the noise term,
a probabilistic analogue to what statisticians call regression error.
Larger inferred noise precision (equivalently, smaller inferred noise variance)
tends to indicate less error and greater model fit,
though we should take caution to avoid too high precision which is prone to overfit.

Operation~\ref{opModelTime} converts the model from discriminative to generative.
Because the last row of our dataset in Table~\ref{tAppleFallPlus} 
contains a missing value for time, we must use a generative model in order to have a way to predict time. 
A generative model also allows us to answer questions such as ``what is the apple elevation'' without specifying time, equivalent to using the posterior distribution (see Section~\ref{sBayesInference}) on time to determine a distribution over elevation.
Alternatively, if we omit the last row of Table~\ref{tAppleFallPlus} and only care about predicting elevation given time, then we could omit Operation~\ref{opModelTime}, leaving time as an input and disallowing missing time values in our dataset.
The advantage of the alternative discriminative model 
is not needing to specify and justify a distribution on time,
which we questionably list as a Gaussian distribution.
In summary, ModelWizard may construct both generative and discriminative versions of the linear model, and both have advantages in different situations.

Figure~\ref{fAppleLinRegResults} shows results of inference\footnote{We use the Expectation Propagation inference algorithm throughout the Apple Freefall example.}
on the linear model.
Inference consists of training a given model on observed (non-missing) values 
in the input dataset to form a posterior model, outputting 
posterior model parameters in the left of Figure~\ref{fAppleLinRegResults},
and outputting marginal distributions from the posterior model on 
missing values, each conditioned on the observed values in the same row,
in the right of Figure~\ref{fAppleLinRegResults}.
Grey-background ``PointMass'' distributions correspond to 
observed values in the original dataset of Table~\ref{tAppleFallPlus},
whereas green-background cells are more meaningful posterior distributions
for missing values in the dataset.

It seems suspicious that the linear model predicts positive elevation at 7 seconds, one second after the apple is known to reach 0m elevation.
To better evaluate the linear model's performance using a method less prone
to overfitting, we use leave-one-out cross-validation 
on the known rows of Table~\ref{tAppleFall} 
with the operation \texttt{CrossValidate\_kFold\_RMSE "tmain" "Elevation" 5}
and similarly for Time.
Resulting root mean square errors 
(not shown in Figure~\ref{fAppleLinRegResults}) 
are 14.996m for elevation and 1.046s for time.
Viewing these numbers alone, we do not know whether the linear model is optimal 
or whether other models could deliver lower error. 

Could a more complicated model like quadratic regression between time and apple elevation add enough predictive power to merit its increase in model complexity?
An answer to this question in either direction---whether quadratic regression is a viable or ill-suited model---is an important clue to understanding the dataset's structure.
Ruling out a model family like the family of quadratic models 
can be just as informative to a data scientist 
as showing a model's plausibility.

We build the quadratic regression model
shown in Figure~\ref{fAppleQuadRegModel} by replacing 
Operation~\ref{opLinReg} with 
\[
\texttt{QuadReg "Elevation" "Time"}
\]
Figure~\ref{fAppleQuadRegResults} tables inference results.

It is heartening to see that the quadratic model 
predicts negative elevation at 7 seconds 
and that it has higher noise precision.
Running the same cross-validation operation returns 
root mean square errors of 13.546m for elevation and 0.759s for time, 
both lower than the linear model.
Comparing cross-validation numbers, we see evidence that 
a quadratic relationship more likely governs the structure
behind our dataset than a linear relationship.

Of course, bringing in knowledge of gravity, the quadratic model parameters make more sense than those of the linear model; 
the inferred parameter \texttt{a} = $-8.625$
is reasonably close to Earth's $-9.8 m/s^2$ true force of gravity. 
The error is a result of the noisy data set,
as the skeptical reader may verify by calculating elevation
$= 200 - \frac{9.8}{2} t^2$ for observed times $t$ in Table~\ref{tAppleFall}
and comparing them to the means of the quadratic model's posterior distributions.

Model evidence, the probability of observing the dataset under a model, 
does not function as well as cross-validation for evaluating model fitness, 
as seen from the quadratic model evidence in Figure~\ref{fAppleQuadRegResults}
being lower than the linear model evidence in Figure~\ref{fAppleLinRegResults}.
We demonstrate inference using model evidence 
in Section~\ref{sApplicationBayesianNetwork} and then switch to (5-fold, i.e. leave-$n/5$-out) cross-validation for other applications.

\subsection{Model Selection Matters: Overfitting and Consistency}
\textit{Overfit} models perform well on training datasets but do not perform well on 
other datasets generated from the same true structure.
Formally, we say a model overfits if it performs better on a training set of data
but worse on the distribution of all possible data sets
than an alternative model \cite{mitchell1997machine}.
The Apple Freefall models, for example, would clearly overfit if their noise precisions are excessively high.  
They would predict tight distributions accurately on their training set that
do not predict well on non-training set data.\footnote{We ignore the case of modeling a perfectly deterministic true structure, in which case a (zero variance) Dirac delta distribution may perfectly capture a deterministic relationship between columns.}

Model comparison methods can be another source of overfitting.
We strive to compare models in a way that does not favor specialized, overfit models.
Luckily for the Apple Freefall example, model comparison by leave-one-out cross-validation is equivalent to model comparison by the 
Akaike information criterion (AIC) \cite{stone1977asymptotic},
which has a desirable model complexity penalty \cite{bozdogan1987model} 
that penalizes models with larger degrees of freedom.
The complexity penalty enables realistic comparison between the simpler linear model and the more adaptive and thereby more likely to overfit quadratic model.

We also strive to compare models \textit{consistently}, 
that is, to compare models with a guarantee that the probability of choosing 
the ``correct'' model closer to the true structure converges to 1 
as the number of data points approaches infinity. 
Unfortunately AIC is an inconsistent model comparison method \cite{shao1993selection},
which means that AIC may not choose the best model even after training on infinite data observations.
We employ leave-one-out cross-validation 
anyway for convenience in the Apple Freefall example.
Consistent model comparison methods include leave-$n_v$-out cross-validation, Bayesian information criterion (BIC) and Bayes factors
for models with a finite number of parameters, and consistent methods exist for 
comparing nonparametric models too \cite{joo2010model}.

\subsection{Bayesian Inference}
\label{sBayesInference}
We obtain cross-validation scores used to compare models via inference,
characterized by its use of \textit{Bayes' Rule}.
Beginning with a prior predictive model\footnote{Vocabulary note: a (prior or posterior) \textit{predictive} model specifies a distribution over a row of data, whereas a (prior or posterior) model specifies a distribution over model parameters. We refer to the latter as parameter models to disambiguate. See \cite{basicBayesian2007} for examples of each.}
 that specifies $\Pr(row)$ for any row of data,
we use Bayes' Rule to iteratively condition on each observed row of a dataset
and form a posterior predictive distribution:
\[
\underbrace{\Pr(row \mid obsRow)}_{\text{posterior pred. model}}
:= \alpha \underbrace{\Pr(obsRow \mid row)}_{\text{obsRow likelihood}}\underbrace{\Pr(row)}_{\text{prior pred. model}}
\]
where $obsRow$ is a row of observed (non-missing) data points and
$row$ is a row of data for which we wish to calculate its probability 
via a model that incorporates the data evidence in $obsRow$.
For example, $obsRow$ may be (time = 1, elevation = 196) and $row$ may be (time = 2, elevation = 0).  We hope our model assigns $\Pr(row \mid obsRow)$ infinitesimal probability.

$\alpha$ is a normalizing constant that ensures the posterior predictive model is a valid probability distribution, i.e., that $\int \Pr(row) \text{ d}row = 1$, integrating over all possible rows:
\[
\alpha = \dfrac{1}{\Pr(obsRow)} 
= \dfrac{1}{\sum_{row} \Pr(obsRow, row)}
= \dfrac{1}{\sum_{row} \Pr(obsRow \mid row) \Pr(row)}
\]
The reasoning is that if $\Pr(obsRow \mid row) > \Pr(obsRow)$, then $row$ increases $obsRow$'s likelihood, which indicates we should raise $\Pr(row \mid obsRow)$ since $obsRow$ is in fact observed.  The `$<$' version similarly decreases $\Pr(row \mid obsRow)$.

In the context of the apple freefall example, 
we tune our model from prior to posterior 
by adjusting the value of regression coefficients---free parameters in the model for which we specify (usually uninformative) prior distributions:
\[
\underbrace{\Pr(param \mid obsRow)}_{\text{posterior param dist.}}
:= \alpha \underbrace{\Pr(obsRow \mid param)}_{\text{param likelihood}}\underbrace{\Pr(param)}_{\text{prior param dist.}}
\]
We use parameters in our model to determine a distribution over possible 
rows of data by $\Pr(row) = \Pr(row \mid param) \Pr(param)$.

The inference algorithms we use in Tabular and ModelWizard
are forms of variational inference,\footnote{e.g., variational message passing, expectation propagation, ...}
a process that fits an easier-to-analyze probability distribution 
as close as possible to an arbitrary probability distribution 
like the one our apple freefall model represents.
In exchange for approximation error in using 
the easier-to-analyze distribution in place of the original distribution, 
we gain computational speedup in estimating the posterior distribution.
Other forms of inference are exact inference,\footnote{e.g., variable elimination, junction tree, conjugate distributions, ...}  explicitly calculating an exact posterior distribution, and 
sampling-based inference,\footnote{e.g., rejection sampling, likelihood weighting, MCMC and its variants, ...}  sampling from a posterior distribution without explicitly calculating it.

Cross-validation calls inference as a subroutine by using the resulting probability distribution from inference to make predictions on held out data.
Well-performing models recover held out data accurately as indicated by low cross-validation error.
Cross-validation performs inference in several rounds, holding out different sets of data randomly at each round.
Thus, a model with high cross-validation score generally maintains performance 
when trained on different subsets of data,
which is an excellent guard against overfitting 
as discussed in Section~\ref{sAppleFreefallExample}.

It is useful to imagine comparing models by cross-validation (or other model scoring measures such as model evidence) as an extension of Bayes' Rule 
where we select the best fitting model at the same time as the best fitting parameters for each model:
\[
\underbrace{\Pr(model, param \mid obsRow)}_{\text{posterior model and param dist.}}
:= \alpha \underbrace{\Pr(obsRow \mid param)}_{\text{param likelihood}}
\underbrace{\Pr(param \mid model)}_{\text{prior param dist.}}
\underbrace{\Pr(model)}_{\text{prior model dist.}}
\]
We select model and its parameters with highest probability. 
Using the cross-validation method, this is the model (with parameters set by usual inference) that scores the least prediction error.
Using the model evidence method, this is the model with greatest likelihood.
Methods such as BIC, Bayes' factors and others discussed in Section~\ref{sAppleFreefallExample} follow the Bayes' rule pattern similarly.
The extension of the distribution over possible rows of data is 
$\Pr(row) = \Pr(row \mid param) \Pr(param \mid model) \Pr(model)$,
though after selection $\Pr(model)$ 
is a point mass distribution on the selected model.

An alternative to model selection is model mixing: 
taking a sum over possible models weighted by likelihood, 
instead of selecting one ``mode'' model with highest likelihood.
Model mixing has more robustness to change in data when two models both explain the data well, since a small shift in data could change a selected model abruptly while a small shift in data will smoothly change a mixture of models
\cite{hoeting1999bayesian}.
We use the model selection approach because understanding is one of our goals.
It is easier to interpret a single model than a mixture of models,
especially when we use our model as a proxy for true structure in nature.

There is a potential vocabulary conflict with statisticians, who define inference as ``the process of deducing properties of an underlying distribution by analysis of data'' \cite{upton2008dictionary}.
Recalling that true structure is a probability distribution over possible data, there is no conflict; true structure is the underlying distribution, whose properties we capture in models we create through model construction.
``Structure learning'' is also equivalent,
as seen from the similar breakdown of a model 
into ``structure'' and ``form'' in \cite{Kemp05082008}.

In this thesis, ``model construction'' refers to the entire process of finding a model close to true structure, 
``inference'' refers to Bayesian inference procedures to answer queries 
and calculate posterior distributions by conditioning on data using Bayes' rule,
and ``model exploration'' refers to determining model family/form, 
using the term ``learning'' whenever a choice is guided by dataset predictive power.

\subsection{Interactive Model Construction}
In Section~\ref{sAppleFreefallExample} we build our quadratic regression model interactively, creating other models like linear regression and comparing them as we naturally posit them through questions about the apple dataset.
In this manner, model construction naturally fits into the scientific method.
To illustrate the benefits of interactivity, 
imagine two other extremes of model construction: manual and automated.

Scientists adopting a \textit{manual} model construction strategy take years to study a scenario well enough to become expert and build a good model. 
The approach works but at high labor cost, whose fruits tend to be one-off, special purpose models applicable to a single problem.

Scientists adopting an \textit{automated} exploration strategy, using algorithms that output good families, face tractability problems trying to search through the space of all models. 
For example, \cite{Ch01} has $O(n^5)$ complexity searching within the family of discrete Bayesian networks, a small subspace of all possible models. Further, many automated algorithms such as multilayer neural network training generate ``black box'' models, which predict accurately but offer no insight into why the models work.

Interactivity delivers a happy medium between the two extremes.
Data scientists will work hand-in-hand
with computers to navigate the space of models, leveraging scientists' creative insight to guide
model search and computers' automation to present statistics, model performance results and
model suggestions as feedback. Scientist and computer together will craft understandable
models on multiple data sources faster than either can alone.

\section{Probabilistic Programming}
In probabilistic programming, machine learning models take the form of programs.  Users write code that samples data points, effectively defining a probability distribution over possible data points.  
The model specified as a program gains power from programming language constructs,
such as functional abstraction and control flow by iteration and conditionals.
Compilers use top level probabilistic program code to synthesize inference code.
See \cite{gordon2014probabilistic}, \cite{de2013probabilistic} and
\cite{goodman2013principles} for different flavors of introduction to probabilistic programming.

One success story demonstrating potential benefits for probabilistic programming is on
creating a seismic monitoring model
to help detect nuclear tests for the UN Preparatory Commission for the Comprehensive Nuclear-Test-Ban Treaty (CTBTO)
One of the CTBTO's models has on the order of 28,000 lines of code in C \cite{PPaMLKickOff13}.
That model was rewritten in the probabilistic programming language BLOG, delivering similar
accuracy at 25 lines of code \cite{russell2014unifying}. 
Reduction in development time and expertise barriers added considerable value.


\subsection{Tabular and Infer.NET}
Tabular is a probabilistic programming language 
where models consist of probabilistic annotations on database schemas 
written inside spreadsheets
\cite{DBLP:conf/popl/GordonGRRBG14, gordon2015probSpreadsheet}. 
Tabular's motivation is the idea that data scientists would find it easier and
more intuitive to specify their model by marking up the schema of their dataset, 
as opposed to specifying their model in a standalone language 
separate from their dataset.
Data scientists do need an understanding of Tabular's primitives, 
which are designed to be easily readable given 
background knowledge in statistics and probability.

The Tabular compiler compiles Tabular models 
into Infer.NET probabilistic programs \cite{InferNET12}, 
which can then be run in an Infer.NET inference engine 
to infer posterior distributions and answer queries. 
Infer.NET supports variational message passing, expectation propagation, 
and Gibbs sampling, all of which are accessible from Tabular.

\section{ModelWizard: Interactive Model Construction for Tabular}
ModelWizard is a domain-specific language, embedded in F\#, for interactively constructing Tabular models. 
Users write scripts in ModelWizard that incrementally construct Tabular models, 
progressing from an initial ``do-nothing'' Tabular model toward an inference-ready Tabular model, one operation at a time. 

ModelWizard is inspired by the difficulty of writing Tabular models, 
just as Tabular is inspired by the difficulty of writing Infer.NET models. 
ModelWizard's goal is to reduce the time it takes users to build a Tabular model from scratch. 
We are especially interested in the case where the user does not know what the final model is and is still exploring the space of models.

In this thesis we present the design and implementation of ModelWizard
as an embedded DSL, together with case studies in ModelWizard's use.
The following subsections highlight key features in ModelWizard's design:

\subsection{Concurrently Refining Data and Model}
ModelWizard's operations modify both a dataset's machine learning model under construction and the dataset itself. 
This is different than the traditional ``pre-processing phase'' approach in machine learning and data mining literature, to perform all dataset-modifying operations prior to constructing a model.

Operations in ModelWizard that traditionally occur during pre-processing are
changing types of columns, creating new tables of unique values, 
remapping values to point to another table with a foreign key relationship, 
and capturing functional dependencies by moving data between tables, 
a key step in table normalization.  
Other operations we imagine as extensions are 
numeric transformations like taking square roots, full table normalization, 
data compression by taking principal components
and outlier detection and cleaning, to name a few.
We also imagine post-processing extensions 
such as model validation and visualization.

While refining data and model at the same time adds flexibility to the model construction process, we are not yet certain whether it is always helpful.  
Data scientists may find performing pre-processing activities prior to modeling
conceptually easier.  
User case studies will help illuminate an answer.

\subsection{Composable Model Primitives} 

Rather than construct complicated machine learning models all at once, we construct models incrementally via sequences of primitive operations. 
Think of primitives as bricks and F\# code as glue.
Composing primitives and F\# together 
cements higher-level, derived operations we think of as pillars and walls.
The beauty of derived operations is that we may abstract them, 
treating them as a reusable single piece 
by forgetting the bricks and glue that compose them.
We construct the mightiest models, machine learning castles indeed, 
composing all available operations in the blueprint of F\#.

We find ModelWizard's primitive operations naturally capture common machine learning paradigms that when composed into derived models, offer easy access to many model families, such as clustering and Naive Bayes. 
Creating models as compositions lends understanding into how the models work, since we can ``open the box'' of a derived operation and inspect its component operations and code to see how it works.
At the same time, abstraction allows analysts to consider derived models without worrying about their low-level details by using the models as ``closed black boxes.''

For example, operations~\ref{opLinReg} and~\ref{opModelTime} 
that construct the Apple Freefall model 
are derived operations, abstracting details of regression and automating choice of
model by deducing that time is a real-valued variable. 
Users who want to "dig in," understanding and modifying the derived operations,
may do so by inspecting their source code.

\subsection{Safety in Model Construction}
Tabular has typing rules in terms of the domains of data columns and distributions to ensure a schema's probabilistic annotations make sense and that data actually conforms to the annotated schema.
A state, consisting of a dataset plus a Tabular model, is \emph{valid} provided it contains no naming conflicts, is well-typed, that its schema and data satisfy 
asserted relational properties like primary/foreign key,
and that no cyclic references are present.

Users developing models directly can easily construct invalid states.
In contrast, ModelWizard ensures only valid states are constructable, throwing exceptions when an operation cannot construct a valid state.

\section{Thesis Outline}
Chapter~\ref{cLanguage} introduces ModelWizard language syntax and design. 
We cover core types and OpMonad, ModelWizard's operation builder, 
followed by a walk through operations grouped by topic.
See Appendix~\ref{AModelWizardAPI} for a concise ModelWizard operation API.

Chapter~\ref{cApplications} opens with simple model construction on small datasets and progresses to advanced model construction on real-world data.
We cover Bayesian networks, Naive Bayes classifiers, a hybrid model with functional dependencies and clustering for collaborative filtering.
Readers seeking further intuition may skip Chapter~\ref{cLanguage} to view Chapter~\ref{cApplications}'s examples.

Chapter~\ref{cDiscussion} concludes by presenting desirable extensions 
and discussing ModelWizard's fit in the data science world.

\chapter{ModelWizard Language}
\label{cLanguage}
We illustrate in this chapter the language syntax and design of 
ModelWizard, a domain-specific language embedded in F\#.  
We cover core types, 
detail how to build operations safely with OpMonad,
survey our current API of operations 
and show how to use them in model construction.
Readers unfamiliar with F\# syntax may find \cite{syme2012expert} a useful reference.

\section{Types: State, Operation and Safety}
ModelWizard's core data type is \texttt{State}: a pair consisting of a dataset and a Tabular model annotating that dataset.  
We call the dataset \texttt{Data} and the Tabular model \texttt{Schema}.

\begin{lstlisting}
type TableName = string
type ColumnName = string
type DataTable = 
    {tablename:TableName; 
    colnames:ColumnName[]; 
    data:System.IComparable[,]} 
type Data = DataTable list
type Schema = //...
type State = Schema * Data
type StateOp<'R> = (State -> 'R * State)
\end{lstlisting}

\texttt{Data} is a relatively thin wrapper around an in-memory array of tables.  A table has a name, an ordered array of column names and a two-dimensional data array,\footnote{The \texttt{System.IComparable[,]} type on \texttt{Data} indicates \texttt{Data} values have a total order.
Do not think deeply into the total ordering; it is a pragmatic requirement that makes working with F\# libraries easier.

The total order may be unrelated to the actual meaning of data values.
For example, string values are compared lexicographically
such that `blue' ``is less than'' `red.'  
For modeling purposes we may still treat color as a nominal column, that is, 
one whose data values have no semantic total order and can be semantically compared only for equality.}
of which the first dimension indicates row number and the second dimension indicates column.
Missing values have value \texttt{null}.

\texttt{Schema} is analogous to the Tabular \texttt{Schema} type \cite{DBLP:conf/popl/GordonGRRBG14}.
Its first component is a representation of a classical relational database schema, composed of 
a list of named tables, each of which has a list of named and typed columns.
Its second component is annotations on columns with probabilistic model expressions
that define distributions over the data entries corresponding to the schema columns.
The Tabular compiler compiles \texttt{Schema}s into code in a probabilistic programming language, 
with which we run inference using the language's inference engine.
Thus, by manipulating \texttt{Schema} in ModelWizard, we manipulate the machine learning model 
that we use in inference.

\texttt{StateOp<'R>} is an operation on \texttt{State}, returning a new \texttt{State} plus additional information of type \texttt{'R}.  
\texttt{'R} could include information about the new state 
such as outlier data points or oddly fitting model components, 
information on other \texttt{State}s worth considering, 
or anything else the user writes in the \texttt{StateOp} function. 
The new \texttt{State} may be the same as the original \texttt{State}.

\subsection{Defining ``Valid'' States}
The \texttt{State} type by itself offers no guarantee that a State is valid.
Before we may provide any notion of validity, let alone the safe construction
of valid states, we must define what we mean by ``valid'' and ``safe.''

We call a State \textit{valid} when its component \texttt{Schema} 
is valid on its own and when the component \texttt{Schema} conforms to the State's 
component \texttt{Data} in the categories listed below.
\textit{Safety} refers to a process of constructing a state 
that preserves state validity, or informally, that 
``valid states in yield valid states out; 
garbage states in yield garbage."

The following categories outline the main kinds of \texttt{Schema} validity
and \texttt{Schema}-\texttt{Data} conformance that we check in the 
ModelWizard implementation:

\begin{enumerate}
\item Naming conflicts.  
Table names must be unique, column names must be unique within each table,
and table and column names present in a \texttt{Schema} must correspond to 
tables and columns in the \texttt{Data} it is paired with.
ModelWizard always checks for naming conflicts, especially when parsing 
arguments that create an operation and 
executing operations that create new columns or tables.

\item Type checking.
Modeled columns in a \texttt{Schema} 
(that is, columns with Tabular model expressions) 
must have a type that can be generated by the model on that column.
It would not make sense, for example, that a column 
with a \texttt{real} type have a \texttt{Discrete} model expression.
\texttt{Data} values must in turn match column types 
in the \texttt{Schema} they are paired with.
ModelWizard performs type checking at runtime when there is a chance that
data may not match a new type, explained further in Section~\ref{sLangTyping}.

\item Primary and foreign key correctness.
ModelWizard guarantees columns with the annotation `pk' 
are unmodeled and have no duplicate values, such that 
each value in a primary key column is unique.
Further, (foreign key) columns whose domain is that of another (primary key)
column may only assume values that are a subset of those in the primary key domain.
Section~\ref{sLangTyping} and~\ref{sExact} introduce the primitive operations 
\texttt{CreateTableUniques}, \texttt{Link} and \texttt{Exact}
that carefully check these requirements.

\item Cyclic dependencies.
Tabular requires that columns and tables only refer to previously
declared columns and tables.  We enforce this requirement in 
Section~\ref{sBookkeeping} via \texttt{Schema}-reordering operations.
\end{enumerate}

Checking state validity at runtime ensures that states are actually valid,
at the sacrifice of scalability to big data or big models.
While unimplemented, we lower barriers to scalability by running checks 
only when necessary, running checks lazily in background threads
or using other techniques discussed in Section~\ref{sBigData}.

The following notions of state validity are outside the scope of ModelWizard:

\begin{enumerate}
\item  
\begin{sloppypar} 
Detecting data valid according to type and syntax 
but questionable according to data likelihood.  
For example, an int-type column with values 
$\{1, 3, 1, 1, 2, 4, 1, 9001, 2, 3, 1\}$ 
is valid syntactically and type-wise, but the value 
9001 is a huge outlier worth questioning.  
However, it would be straightforward to write an operation 
that returns unlikely \texttt{Data} values.
\end{sloppypar}

\item Suggesting corrections to data values with syntax or type errors.
For example, suppose we have the data value `15B'.  
If present in a column containing with many other integers, 
we could propose that the `B' is a typo, accidentally inserted.  
We may also interpret the value as hexadecimal and propose 
replacing `15B' with the decimal `347'.  
With supporting evidence from other columns,
a third plausible alternative 
is that `15B' is the concatenation of two columns, 
such as an airline seat reservation with row number 15 
and seat position B for ``aisle.'' 

We refrain from heroically suggesting remedies for invalid states.  
More extensive data cleaning is the subject of active research, 
as seen in the DataTamer \cite{stonebraker2013data} 
and Potter's Wheel \cite{raman2001potter} systems among others.  
We would welcome integrating data cleaning techniques with ModelWizard 
in order to create a better end-to-end 
data analytics and model construction platform.

\item Detecting models valid according to type and syntax 
but questionable according to data likelihood.  
For example, nothing in our system prevents a user from placing 
a Gaussian distribution with mean 10,000 and variance 1 on a column with values 
$\{1, 5, 2, 2\}$.  
However, we do provide features that make model scoring and comparison easier, 
such as running inference, computing data likelihood scores 
and running cross-validation.  
We welcome contributions that perform model scoring in the background,
alerting the user if he performs an operation that leads to a very unlikely model.

\item Guaranteeing models are inferable by a particular back-end inference engine.
See Section 4.2 for a discussion on model inferability.
\end{enumerate}

Defining and enforcing validity as outlined above 
allows us to make the guarantee that
\textit{every model constructed by ModelWizard 
has a valid Tabular counterpart and will pass Tabular's type checking},
a strong enough guarantee to be useful in practice
but not so strong as to result in intractable model checking.

\subsection{ValidState and ValidOp Types}
We use the \texttt{ValidState} type, presented below, to indicate a \texttt{State} ModelWizard guarantees valid.
The below code is part of an F\# signature file, 
whose purpose is to declare the types an F\# library implementation file exports.
Because \texttt{ValidState}'s constructor is missing from the signature file,
there is no direct way to construct \texttt{ValidState}s 
except through other methods that handle \texttt{ValidState}s in a controlled manner, to guarantee preservation of \texttt{State} validity.
The function \texttt{unwrapVS} enables de-construction: 
retrieving encapsulated \texttt{State} from a \texttt{ValidState}.

\begin{lstlisting}
type ValidState  // Constructor hidden!!
val unwrapVS : ValidState -> State
val UNSAFE_ValidState : Schema * Data -> ValidState
\end{lstlisting}

For development convenience and expert users, we include an extra function \texttt{UNSAFE\_ValidState} which escapes \texttt{ValidState}'s safety guarantee and allows direct construction of \texttt{ValidState}s. 
Writing \texttt{UNSAFE} in all capitals flags users to be especially careful to only construct truly valid \texttt{ValidState}s when calling this function.

\texttt{ValidOp} is a type, presented below, for a \texttt{StateOp} 
guaranteed to preserve state validity.
In fact, \texttt{ValidOp}s throw exceptions when run in a way that would
create an invalid state and possibly corrupt the user's data. 
The user may run \texttt{ValidOp}s on \texttt{ValidState}s by \texttt{runValidOp} and on plain old \texttt{State}s by \texttt{unwrapVOP}.
\texttt{UNSAFE\_ValidOp} is similar to \texttt{UNSAFE\_ValidState}.

\begin{lstlisting}
type ValidOp<'R> // Constructor hidden!!
val runValidOp : ValidOp<'R> -> ValidState -> 'R * ValidState
val unwrapVOP : ValidOp<'R> -> StateOp<'R>
val UNSAFE_ValidOp : StateOp<'R> -> ValidOp<'R>
\end{lstlisting}

ModelWizard's principal workflow is to load a \texttt{ValidState} 
from an original \texttt{Data} source (such as an Excel workbook)
with a default ``do-nothing'' \texttt{Schema}, 
perform a series of \texttt{ValidOp}s that refine \texttt{ValidState} incrementally, and end with a final \texttt{ValidState} guaranteed valid.
We expect users will create many intermediary ModelWizard scripts before 
settling on a final script that creates a final model.
Intermediary scripts may return extra information \texttt{'R}, such as 
inference performance or the names of columns in the range of a functional dependency.
See Section~\ref{sExtendUsable} for a sketch of how we envision users editing scripts interactively.

\subsection{OpMonad: ValidOp Computation Expression}
\texttt{OpMonad} is an F\# computation expression class to create an intuitive syntax for chaining together \texttt{ValidOp}s into a compound, derived \texttt{ValidOp}.
As required of a proper monad, 
OpMonad respects the monad laws of left and right identity and associativity
\cite[Section 2.10]{wadler1992essence}.
Code Listing~\ref{fOpMonad} lists selected type signatures and implementations.


\begin{code}
\begin{lstlisting}
type OpMonad =
	member Bind:      ValidOp<'R> * ('R -> ValidOp<'N>)->ValidOp<'N>
	member Return:    'R -> ValidOp<'R>
	member Zero:      unit -> ValidOp<unit>
	member Combine:   ValidOp<unit> * ValidOp<'R> -> ValidOp<'R>
	member Delay:     (unit -> ValidOp<'R>) -> ValidOp<'R>
	member ReturnFrom:ValidOp<'R> -> ValidOp<'R>
	member For:       seq<'a> * ('a -> ValidOp<unit>)->ValidOp<unit>
val OPM : OpMonad 	  // shorthand for an OpMonad instance
// Implementation:
member __.Bind(vop, f) = VOP <| fun s -> 
								let res,newst = (unwrapVOP vop) s
								in  unwrapVOP (f res) newst 
member __.Zero()    		    = VOP <| fun s -> ((),s)
member __.Return(x) 		    = VOP <| fun s -> (x,s)
member __.ReturnFrom(x)         = x
member this.Combine(vop1, vop2) = this.Bind(vop1, fun () -> vop2)
member this.Delay(f)  			= this.Bind(this.Return (), f)
\end{lstlisting}
\caption{Selected \texttt{OpMonad} type signatures and implementations}
\label{fOpMonad}
\end{code}


The key notion to understand OpMonad is how \texttt{let!} statements 
translate to monadic bind. The following guidelines, in which \texttt{vop} 
is a placeholder for any \texttt{ValidOp}, outline the translation:
\begin{itemize}
\item \hspace{-0.5ex}\lstinline♠let!  r = vop♠
$\,$translates to \lstinline[mathescape=true]♠$ $let r,nextState = vop curState♠ %[
\item \lstinline♠do! vop♠ $\,$is a synonym for\lstinline♠let! () = vop♠ for a ValidOp encapsulating \lstinline♠StateOp<unit>♠.
\item \lstinline[mathescape=true]♠$ $return x♠ $\,$creates a ValidOp that returns x without modifying state. %[
\item \lstinline♠return!  vop♠ $\,$evaluates a ValidOp directly.
\item \lstinline♠|>♠ and \hspace{-1ex}\lstinline♠<|♠ are F\# operators for function application.
\end{itemize}
Please see \cite{petricek2012syntax} for a deeper review of monads as F\# computation expressions.
Refer also to Appendix~\ref{AOpMonadQuote} for an example translation of an OpMonad expression into 
an F\# quotation calling \texttt{OpMonad}'s computation expression components.


\section{Escape to F\#: GetState}
\texttt{GetState} links ModelWizard operations to F\# code
by returning a copy of an executing operation's current schema and data, 
exposing them to F\# code.
Operations may subsequently inspect column names, types and models 
as well as their underlying data.
Nearly every derived operation includes a call to \texttt{GetState}.

There is no corresponding \texttt{WriteState} operation except via \texttt{UNSAFE} operations. 
Including one would allow a user writing derived operations 
to easily create an invalid operation, destroying OpMonad's guarantees.
Thus, user code may inspect state but not alter state, 
except through ModelWizard's suite of primitive operations.

\section{Machine Learning Base Models}
\label{sBaseModels}
ModelWizard includes two base machine learning models: 
Gaussian and Discrete\footnote{Throughout this thesis, ``Discrete'' refers to a Categorical distribution, also known as a 
generalized Bernoulli distribution. The range of a Discrete distribution are the integers 1 to N for fixed distribution parameter N. An integer is sampled from a Discrete distribution with probability determined by an N-dimensional probability vector, stored as a distribution parameter.} distributions. 
Both have conjugate priors: Gamma on the precision of a Gaussian, 
another Gaussian on the mean of a Gaussian, 
and Dirichlet on the probability vector of a Discrete.
Conjugate priors allow inference of a Gaussian and Discrete distribution's parameters.

Tabular represents Gaussian and Discrete conjugate distributions 
by the column markup CGaussian and CDiscrete.
Figure~\ref{fCDist} presents the markup's meaning 
as a procedure to draw from their represented probability distributions.
\textit{N} and \textit{MeanPrec} are hyperparameters that may be specified 
in the Tabular model referencing the conjugate models, which affect the 
prior distribution from which conjugate prior distribution itself is drawn.
We typically assign uninformative priors to reduce bias by choice of prior in inference,
though one can assign informative priors just as easily.  
CGaussian has in fact, another hyperparameter MeanMean that is the table-level $\mu$ 
in Figure~\ref{fCDist} which comes into play when MeanPrec is sufficiently large.

\begin{figure}
\begin{subfigure}[t]{.43\linewidth} 
CDiscrete(N=\_):
\begin{enumerate}[leftmargin=*] \itemsep1pt \parskip1pt
\item Draw table-level $\vec{p}$ from an \\N-dimensional Dirichlet distribution with concentration $\vec{\alpha} = \vec{1}$. 
\item For each row, draw from a discrete distribution with probability vector $\vec{p}$.
\end{enumerate}
\end{subfigure}
\begin{subfigure}{.025\linewidth}
\end{subfigure}
\begin{subfigure}[t]{.545\linewidth}
CGaussian(MeanPrec=\_):
\begin{enumerate}[leftmargin=*] \itemsep1pt \parskip1pt
\item Draw table-level $\mu$ from a Gaussian distribution with mean 0 and precision (inverse of variance) MeanPrec.
\item Draw table-level $\tau$ from a Gamma distribution with shape and scale 1.0.
\item For each row, draw from a Gaussian distribution with mean $\mu$ and precision $\tau$.
\end{enumerate}
\end{subfigure}
\caption{Tabular Conjugate Distributions}
\label{fCDist}
\end{figure}

We create these distributions on a column, independent of all other columns, 
in ModelWizard by the primitive operations
\texttt{ModelDiscrete} and \texttt{ModelGaussian}.
\texttt{ModelDiscrete} may be performed on any column of type 
upto(N) or link(T). Link(T) columns are essentially upto columns with N set to the number of rows in table T.
\texttt{ModelGaussian} may be performed on columns of type 
real or int, converting the column to type real in the int case.
We enforce type constraints by throwing an exception if one operates
on a mismatching type.

To make it easier to independently model a column, 
we use the derived operation \texttt{Model} 
to place a ``default'' distribution appropriate for the column's type
by means of a call to \texttt{ModelDiscrete} or \texttt{ModelGaussian}.
Columns of type string, the raw text type, 
are handled by a \texttt{TypeNominal} operation described in the next section.

Future work may extend the ``two-level'' Gaussian and Discrete conjugate distributions to higher levels of hierarchical models. 
This can be done partially for the prior on the mean of a Gaussian distribution, since the prior is itself another Gaussian distribution.
A Gaussian distribution's precision (alternatively, variance) is more tricky.
As for the Dirichlet distribution, the prior of a Discrete distribution, placing a prior on a Dirichlet requires the nonparametric Dirichlet process \cite{teh2010dirichlet},
which is historically hard to perform variational inference on 
because the number of variational parameters is nonconstant.
Recent work has had more success \cite{blei2004variational}.

We also leave integrating other base distributions 
into ModelWizard as future work.
Good candidates are Gamma, Wishart, Poisson and Binomial, 
as these distributions all have existing representations in Tabular.
Multivariate Gaussian is a particularly excellent candidate as 
it creates a new way to couple continuous columns alongside 
polynomial regression.
We do not include the Beta and Bernoulli distributions as they are special cases of the Dirichlet and Discrete distributions, respectively.
It may also be worth investigating other prior distributions; 
for example, Gelman argues that a Gaussian distribution truncated to only place
probability mass on positive values tends to perform better as a prior on the 
variance of a Gaussian than the inverse-Gamma we use in CGaussian \cite{gelman2006}.

A challenge to additional base distributions 
is their level of inference support in Infer.Net.
Using other back-end inference engines 
such as Stan \cite{stan-software:2014} or R2 \cite{nori2014r2}
may add diversity to other base distributions' level of inference support.

\section{Typing Data, Tables as Nominal Column Domains}
\label{sLangTyping}
Columns in Tabular can have type int, real, link(T), upto(N) and string.
The domain of upto(N) is the integers $[0,N-1]$.
One can imagine a bool type as upto(2).
The domain of link(T) is all rows of table T.

\begin{sloppypar}
We change column types by the operations
\texttt{TypeUpto}, \texttt{TypeReal} and \texttt{TypeNominal}.
These operations are checked such that they will not place an improper type on a column, 
and they are only allowed to act on unmodeled input columns.
For example, \texttt{TypeReal} will check the data to ensure that the target column's data 
really are real-valued when there is a chance the data may not be, 
say when converting from type string as opposed to type int. 
\texttt{TypeUpto} and \texttt{TypeLink} act similarly.
\texttt{TypeInt} is also implemented but not discussed here or in the API 
as no machine learning models use the int type, taking upto or real in its place.
\end{sloppypar}


String types have special handling. One could use the former three type operations to convert a string column that contains numeric data, but it is more common that a string column has \textit{nominal} data, that is, data whose values are only comparable by equality like \{red, blue, red, green\}.

\begin{sloppypar}
We handle a nominal column C in table T 
with the primitive operation \texttt{CreateTableUniques T [C]}.
It creates a new Tabular table (call it NT) whose rows contain the unique values of column C.
Thus, NT's rows form the domain of C.
We bind the domain of C in T to NT by the operation \texttt{Link NT [C,C] T C},
which changes column C to type link.
Read the \texttt{Link} operation as 
``map column C in foreign table T to column C in primary table NT and store the corresponding row matches in foreign table T's column C.''
\end{sloppypar}

\begin{sloppypar}
Composing \texttt{CreateTableUniques} with \texttt{Link} 
forms the derived operation \texttt{TypeNominal}, 
an operation that effectively types a string column as a nominal column.
It is also possible to run \texttt{TypeNominal} on int and real columns, effectively treating the numbers in those columns as nominal labels instead of numeric values. 
Running \texttt{TypeNominal} on an upto column is redundant and disallowed.
\end{sloppypar}

A promising future direction is handling ordinal columns.
Ordinal data have ranking in addition to equality comparison,
like the data \{first, second, third\}.
One way to represent ordinal data is by logistic regression, 
supposing that ordinal values are delineated by continuous 
thresholds and then more easily inferring a continuous number and seeing which ``bucket'' between thresholds the continuous number lies in.
``Indicator'' columns perform the logistic regression in Tabular to indicate when a column is ``greater than'' its threshold value.
While we have built this model in Tabular, 
we have not yet represented it as an operation due to 
inference troubles related to algorithm support in Infer.NET.

\subsection{Type Inference}
Having access to the data at the time of modeling and
interleaving F\# code with operations enables typing automation.
We can create \texttt{TypeInfer}, a derived operation that infers the type of a column from its data.

We design \texttt{TypeInfer} to first apply a heuristic that columns
with less than 5\% unique values are likely to be nominal columns, because they 
contain significantly high redundancy. 
For example the data \{1, 1, 1, 0, 1, 0, 0\} is more likely nominal 
than continuous, with 0 and 1 indicating 
presence or absence of some label or property. 
Users may call \texttt{TypeReal} to override this heuristic.

If a column contains greater than 5\% unique values,
then \texttt{TypeInfer} attempts to type as an int,
then to type as a real if the data contains a non-integer,
then to type as a nominal as a last resort.

\begin{sloppypar}
We extend \texttt{TypeInfer} to automatically type all of a table's input columns
by the derived operation \texttt{TypeInferTable}.
We show a simplified version of \texttt{TypeInferTable}'s implementation
in the \texttt{OpMonad} expression of Appendix~\ref{AOpMonadQuote}.
\end{sloppypar}

\section{Coupling Columns}

The previous two sections introduce how to type columns and 
place base machine learning models on a column independent of all other columns.
We now consider coupling models of columns such that one column depends on another.

\subsection{Continuous Coupling: Noisy Regression}

Our prelude apple freefall example in Section~\ref{sAppleFreefallExample}
introduces \texttt{LinReg} and \texttt{QuadReg} as two operations to couple 
continuous columns by polynomial regression. 
The user specifies the domain and the range of
the regression as input to the operations.
Polynomial regression is a generic, ``default'' way to couple continuous
columns without special knowledge of the structure of a dataset.

We add noise to polynomial regression in order to place a non-point-mass 
distribution over the range column when the domain column is deterministic
(e.g., when the domain column is unmodeled with markup input).
The noise has fixed mean 0 so that we may infer the constant term properly
(`b' in linear regression, `c' in quadratic),
or else the noise and constant term would conflict in explaining regression bias.
Inferred noise variance corresponds to regression error magnitude.

\texttt{LinReg} and \texttt{QuadReg} rely on the function tables in Figure~\ref{fAppleLinRegModel} and Figure~\ref{fAppleQuadRegModel} defined above the main tables to implement regression in an abstracted manner.
Function tables use a special syntax in Tabular to implement a generic function 
that involves table-level inputs with default parameters (``hyper'' columns),
additional table-level columns determined by the inputs (``param'' columns),
concrete row-level inputs from a table (``input'' columns), 
additional row-level columns that depend on the inputs (``latent'' columns), 
and a row-level ``output'' column that returns a function's result 
to the table that calls it.
``Table-level'' refers to columns that have the same value for every row in a table, whereas ``row-level'' refers to columns whose value may differ for each row of data.
Tabular's compiler inserts function tables inline into the non-function tables that call them, a process similar to macro expansion. The insertion should occur in a way that does not conflict with names in the calling table.

Unfortunately, Tabular syntax forbids function tables that reference other function tables, recursively or not.
If we had the ability to conduct recursive calls, then we could 
implement polynomial regression in a recursive, more concise manner.
Imagine the operation \texttt{PolyReg N rngcol domcol} to do N-ary polynomial regression between domain column domcol and range column rngcol.
We may implement \texttt{PolyReg N} by adding an N\textsuperscript{th} degree term in the function and then recursively calling \texttt{PolyReg (N-1)} for the rest of the regression, down to \texttt{PolyReg 0} base case.
Even without the recursive implementation, 
it is possible to implement \texttt{PolyReg N} without individually using the operations \texttt{LinReg}, \texttt{QuadReg}, and so on by 
using a symbolic algebra algorithm to create an N\textsuperscript{th} degree
regression as a function table in F\# code, though we have not presently implemented this in ModelWizard.

The regression operations place a distribution over the range column
determined by the domain column, 
effectively meaning that the domain column value determines the range column.
If the domain column is modeled 
(i.e., has a probability distribution as opposed to an input column), 
then the regression on the range column distribution 
is a function of the domain column distribution.

\subsection{Discrete Coupling: Index}
Indexed models are a natural way to couple Discrete machine learning models.
We \texttt{Index} a (discrete or continuous) modeled range column's distribution 
by a discrete (modeled or unmodeled) dimension D domain column
by creating an array of 
copies of the range column's distribution, one for each of the D domain values,
and selecting the copy according to the domain column's corresponding value.
The copy distributions are identical to the original range column distribution,
except that the copies have their own parameters.

\begin{sloppypar}
The indexed range column may have any model.
We use the notation \texttt{Model[DomCol]} to indicate
indexing the range column's model by discrete domain column DomCol.
Section~\ref{sApplicationBayesianNetwork} and~\ref{sApplicationNaiveBayes} show models where a Discrete and Gaussian distribution, respectively, are indexed by a discrete variable.
It is also possible to index a \texttt{LinReg} or \texttt{QuadReg} model by a discrete variable. Such an indexing would make sense in the context of 
Section~\ref{sAppleFreefallExample}'s apple freefall example if we had data
for apples on different planets (each with their own force of gravity). 
We could index the regression on apple elevation by the planet the apple is on.
\end{sloppypar}

\begin{figure}
\centering
\begin{subfigure}[t]{.43\linewidth} 
\begin{tikzpicture}[auto,node distance=2.2cm,>=stealth',thick,
state/.style={minimum size=1.3cm,draw,circle,inner sep=1}]
\node [state] (cloudy) at (0,0) {cloudy};
\node [draw,inner sep=7,line width=1, 
fill=orange!20] (box) [right of =cloudy] {};
\node [state, node distance=1.2cm] (PVec) [above of =box] {PVec};


\node [state,inner sep=1] (Alpha) [above of =PVec, node distance=2.4cm] {Alpha};
\node [state] (rain) [below of =box] {rain};
\draw [dashed](1,2.25) node (v1) {} -- (3.5,2.25) -- (3.5,-0.75) -- (1,-0.75) -- cycle;
\draw [->] (cloudy) to (1,0);
\draw [->] (PVec) to node {} (box);
\draw [->,above left] (Alpha) to node {Dirichlet} (PVec);
\draw [->,below left] (box) to node {Discrete} (rain);
\end{tikzpicture}
\caption{Factor Graph with Gate}
\label{fGenderGates1}
\end{subfigure}
\begin{subfigure}[t]{.43\linewidth} 
\begin{tikzpicture}[auto,node distance=2.5cm,>=stealth',thick,
state/.style={minimum size=1.2cm,draw,circle,inner sep=1}]
\node [state] (cloudy) at (0,0) {cloudy};
\node [draw,inner sep=7,line width=1, 
fill=orange!20] (box1) [right of =cloudy] {};
\node [draw,inner sep=7,line width=1, 
fill=orange!20] (box2) [right of =box1] {};
\node (mid) at ($(box1)!0.5!(box2)$) {};
\node [state, node distance=1.5cm] (PVec1) [above of =box1] {$\text{PVec}_1$};
\node [state, node distance=1.5cm] (PVec2) [above of =box2] {$\text{PVec}_2$};
\node [state,inner sep=1] (Alpha) [above of =mid,node distance=3.2cm] {Alpha};
\node [state] (rain) [below of =mid, node distance=2.0cm] {rain};
\draw [dashed](3.75,2.25) node (v1) {} -- (6,2.25) -- (6,-0.5) -- (3.75,-0.5);
\draw [dashed](1.5,2.25) node (v1) {} -- (3.75,2.25) -- (3.75,-0.5) -- (1.5,-0.5) -- cycle;
\draw [->] (cloudy) -> (1.5,0);
\draw [->] (PVec1) to node {} (box1);
\draw [->] (PVec2) to node {} (box2);
\draw [->,above left] (Alpha) to node {Dirichlet} (PVec1);
\draw [->,above left] (Alpha) to node {} (PVec2);
\draw [->,below left] (box1) to node {Discrete} (rain);
\draw [->,below left] (box2) to node {} (rain);
\end{tikzpicture}
\caption{Expanded Factor Graph
}
\label{fGenderGates2}
\end{subfigure}
\caption{Factor graph of \texttt{CDiscrete(N=2)[cloudy]} on column rain}
\label{fGenderGates}
\end{figure}

Figure~\ref{fGenderGates1} graphically depicts one of the indexed distributions from Section~\ref{sApplicationBayesianNetwork}'s Bayesian network in terms of 
gates \cite{MinkaWinnNIPS2008_3379}
and plates \cite{buntine1994operations}.
PVec is the probability vector governing sampled outcomes of rain on each row.
Alpha is the concentration vector for the Dirichlet distribution used as the prior for the Discrete distribution.
Alpha is a vector of 1s when left unspecified.  
\texttt{N=2} sets the dimension of the vectors at 2, which is equal to the number of states of rain (a boolean-valued column, typed as upto(2)).

The dashed box is a plate, which means that its contents are replicated; there is a distinct PVec for every value of cloudy (also of type upto(2), though the dimension of the domain column need not equal the dimension of the range column in general).
Inputs to the plate (Alpha) are sent to each replicated PVec.
The square inside a box is a gate, which determines the output of the plate (rain) by the indexing variable cloudy.
Figure~\ref{fGenderGates2} shows the ``unfolded'' version of the gate and plate.

A helpful way to think about the notation is that there are distinct PVec parameters for the Discrete distribution governing rain.
The distribution of cloudy determines the weights with which we use each parameter.  Thus, we say that cloudy determines the distribution over rain.

In terms of operations, \texttt{Index} is a simple primitive operation.  
\texttt{Index} verifies that the domain column is discrete and the range column is modeled, throwing an exception if either condition does not hold, and replaces the range column's distribution by an indexed version.
Syntactically, however, \texttt{Index} is more complicated due to the occasional need to index distributions across table links. 
The user may specify a list of table links to traverse.
See Figure~\ref{fInfernoModel} in the Applications chapter for an example where movie ratings are indexed across a table link to a cluster column in a user table and a title table.

\texttt{Index} and \texttt{LinReg}/\texttt{QuadReg} offer venues for a discrete column to determine a discrete column, for a discrete column to determine a continuous column, and for a continuous column to determine a continuous column.
We do not have a way for a continuous column to determine a discrete column.
As mentioned in Section~\ref{sLangTyping} above, we may create a way in the future by implementing a logistic regression operation.
With that future addition, we would have ``default'' ways to couple columns of any type, a valuable ability for users experimenting with dependencies in a dataset.

\subsection{Bookkeeping: Reordering Tables and Columns}
\label{sBookkeeping}
With the power of the coupling operations \texttt{LinReg}, \texttt{QuadReg}
and \texttt{Index} to create dependencies between columns, 
we run the risk of creating cyclic dependencies: 
one column depending on another that transitively depends on the first.
Cyclic dependencies are inexpressible in Tabular; 
columns may only refer to columns declared above them in a Tabular model.
In other words, the overall model must fit into a directed acyclic graph.

We guard against cyclic dependencies by checking the set of columns that a column depends on at operation-runtime.
There are three scenarios.
If a dependency can be created without changing column order, it is done.
If a dependency can be created that requires changing column order, 
then a \texttt{ReorderColumns} operation is invoked internally.
If a dependency cannot be created due to cyclic dependencies (sometimes involving many columns), then an exception is thrown.

The same caution applies to tables.
The \texttt{Link} operation, for example, creates a table dependence that may require reordering tables.  
Cyclic table references are not allowed.

\section{Latent Columns}
A common modeling paradigm is to create latent columns not present in the original dataset in addition to the concrete columns present with data values.
Derived from concrete columns, 
latent columns expose intermediary states of probabilistic modeling to the user,
often increasing model understandability.
For example, latent columns inside the function tables of \texttt{LinReg} and \texttt{QuadReg} increase their interpretability by delineating the addition of noise from the regression.

\texttt{NewColumn} is an operation that may create an \texttt{UptoColumn} or a \texttt{LinkColumn}.  
Section~\ref{sApplicationInferno} demonstrates \texttt{UptoColumn}
by creating new cluster columns for tables. The number of clusters is given as an argument to \texttt{UptoColumn}.
We see these columns as output in Figure~\ref{fInfernoModel}, since we can treat new columns as concrete columns with all missing data.
Section~\ref{sInternetPlantSales} demonstrates \texttt{LinkColumn}
in the \texttt{ExactInfer} operation, where links are created from tables with range columns to tables with domain columns that exactly determine the range columns.

%

\section{Excel, Inference and Advanced Operations}
\label{sAdvancedOps}

Operations interfacing with Excel are essential 
to every application so that we can load and save data and models.
ModelWizard reads data from an Excel workbook data model 
using bindings in the .NET framework API.
Luckily, Excel's data model holds types on columns that are enforced
by Excel's processing model.
ModelWizard therefore trusts the type of a column indicated by Excel
by using an \texttt{UNSAFE\_ValidState} operation when constructing a state 
directly from an Excel workbook.
Writing data is similar, except that data is first written to a 
user-named Excel spreadsheet and then added to the containing workbook's data model.
Tabular models are read from and written to spreadsheets directly.


Inference operations call the Tabular compiler. 
We implement them by calling \texttt{GetState}, 
converting ModelWizard \texttt{Data} and \texttt{Schema} into Tabular \texttt{Data} and \texttt{Schema}
and executing the Tabular compiler to create an Infer.NET probabilistic program. 
The Infer.NET engine returns inference results, which we parse and return to the caller.

More advanced machine learning operations including \texttt{NaiveBayes}, 
\texttt{Exact}, \texttt{ExactInfer} and \texttt{Inferno}
are described in the next chapter: Applications.

\chapter{Applications}
\label{cApplications}
In this chapter we illustrate the model construction process with ModelWizard 
through progressively more complex examples.

\section{Small Model Search: Discrete Bayesian Networks}
\label{sApplicationBayesianNetwork}
\begin{figure}
\begin{subfigure}{0.34\linewidth}
\begin{tikzpicture}[
auto,shorten >=1pt,node distance=2.2cm,>=stealth',thick,
state/.style={draw,circle,text width={6.7ex},align=center,inner sep=0pt,font={\fontsize{7.8}{9.36}\selectfont}} 
]
\node[state] (cloudy) {Cloudy};
\node[state] (sprinkler) [below left of =cloudy] {Sprinkler};
\node[state] (rain) [below right of =cloudy] {Rain};
\node[state] (wetGrass) [below right of =sprinkler] {Wet Grass};
\draw [->] (cloudy) to node {} (sprinkler);
\draw [->] (cloudy) to node {} (rain);
\draw [->] (sprinkler) to node {} (wetGrass);
\draw [<->] (rain) to node [midway,fill=white,anchor=center,text height=1.0ex] {?} (wetGrass); 
\end{tikzpicture}
\end{subfigure}
\begin{subfigure}{0.66\linewidth}
  	\includegraphics[width=0.48\linewidth]{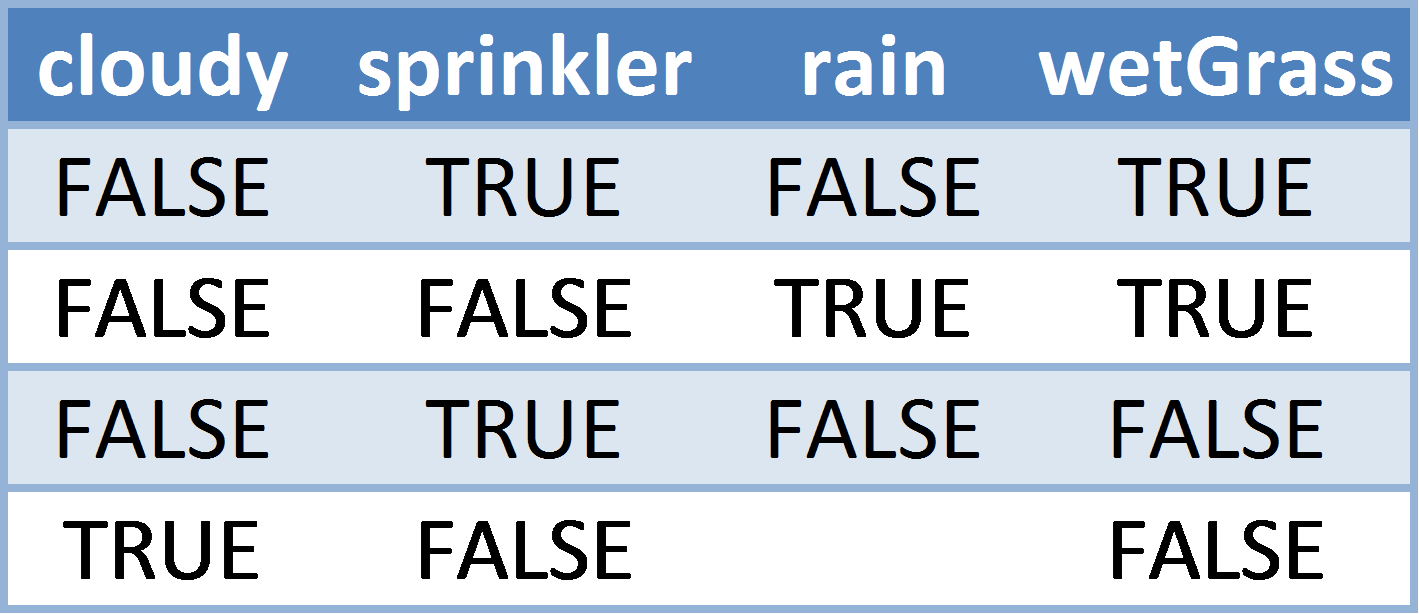}
    \includegraphics[width=0.48\linewidth]{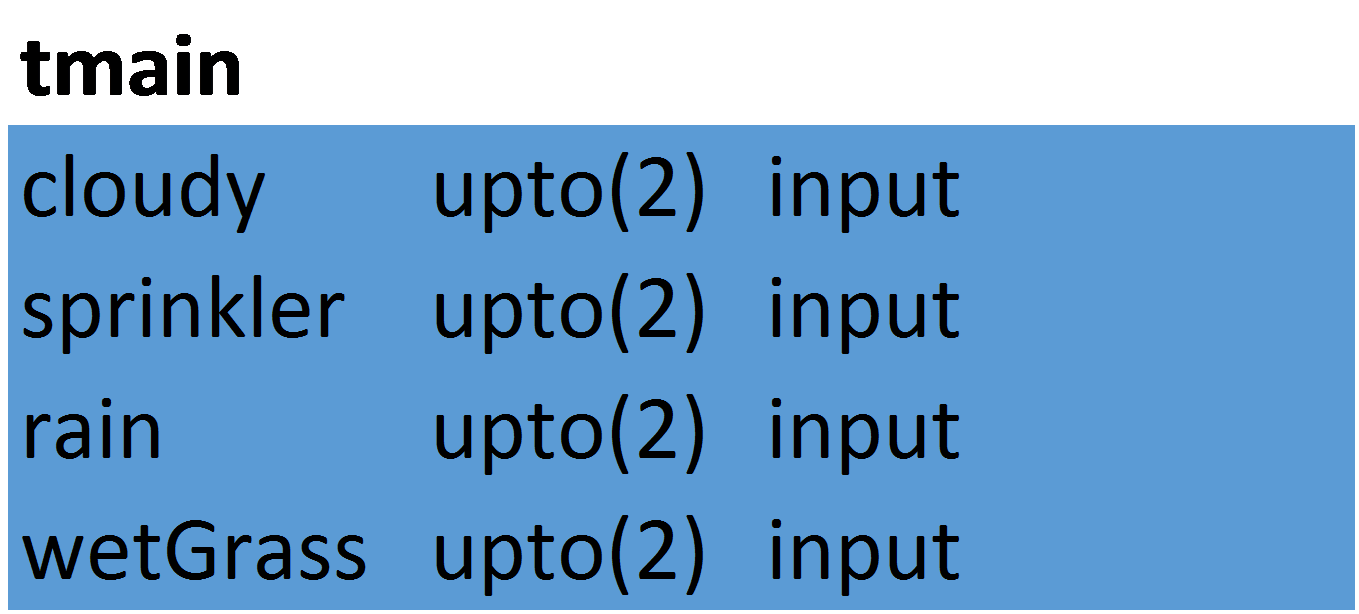}
  \caption{``No-predict'' default State: Data and Schema.\newline\noindent
  `upto(2)' is equivalent to `bool'.}
  \label{fDefaultBayNet}
\end{subfigure}
\caption{Uncertain Bayesian Network}
\label{fUBN}
\end{figure}
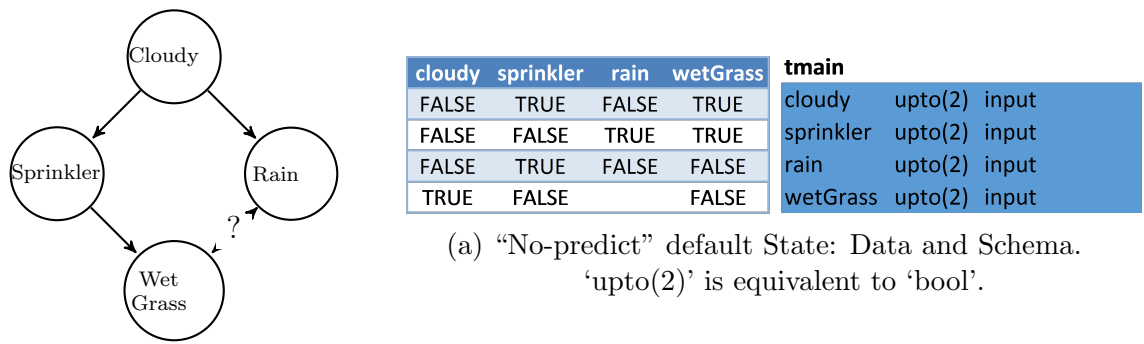

In this section we demonstrate how one constructs an acyclic discrete Bayesian network in ModelWizard, and how to search over and select different models using inference to score them.  
Our model search example is small, selecting between two Bayesian networks that differ in one arrow. One may extend it to larger search spaces 
bearing in mind computational feasibility.

Suppose we want to model Figure~\ref{fUBN}'s Bayesian network, modified from \cite{murphy2001introduction} adding uncertainty in the arrow between wet grass and rain. 
Does wet grass determine rain's distribution, rain determine wet grass's distribution, or are wet grass and rain independent? 
We answer by building and comparing each candidate model, using data likelihood 
as selection criterion.

\begin{figure}
\includegraphics[width=0.51\linewidth,clip=true,trim=0 0 150pt 0]{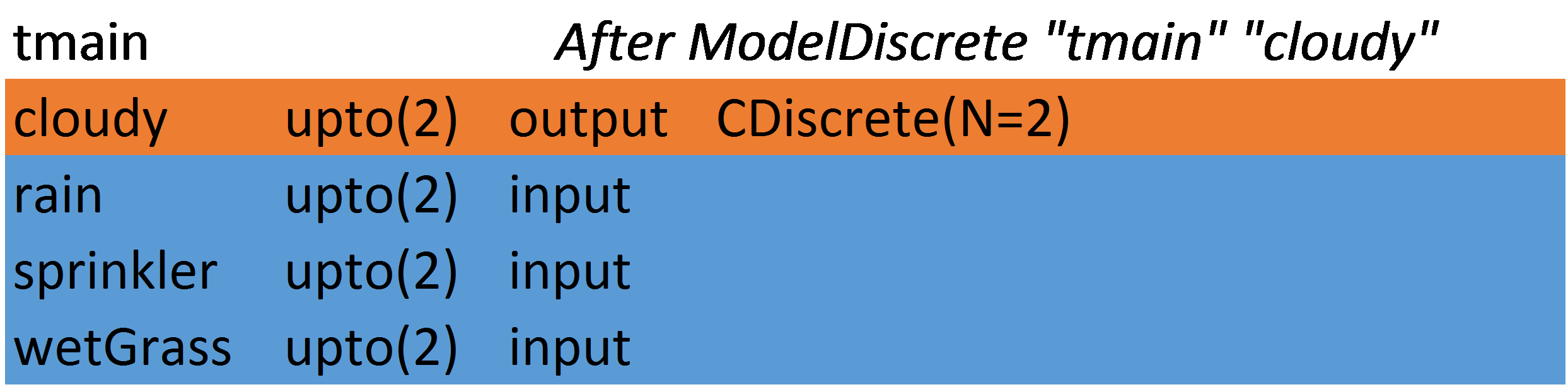}
\includegraphics[width=0.51\linewidth,clip=true,trim=0 0 150pt 0]{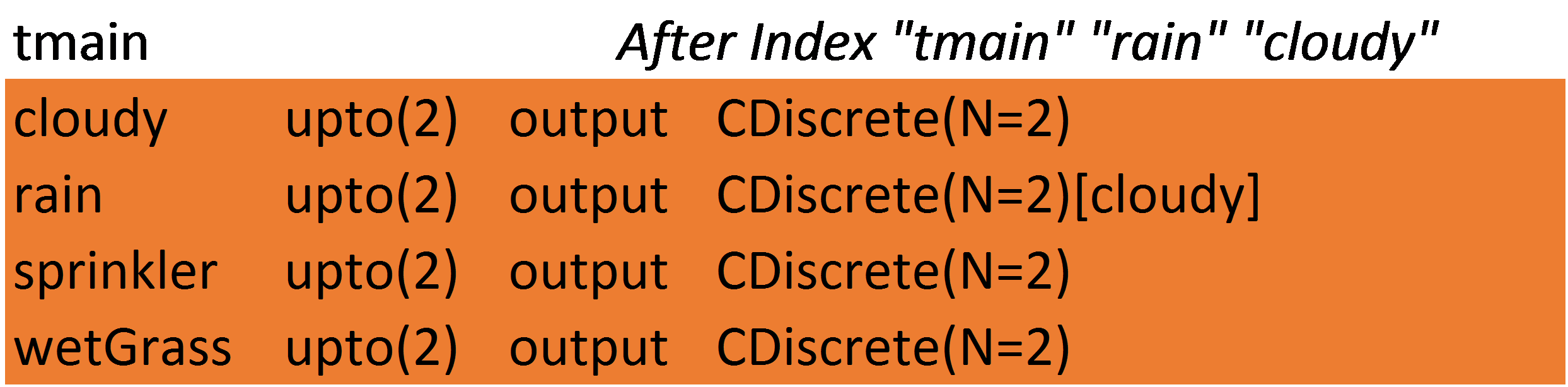}
\includegraphics[width=0.51\linewidth,clip=true,trim=0 0 150pt 0]{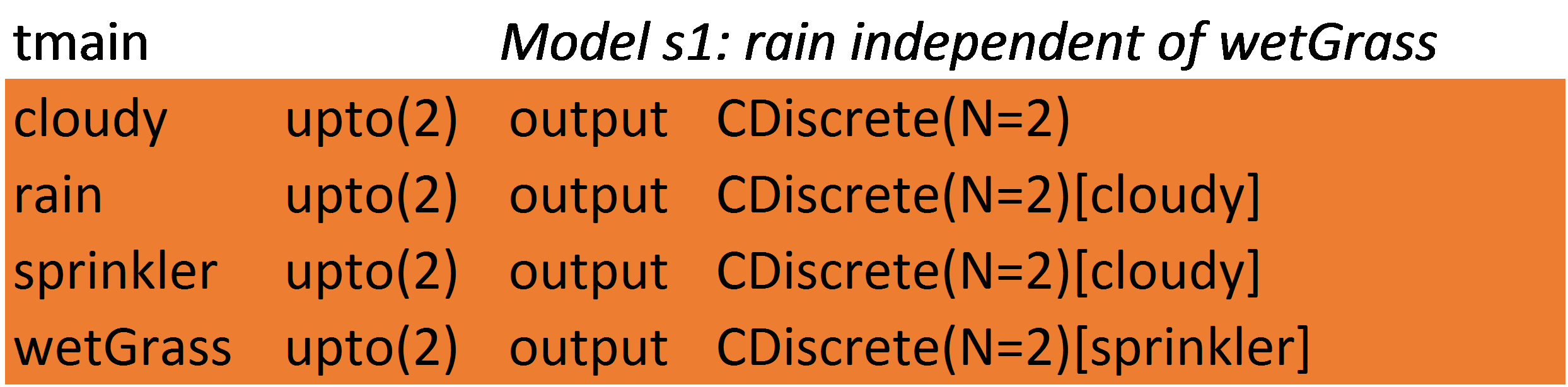}
\includegraphics[width=0.51\linewidth,clip=true,trim=0 0 152pt 0]{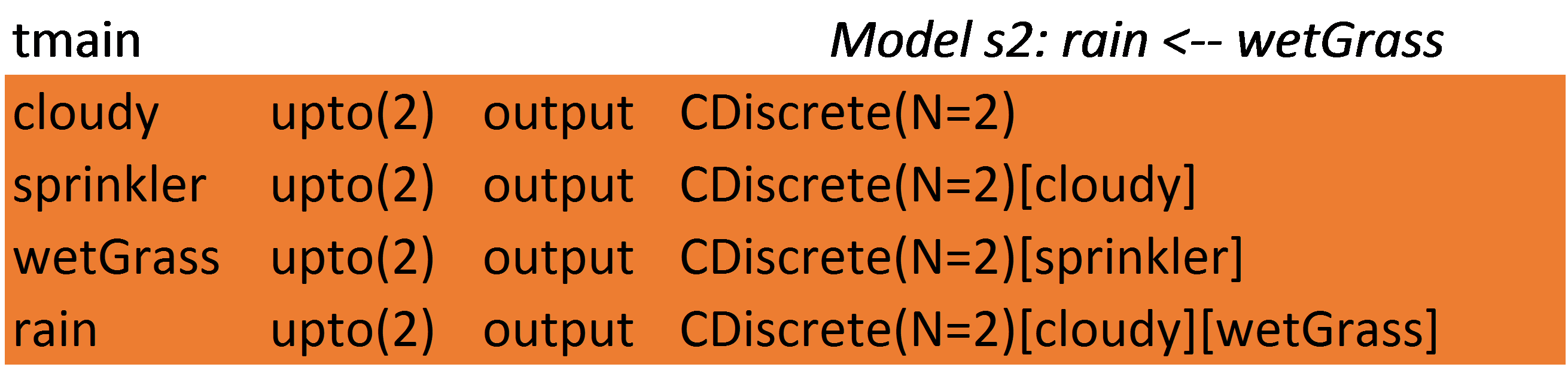}
\includegraphics[width=0.51\linewidth,clip=true,trim=0 0 150pt 0]{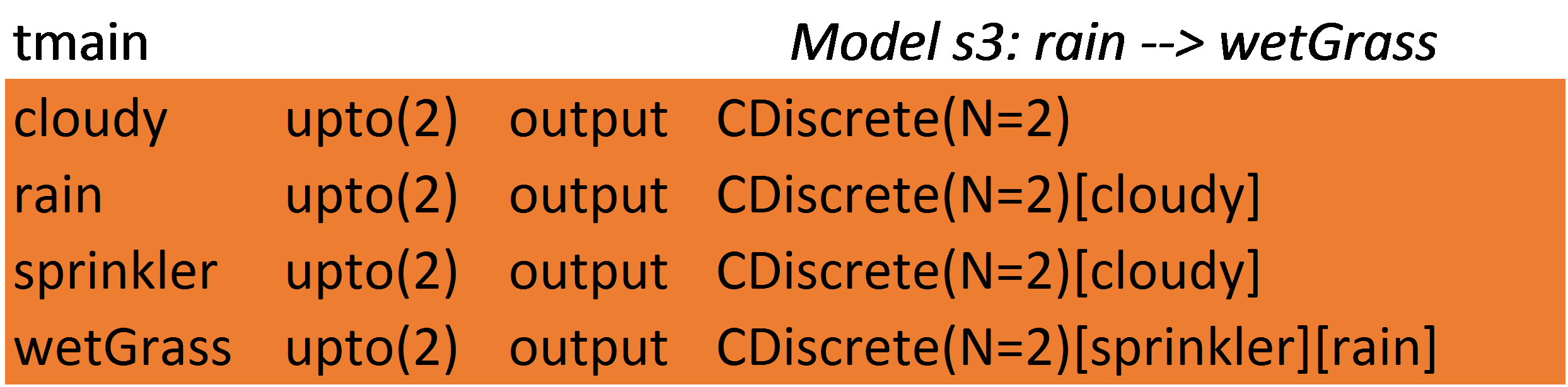}
\caption{Intermediate states during execution of Figure~\ref{fBayesNetSearch}}
\label{fBayesNetSearchStates}
\end{figure}

We begin with a default all-input model and dataset that performs no prediction, shown in Figure~\ref{fDefaultBayNet}, and walk through the operations 
in Figure~\ref{fBayesNetSearch} 
to construct a predictive model. 
Figure~\ref{fBayesNetSearchStates} tables intermediate models.

\begin{code}
\begin{lstlisting}[linerange=bayesnet_search]
namespace ModelWizard.Test

//#nowarn "20" // Eliminate compiler warnings for expressions that return a value that is discarded.  Used to avoid putting "ignore" everywhere.

open Xunit;
open Swensen.Unquote.Assertions // https://code.google.com/p/unquote/wiki/GettingStarted
//module W = ModelWizard.Drive.WizDrive //ModelWizard.Demo.WizLive

module TestDemoSimple =
    open ModelWizard.Merlin.DemoSimple
    open ModelWizard.OpMonad
    open ModelWizard.WizOpSimple
    open ModelWizard.WizOpCompound
    open ModelWizard.Drive.DMLoadState
    open ModelWizard.Excel.WriteOps
    open ModelWizard.Drive.CallTabular

    [<Fact>]
    let ``simple BayesNet_Rain`` () =
        let BayesNet_VS_init = readValidState BayesNet.BayesNet_file_input
        let (),BayesNet_VS_after = runValidOp BayesNet.BayesNet_Rain_op BayesNet_VS_init
        printfn "%A" (unwrapVS BayesNet_VS_after)

    [<Fact>]
    let ``simple BayesNet01`` () =
        let BayesNet01_VS_init = readValidState BayesNet01.BayesNet01_file_input
        let (),BayesNet01_VS_after = runValidOp BayesNet01.BayesNet01_Rain_op BayesNet01_VS_init
        printfn "%A" (unwrapVS BayesNet01_VS_after)

///    Score1: -2661.411073
///    Score2: -1963.565170
    [<Fact>]
    let ``simpleScore BayesNet_Rain`` () =
        let BayesNet_VS_init = readValidState BayesNet.BayesNet_file_input
        let le1,m1 = runValidOp BayesNet.BayesNet_Rain_step1 BayesNet_VS_init
        printfn "Score1: %f" le1
        let le2,m2 = runValidOp BayesNet.BayesNet_Rain_step2 m1
        printfn "Score2: %f" le2


    [<Fact>]
    let ``simpleScore reverse BayesNet``() =
        let inputFile = BayesNet01.BayesNet01_file_input
        //LATEX{ bayesnet_search
        let s0 = readValidState inputFile
        let le1,s1 =  
            s0 |> runValidOp (OPM {
            do! ModelDiscrete "tmain" "cloudy"
            do! ModelDiscrete "tmain" "rain"
            do! ModelDiscrete "tmain" "sprinkler"
            do! ModelDiscrete "tmain" "wetGrass"
            do! Index "tmain" "rain" "cloudy" []
            do! Index "tmain" "sprinkler" "cloudy" []
            do! Index "tmain" "wetGrass" "sprinkler" []
            return! ScoreLogEvidence}) //Ok:     -2286.50
        let le2,s2 =  
            s1 |> runValidOp (OPM {
            do! Index "tmain" "rain" "wetGrass" []
            return! ScoreLogEvidence}) //Better: -2128.71
        let le3,s3 =  
            s1 |> runValidOp (OPM {
            do! Index "tmain" "wetGrass" "rain" []
            return! ScoreLogEvidence}) //Best:   -1963.57
        let candidates= List.zip [le1;le2;le3] [s1;s2;s3]
        let bestle,bestmodel = List.maxBy fst candidates
        //LATEX} bayesnet_search
        // model: rain -|- wetGrass
        // model: rain <-- wetGrass
        // model: rain --> wetGrass

        printfn "candidates: %A" candidates
        printfn "best: %f\n%A" bestle bestmodel

//    [<Fact>]
//    let ``simpleScore reverse BayesNet 2``() =
//        let inputFile = BayesNet01.BayesNet01_file_input
//        //LATEX{ bayesnet_search
//        let le1,s1 = 
//            runValidOp (readValidState inputFile) <| OPM {
//            do! ModelDiscrete "tmain" "cloudy"
//            do! ModelDiscrete "tmain" "sprinkler"
//            do! ModelDiscrete "tmain" "rain"
//            do! ModelDiscrete "tmain" "wetGrass"
//            do! Index "tmain" "sprinkler" "cloudy" []
//            do! Index "tmain" "rain" "cloudy" []
//            do! Index "tmain" "wetGrass" "sprinkler" []
//            return! ScoreLogEvidence}  // Ok:     -2286.50
//        [   (le1,s1);
//            runValidOp s1 <| OPM {
//            do! Index "tmain" "rain" "wetGrass" []
//            return! ScoreLogEvidence}; // Better: -2128.71
//            runValidOp s1 <| OPM {
//            do! Index "tmain" "wetGrass" "rain" []
//            return! ScoreLogEvidence}  // Best:   -1963.57
//        ]
//        |> List.maxBy fst 
//        //LATEX} bayesnet_search




    [<Fact>]
    let ``simple NaiveBayes_HeightWeight three`` () =
        let NBHW_VS_init = readValidState NaiveBayes_HeightWeight.NBHW_file_input
        let (),NBHW_opManual_result = runValidOp NaiveBayes_HeightWeight.NBHW_opManual NBHW_VS_init
        printfn "%A" (unwrapVS NBHW_opManual_result)
        let (),NBHW_opNB_result = runValidOp NaiveBayes_HeightWeight.NBHW_opManual NBHW_VS_init
        NBHW_opNB_result =? NBHW_opManual_result
        let (),NBHW_opCore_result = runValidOp NaiveBayes_HeightWeight.NBHW_opCore NBHW_VS_init
        NBHW_opNB_result =? NBHW_opCore_result

        runValidOp NaiveBayes_HeightWeight.MySaveStateToWokbook NBHW_opNB_result

    [<Fact>]
    let ``simple ExactFDs_MadagascarPlants`` () =
        let EMP_VS_init = 
            readValidState ExactFDs_MadagascarPlants.EMP_file_input
            |> runValidOp (OPM {
                do! DeleteColumn "tmain" "redord_ID_"
                do! DeleteColumn "tmain" "conserve_concern"
            })
            |> snd
        let (),result = runValidOp ExactFDs_MadagascarPlants.EMP_op EMP_VS_init
        printfn "%A" (unwrapVS result)
    
    #if MRD
    module MR = ModelWizard.Merlin.DemoSimple.MovieRec_DeNormed
    [<Literal>]
    let MRS = "MRD"
    #else
    module MR = ModelWizard.Merlin.DemoSimple.MovieRec_Inferno
    [<Literal>]
    let MRS = "MR"
    #endif

    [<Fact>]
    let ``simple MovieRec`` () =
        let MRD_VS_init = readValidState MR.MRD_file_input
        let (),result = runValidOp MR.MRD_op MRD_VS_init
        printfn "%A" (unwrapVS result)

    [<Fact>]
    let ``CV Rating MovieRec``() =
//        use wr = new System.IO.StreamWriter("test.txt", false)
//        wr.AutoFlush <- true
//        System.Console.SetOut(wr)
//        System.Console.Out.Flush()
//        System.Diagnostics.Debug.AutoFlush <- true
//        ignore <| System.Diagnostics.Debug.Listeners.Add(new System.Diagnostics.TextWriterTraceListener(System.Console.Out))
        let MRD_VS_init = readValidState MR.MRD_file_input
        let score,result = runValidOp MR.MRD_opCV MRD_VS_init
        printfn "RMSE: %f\n%A\nRMSE: %f" score (unwrapVS result) score // 1.0145

    [<Fact>]
    let ``LEsweep Rating MovieRec``() =
        let MRD_VS_init = readValidState MR.MRD_file_input
        let score,result = runValidOp MR.MRD_opCVsweep MRD_VS_init
        printfn "LE: %f\n%A\nLE: %f" score (unwrapVS result) score // 

    [<Fact>]
    let ``LEsweepIter Rating MovieRec``() =
        let MRD_VS_init = readValidState MR.MRD_file_input
        let score,result = runValidOp MR.MRD_opCVsweepIterLE MRD_VS_init
        printfn "LE: %f\n%A\nLE: %f" score (unwrapVS result) score // 

    open ModelWizard.Drive.MultiCV
    open ModelWizard.Drive.SaveLoadSchemaFile

    [<Fact>]
    let ``CVsweepIter Rating MovieRec``() =
        let MRD_VS_init = readValidState MR.MRD_file_input
        let score,result = runValidOp MR.MRD_opCVsweepIterCV5 MRD_VS_init
        let (),_ =
            runValidOp
                (OPM { do! SaveSchema (__SOURCE_DIRECTORY__  + @"\..\ModelWizard.Demo\MovieRec_Inferno_output_schema.bin") })
                result
        printfn "RMSE: %f\n%ARMSE: %f" score (unwrapVS result) score // 

    open ModelWizard.WizOpCompound
    open ModelWizard.Drive.CrossValidate

    [<Fact>]
    let ``loadAndPlot LEsweepIter``() =
        let a = loadRes (MRS+"_LE.bin")
        printRes a
        plotRes a (MRS+"_LE")
    [<Fact>]
    let ``loadAndPlot CV5sweepIter``() =
        let a = loadRes (MRS+"_CV5.bin")
        printRes a
        plotRes a (MRS+"_CV5")

    module M = ModelWizard.Drive.MultiCV

    [<Literal>]
    let NUM_REPEATS = 5
    type MyChartProps =
        {ymin:int ;ymax:(int*int) option;ylog:bool;}
    let chartPropMap = 
        [   ("T_User","Age"),       {ymin= 0; ymax=Some (60,10); ylog=false};
            ("T_User","Gender"),    {ymin= 0; ymax=None; ylog=false}; 
            ("T_Title","Category"), {ymin= 0; ymax=None; ylog=false}; 
            ("T_Title","Year"),     {ymin= 1; ymax=Some (10000,0); ylog=true}; 
            ("tmain","Rating"),     {ymin= 0; ymax=Some (8,1); ylog=false}]
        |> Map.ofList

    [<Fact>]
    let ``missingSweep Inferno``() =
        let MRD_VS_init = readValidState (__SOURCE_DIRECTORY__  + @"\..\ModelWizard.Demo\MovieRec_Inferno_output_save.xlsx") 
        let (map), modWithOutputSchema = 
            runValidOp 
                (OPM { 
                    do! UNSAFE_LoadSchema (__SOURCE_DIRECTORY__  + @"\..\ModelWizard.Demo\MovieRec_Inferno_output_schema.bin") 
                    return! RunMissingDataAnalysis 
                                [("T_User","Age"); ("T_User","Gender"); ("T_Title","Category"); ("T_Title","Year"); ("tmain","Rating")]
                                [("T_User","Age"); ("T_User","Gender"); ("T_Title","Category"); ("T_Title","Year"); ("tmain","Rating")]
                                NUM_REPEATS // Number of Repeats to average for each data point
                })
                MRD_VS_init
        M.saveRes map (@"MovieRec_Inferno_output_MissingMap_"+NUM_REPEATS.ToString()+".bin")
        //Map.map M.plotRes map 

    let sgMap = SinghGraepelData.singhGraepelMap

    [<Fact>]
    let ``LOAD missingSweep Inferno``() =
        let map = M.loadRes (@"MovieRec_Inferno_output_MissingMap_"+NUM_REPEATS.ToString()+".bin")
        Map.map (fun tc datalist ->
            let {ymin=min; ymax=max; ylog=log} = Map.find tc chartPropMap
            match Map.tryFind tc sgMap with
            | None -> M.plotRes tc datalist NUM_REPEATS min max log
            | Some sgData -> M.plotResCompare tc datalist NUM_REPEATS min max log (List.ofSeq sgData)
        ) map





    let AFF_file_input  = __SOURCE_DIRECTORY__  + @"\..\ModelWizard.Demo\AppleFall_mockup_D2.xlsx"
    let AFF_file_output = __SOURCE_DIRECTORY__  + @"\..\ModelWizard.Demo\AppleFall_mockup_D2_output.xlsx"
    let AFF_SaveStateToWokbook = SaveStateToWokbook AFF_file_output ws_output_schema ws_output_data

    [<Fact>]
    let ``Apple LinRegTest``() =
        let (errorTime,errorElevation),result =
            readValidState AFF_file_input
            |> runValidOp (OPM {
                do! LinReg "tmain" "Elevation" "Time"
                do! Model "tmain" "Time"
                do! AFF_SaveStateToWokbook
                let! errorTime = CrossValidate_kFold_RMSE "tmain" "Time" 5
                let! errorElevation = CrossValidate_kFold_RMSE "tmain" "Elevation" 5
                return errorTime,errorElevation
            })
        printfn "errorTime     : %f\nerrorElevation: %f\n%A" errorTime errorElevation (unwrapVS result)

    [<Fact>]
    let ``Apple QuadRegTest``() =
        let (errorTime,errorElevation),result =
            readValidState AFF_file_input
            |> runValidOp (OPM {
                do! QuadReg "tmain" "Elevation" "Time"
                do! Model "tmain" "Time"
                do! AFF_SaveStateToWokbook
                let! errorTime = CrossValidate_kFold_RMSE "tmain" "Time" 5
                let! errorElevation = CrossValidate_kFold_RMSE "tmain" "Elevation" 5
                return errorTime,errorElevation
            })
        printfn "errorTime     : %f\nerrorElevation: %f\n%A" errorTime errorElevation (unwrapVS result)
    
    
\end{lstlisting}
\caption{
Script to build and select Figure~\ref{fUBN}'s most likely model}
\label{fBayesNetSearch}
\end{code}

A natural first step is to model each column independently, in this case using \texttt{ModelDiscrete} since all columns are nominal. 
\texttt{ModelDiscrete} allows inference of each nominal value's probability (see Section~\ref{sBaseModels} for details) and prediction of missing values, as with rain's missing value in Figure~\ref{fDefaultBayNet}. 
Recall that Tabular inference consists of training a given model on observed (non-missing) values in the input dataset to form a posterior model, outputting 
posterior model parameters that in this case form conditional probability tables,
and outputting marginal distributions from the posterior model on each missing value, conditioned on the observed values in the same row of that missing value.

\texttt{Index} correlates two columns. 
Taking \texttt{Index "tmain" "rain" "cloudy"} (read ``Index the distribution of rain by cloudy''), for example, \texttt{Index} creates separate distributions on rain for each value of cloudy, effectively creating a conditional probability table where rain depends on cloudy. Rain's Tabular model suffix `[cloudy]' indicates the new dependence.

Model candidates \texttt{s1}, \texttt{s2} and \texttt{s3} follow from further indexing. 
Rain and wetGrass are independent in model \texttt{s1},
rain indexes the distribution of wetGrass in model \texttt{s2},
and wetGrass indexes the distribution of rain in model \texttt{s3}.
We use the \texttt{ScoreLogEvidence} procedure to compile the current Tabular model into Infer.NET inference code, execute inference and return model log evidence scores under 1000 rows of input data, storing scores in \texttt{le1} through \texttt{le3}.  

An unusual paradigm in the example is the reuse of State \texttt{s1} in the construction 
of \texttt{s2} and \texttt{s3}.  Many monads are ``single-track'' in the sense that they 
disallow storing and reusing internal state because it 
could lead to undesirable operations, such as saving a pointer to a file, then deleting the file,
then attempting to access the file through the pointer. 
This kind of behavior is also disallowed in \texttt{OpMonad} since \texttt{OpMonad} expressions 
evaluate to a single ValidState.  We had to jump outside the \texttt{OpMonad} construct into general F\#
let bindings in order to save state \texttt{s1} and use it in the construction of two different states
bound at the top level.
We retain ``single-track'' enforcement in \texttt{OpMonad} for efficiency, since 
copying State in the current implementation entails copying \texttt{Data} within that State eagerly.
Section~\ref{sBigData} discusses delayed approaches to updating \texttt{Data}.

We find model \texttt{s3} is most likely because \texttt{le3} is greatest, 
fulfilling our intuition that rain more likely causes wet grass 
(assuming conditions necessary for causal interpretation; see \cite{Ko05}).
We also recover the same conditional probability tables (not shown) 
from Tabular inference as the Infer.NET program, 
proof that Tabular succeeded in learning the true model for this simple scenario.
In fact, we generated the 1000 rows of input data from an Infer.NET probabilistic program 
with model \texttt{s3} as ground truth and fixed conditional probability tables,\footnote{Adapted from \url{http://research.microsoft.com/en-us/um/cambridge/projects/infernet/docs/discrete\%20bayesian\%20network.aspx}} 
and so model \texttt{s3} is the true model.

We call our procedure interactive model search because the user 
specifies the search space through scripts like Figure~\ref{fBayesNetSearch}. 
The user may initiate a larger search, 
perhaps over the space of all Bayesian networks as an automated search algorithm would,
by searching over the presence and direction of every arrow. 
See \cite{cheng2001learning} for a survey of learning algorithms 
on several variants of Bayesian networks.
In short, ModelWizard opens a design space 
in which the user insightfully scripts ad hoc searches through model subspaces, 
writing in automated selection or prioritization as needed to 
search spaces large enough to be useful 
without resorting to infeasible, exhaustive search.

Model selection by data likelihood is prone to overfitting. 
In remaining examples we switch to the cross-validation inference scoring method: 
holding out a test set of data and measuring test set predictive accuracy 
according to a user-defined error function, such as the 0-1 loss function on predicting rain.

\section{Derived Model Family: Naive Bayes}
\label{sApplicationNaiveBayes}
We now illustrate data-transforming and derived operations. 
Suppose we want to construct a Naive Bayes model to classify gender given height, weight and footsize as features, for a dataset taken from \cite{wikiNaiveBayes}. 
Naive Bayes is a machine learning family of classifiers 
where a class column determines the distribution behind feature columns, 
and the features are assumed independent given class.

\begin{figure}
\centering
\includegraphics[width=.3\linewidth]{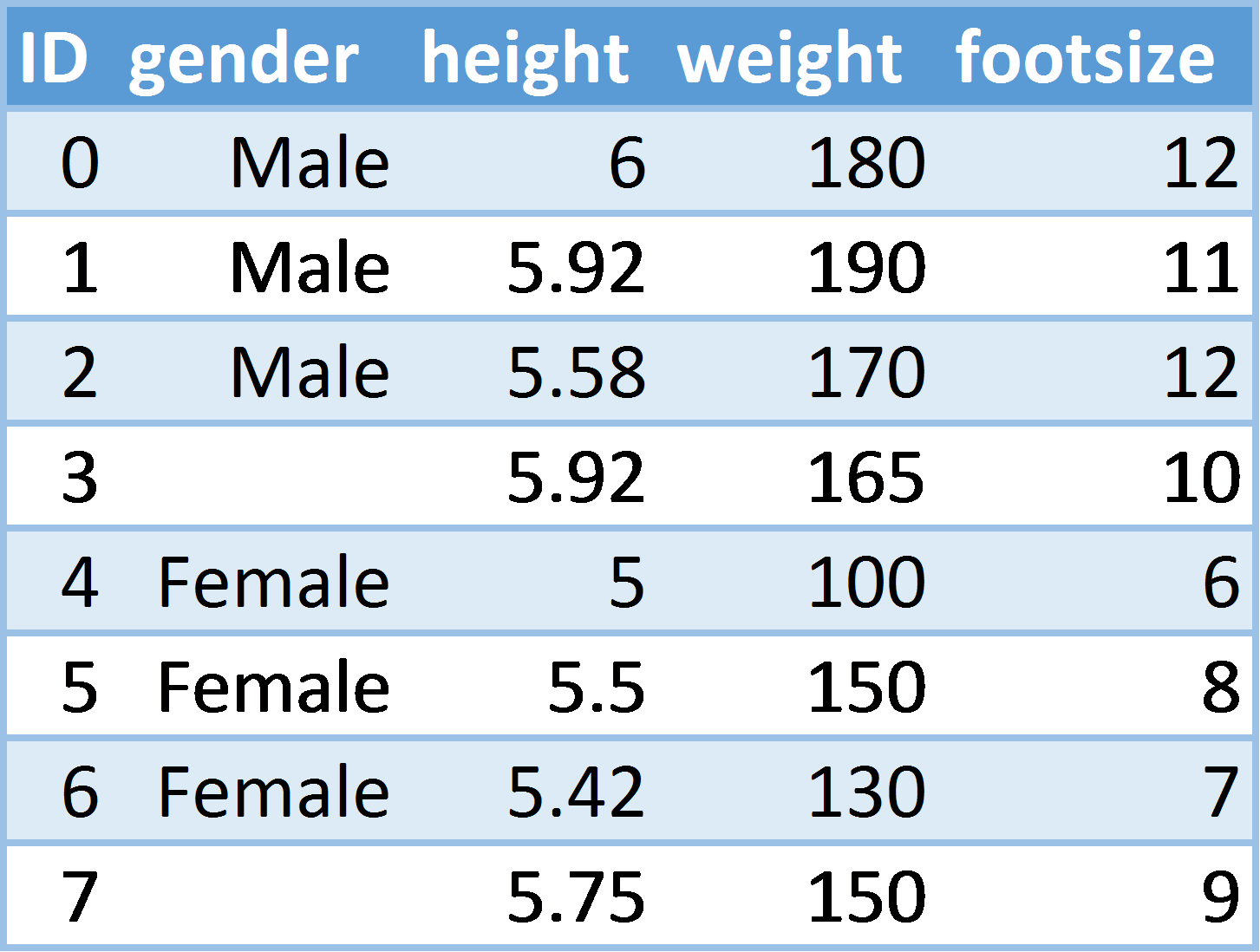} $\quad$
\includegraphics[width=.2\linewidth,clip=true,trim=0 0 100pt 0]{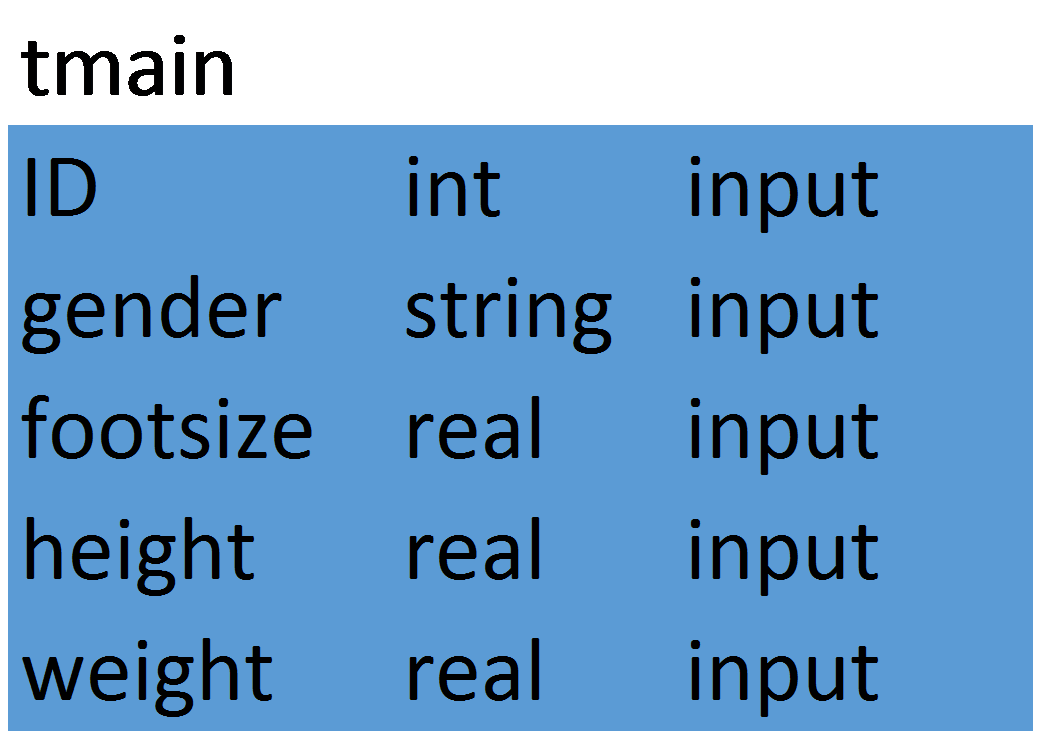}
\caption{Initial State: Data and Schema before Naive Bayes}
\label{fNBs0}
\end{figure}

Figure~\ref{fNBs0} shows our beginning all-input state. 
A new difficulty is that our data source has string type for gender, 
a type which Tabular cannot perform inference over. 
\texttt{CreateTableUniques} and \texttt{Link} in Code Listing~\ref{fNBPrim} overcome this difficulty, the former by creating a new table of gender's unique values called T\_gender, and the latter by changing tmain's data such that gender is an integer link to the appropriate ID column of T\_gender.  
The annotation `pk' in the gender column of T\_gender enforces that every row of T\_gender must have a unique value. Because T\_gender has two rows, Gender's new `link(T\_gender)' type is synonymous with `upto(2).'

\begin{code}
\centering
\begin{lstlisting}[linerange=nbhw_core]
namespace ModelWizard.Merlin

module DemoSimple =
    open ModelWizard.OpMonad
    open ModelWizard.WizOpCompound
    open ModelWizard.Excel.WriteOps
    open ModelWizard.Drive.DMLoadState
    open ModelWizard.Drive.CallTabular

    let ws_output_schema = "Tabular_Run"
    let ws_output_data = "datasheet"


    module BayesNet =
        let BayesNet_file_input = __SOURCE_DIRECTORY__  + @"\..\ModelWizard.Demo\wetGrassDemo.xlsx"
        let BayesNet_file_output = __SOURCE_DIRECTORY__  + @"\..\ModelWizard.Demo\wetGrassDemo_output.xlsx"
        //let BayesNet_VS_init = readValidState BayesNet_file_input
        let MySaveStateToWokbook = SaveStateToWokbook BayesNet_file_output ws_output_schema ws_output_data

        let BayesNet_Rain_op =
            OPM {
            do! TypeInferTable "tmain" 
            do! Model "tmain" "cloudy"
            do! Model "tmain" "sprinkler"
            do! Model "tmain" "rain"
            do! Model "tmain" "wetGrass"
            do! Index "tmain" "sprinkler" "cloudy" []
            do! Index "tmain" "rain" "cloudy" []
            do! Index "tmain" "wetGrass" "sprinkler" []
            do! Index "tmain" "wetGrass" "rain" []
            do! MySaveStateToWokbook
            }

        let BayesNet_Rain_step1 =
            OPM {
            do! TypeInferTable "tmain" 
            do! Model "tmain" "cloudy"
            do! Model "tmain" "sprinkler"
            do! Model "tmain" "rain"
            do! Model "tmain" "wetGrass"
            return! ScoreLogEvidence   //    Score1: -2661.411073
            }
        let BayesNet_Rain_step2 =
            OPM {
            do! Index "tmain" "sprinkler" "cloudy" []
            do! Index "tmain" "rain" "cloudy" []
            do! Index "tmain" "wetGrass" "sprinkler" []
            do! Index "tmain" "wetGrass" "rain" []
            return! ScoreLogEvidence   //    Score2: -1963.565170 = Improvement!
            }

        //let (), BayesNet_VS_after = printfn "%s" __SOURCE_DIRECTORY__; runValidOp BayesNet_Rain_op  BayesNet_VS_init

    module BayesNet01 =
        let BayesNet01_file_input = __SOURCE_DIRECTORY__  + @"\..\ModelWizard.Demo\wetGrassDemo01.xlsx"
        let BayesNet01_file_output = __SOURCE_DIRECTORY__  + @"\..\ModelWizard.Demo\wetGrassDemo01_output.xlsx"
        //let BayesNet01_VS_init = readValidState BayesNet01_file_input
        let MySaveStateToWokbook = SaveStateToWokbook BayesNet01_file_output ws_output_schema ws_output_data

        let BayesNet01_Rain_op =
            OPM {
            do! ModelDiscrete "tmain" "cloudy"
            do! ModelDiscrete "tmain" "sprinkler"
            do! ModelDiscrete "tmain" "rain"
            do! ModelDiscrete "tmain" "wetGrass"
            do! Index "tmain" "sprinkler" "cloudy" []
            do! Index "tmain" "rain" "cloudy" []
            do! Index "tmain" "wetGrass" "sprinkler" []
            do! Index "tmain" "wetGrass" "rain" []
            do! MySaveStateToWokbook
            }

        let BayesNet01_Rain_step1 =
            OPM {
            do! Model "tmain" "cloudy"
            do! Model "tmain" "sprinkler"
            do! Model "tmain" "rain"
            do! Model "tmain" "wetGrass"
            return! ScoreLogEvidence   //    Score1: -2661.411073
            }
        let BayesNet01_Rain_step2 =
            OPM {
            do! Index "tmain" "sprinkler" "cloudy" []
            do! Index "tmain" "rain" "cloudy" []
            do! Index "tmain" "wetGrass" "sprinkler" []
            do! Index "tmain" "wetGrass" "rain" []
            return! ScoreLogEvidence   //    Score2: -1963.565170 = Improvement!
            }

        

        //let (), BayesNet01_VS_after = printfn "%s" __SOURCE_DIRECTORY__; runValidOp BayesNet01_Rain_op  BayesNet01_VS_init

    module NaiveBayes_HeightWeight =
        let NBHW_file_input = __SOURCE_DIRECTORY__  + @"\..\ModelWizard.Demo\NaiveBayes_HeightWeight.xlsx"
        let NBHW_file_output = __SOURCE_DIRECTORY__  + @"\..\ModelWizard.Demo\NaiveBayes_HeightWeight_output.xlsx"
        //let NBHW_VS_init = readValidState NBHW_file_input
        let MySaveStateToWokbook = SaveStateToWokbook NBHW_file_output ws_output_schema ws_output_data

        let NBHW_opCore =
            OPM {
            //LATEX{ nbhw_core
            let! tn = CreateTableUniques "tmain" ["gender"]
            do! Link tn ["gender","gender"] "tmain""gender"
            do! ModelDiscrete "tmain" "gender"
            do! ModelGaussian "tmain" "height"
            do! ModelGaussian "tmain" "weight"
            do! ModelGaussian "tmain" "footsize"
            do! Index "tmain" "height"   "gender" []
            do! Index "tmain" "weight"   "gender" []
            do! Index "tmain" "footsize" "gender" []
            //LATEX} nbhw_core
            }

        let NBHW_opManual =
            OPM {
            do! TypeNominal "tmain" "gender"
            do! Model "tmain" "gender"
            do! Model "tmain" "height"
            do! Model "tmain" "weight"
            do! Model "tmain" "footsize"
            do! Index "tmain" "height"   "gender" []
            do! Index "tmain" "weight"   "gender" []
            do! Index "tmain" "footsize" "gender" []
        
            }

        let NBHW_opNB =
            NaiveBayes "tmain" "gender"  // single line; no need for do! notation
        
//        let (), NBHW_opManual_result = runValidOp NBHW_opManual  NBHW_VS_init
//        let (), NBHW_opNB_result = runValidOp NBHW_opNB  NBHW_VS_init
//        do assert (NBHW_opManual_result = NBHW_opNB_result)

        //ignore <| runValidOp MySaveStateToWokbook NBHW_opNB_result

    module ExactFDs_MadagascarPlants =
        let EMP_file_input = __SOURCE_DIRECTORY__  + @"\..\ModelWizard.Demo\ExactFDs_MadagascarPlants.xlsx"
        let EMP_file_output = __SOURCE_DIRECTORY__  + @"\..\ModelWizard.Demo\ExactFDs_MadagascarPlants_output.xlsx"
//        let EMP_VS_init = readValidState EMP_file_input
        let MySaveStateToWokbook = SaveStateToWokbook EMP_file_output ws_output_schema ws_output_data

        //do! Delete "tmain" "redord_ID_"
        let EMP_op =
            OPM { // ["conserve_priority"; "conserve_concern"; "On_CITES"]
            //LATEX{ EFDex
            do! TypeInferTable "tmain"
            let! efd_rngcols = ExactInfer "tmain" ["Genus";"Species"]
            do! NaiveBayes "tmain" "Wild_Propagate"
            for col in efd_rngcols do 
                do! Model "T_Genus_Species" col
            //LATEX} EFDex
            do! MySaveStateToWokbook
            }

        //let (), EMP_VS_after = runValidOp EMP_Rain_op  EMP_VS_init

    module MovieRec_DeNormed =
        let MRD_file_input = __SOURCE_DIRECTORY__  + @"\..\ModelWizard.Demo\MovieRec_DeNormed.xlsx"
        let MRD_file_output = __SOURCE_DIRECTORY__  + @"\..\ModelWizard.Demo\MovieRec_DeNormed_output.xlsx"
//        let MRD_VS_init = readValidState MRD_file_input
        let MySaveStateToWokbook = SaveStateToWokbook MRD_file_output ws_output_schema ws_output_data

        let MRD_op =
            OPM {
            //do! TypeInferTable "tmain" // bug with changing type of year, and we don't want to type age as Nominal
            for c in ["User";"IsMale";"Title";"Category"] do //;"Age";"Year";"Rating"] do
                do! TypeNominal "tmain" c
            for c in ["Age";"Year";"Rating"] do
                do! TypeReal "tmain" c
            let! _ = ExactInfer "tmain" ["User"]
            let! _ = ExactInfer "tmain" ["Title"]
            let! _ = Inferno "tmain"
            do! MySaveStateToWokbook
            }

        open ModelWizard.Drive.CrossValidate
        let MRD_opCV =
            OPM {
            //do! TypeInferTable "tmain" // bug with changing type of year, and we don't want to type age as Nominal
            for c in ["User";"IsMale";"Title";"Category"] do //;"Age";"Year";"Rating"] do
                do! TypeNominal "tmain" c
            for c in ["Age";"Year";"Rating"] do
                do! TypeReal "tmain" c
            let! _ = ExactInfer "tmain" ["User"]
            let! _ = ExactInfer "tmain" ["Title"]
            let! a = Inferno "tmain"
            printfn "Inferno Result: %A" (List.map fst a)
            return! CrossValidate_kFold_RMSE "tmain" "Rating" 5
            }
        
        let MRD_opCVsweep =
            OPM {
            //do! TypeInferTable "tmain" // bug with changing type of year, and we don't want to type age as Nominal
            for c in ["User";"IsMale";"Title";"Category"] do //;"Age";"Year";"Rating"] do
                do! TypeNominal "tmain" c
            for c in ["Age";"Year";"Rating"] do
                do! TypeReal "tmain" c
            let! _ = ExactInfer "tmain" ["User"]
            let! _ = ExactInfer "tmain" ["Title"]
            let! a = Inferno "tmain"
            printfn "Inferno Result: %A" a
            for (cnl,clusterFun) in a do
                printfn "Now Sweeping # of clusters on: %A" cnl
                let! (bestK,score,_) = sweepNumberClusters clusterFun ScoreLogEvidence true
                printfn "Best # of clusters on %A is %d with score %f" cnl bestK score
            return! ScoreLogEvidence
            }

        let MRD_opCVsweepIterLE =
            OPM {
            //do! TypeInferTable "tmain" // bug with changing type of year, and we don't want to type age as Nominal
            for c in ["User";"IsMale";"Title";"Category"] do //;"Age";"Year";"Rating"] do
                do! TypeNominal "tmain" c
            for c in ["Age";"Year";"Rating"] do
                do! TypeReal "tmain" c
            let! _ = ExactInfer "tmain" ["User"]
            let! _ = ExactInfer "tmain" ["Title"]
            let! infernoResult = Inferno "tmain"
            printfn "Inferno Result: %A" infernoResult
            let! _ = SaveDoAllSweep infernoResult ScoreLogEvidence true "MRD_LE"
            return! ScoreLogEvidence
            }

        let MRD_opCVsweepIterCV5 =
            OPM {
            //do! TypeInferTable "tmain" // bug with changing type of year, and we don't want to type age as Nominal
            for c in ["User";"IsMale";"Title";"Category"] do //;"Age";"Year";"Rating"] do
                do! TypeNominal "tmain" c
            for c in ["Age";"Year";"Rating"] do
                do! TypeReal "tmain" c
            let! _ = ExactInfer "tmain" ["User"]
            let! _ = ExactInfer "tmain" ["Title"]
            let! infernoResult = Inferno "tmain"
            printfn "Inferno Result: %A" infernoResult
            let! _ = SaveDoAllSweep infernoResult (CrossValidate_kFold_RMSE "tmain" "Rating" 5) false "MRD_CV5"
            return! CrossValidate_kFold_RMSE "tmain" "Rating" 5
            }


    module MovieRec_Inferno =
        let MRD_file_input = __SOURCE_DIRECTORY__  + @"\..\ModelWizard.Demo\MovieRec_Inferno.xlsx"
        let MRD_file_output = __SOURCE_DIRECTORY__  + @"\..\ModelWizard.Demo\MovieRec_Inferno_output.xlsx"
//        let MRD_VS_init = readValidState MRD_file_input
        let MySaveStateToWokbook = SaveStateToWokbook MRD_file_output ws_output_schema ws_output_data

        let MRD_op =
            OPM {
            do! TypeNominal "T_Title" "Category"
            do! TypeNominal "T_User" "Gender"
//            do! Link "T_User" ["ID","User"] "tmain" "User"
//            do! Link "T_Title" ["ID","Title"] "tmain" "Title"
            let! _ = Inferno "tmain"
            do! MySaveStateToWokbook
            }

        open ModelWizard.Drive.CrossValidate
        let MRD_opCV =
            OPM {
            do! TypeNominal "T_Title" "Category"
            do! TypeNominal "T_User" "Gender"
            let! a = Inferno "tmain"
            printfn "Inferno Result: %A" (List.map fst a)
            return! CrossValidate_kFold_RMSE "tmain" "Rating" 5
            }
        
        let MRD_opCVsweep =
            OPM {
            do! TypeNominal "T_Title" "Category"
            do! TypeNominal "T_User" "Gender"
            let! a = Inferno "tmain"
            printfn "Inferno Result: %A" a
            for (cnl,clusterFun) in a do
                printfn "Now Sweeping # of clusters on: %A" cnl
                let! (bestK,score,_) = sweepNumberClusters clusterFun ScoreLogEvidence true
                printfn "Best # of clusters on %A is %d with score %f" cnl bestK score
            return! ScoreLogEvidence
            }

        let MRD_opCVsweepIterLE =
            OPM {
            do! TypeNominal "T_Title" "Category"
            do! TypeNominal "T_User" "Gender"
            let! infernoResult = Inferno "tmain"
            printfn "Inferno Result: %A" infernoResult
            let! _ = SaveDoAllSweep infernoResult ScoreLogEvidence true "MR_LE"
            return! ScoreLogEvidence
            }

        let MRD_opCVsweepIterCV5 =
            OPM {
            do! TypeNominal "T_Title" "Category"
            do! TypeNominal "T_User" "Gender"
            let! infernoResult = Inferno "tmain"
            printfn "Inferno Result: %A" infernoResult
            let! _ = SaveDoAllSweep infernoResult (CrossValidate_kFold_RMSE "tmain" "Rating" 5) false "MR_CV5"
            do! MySaveStateToWokbook
            return! CrossValidate_kFold_RMSE "tmain" "Rating" 5
            }

        
\end{lstlisting}
\caption{Naive Bayes script expressed in primitives}
\label{fNBPrim}
\end{code}
\begin{figure}
\centering
\begin{subfigure}[b]{.09\linewidth}
\includegraphics[width=\linewidth]{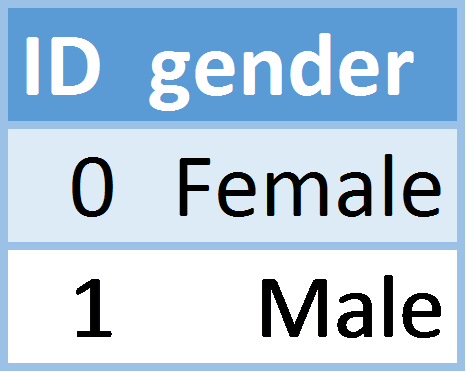}
\end{subfigure}
\hfil
\begin{subfigure}[b]{.27\linewidth}
\includegraphics[width=\linewidth]{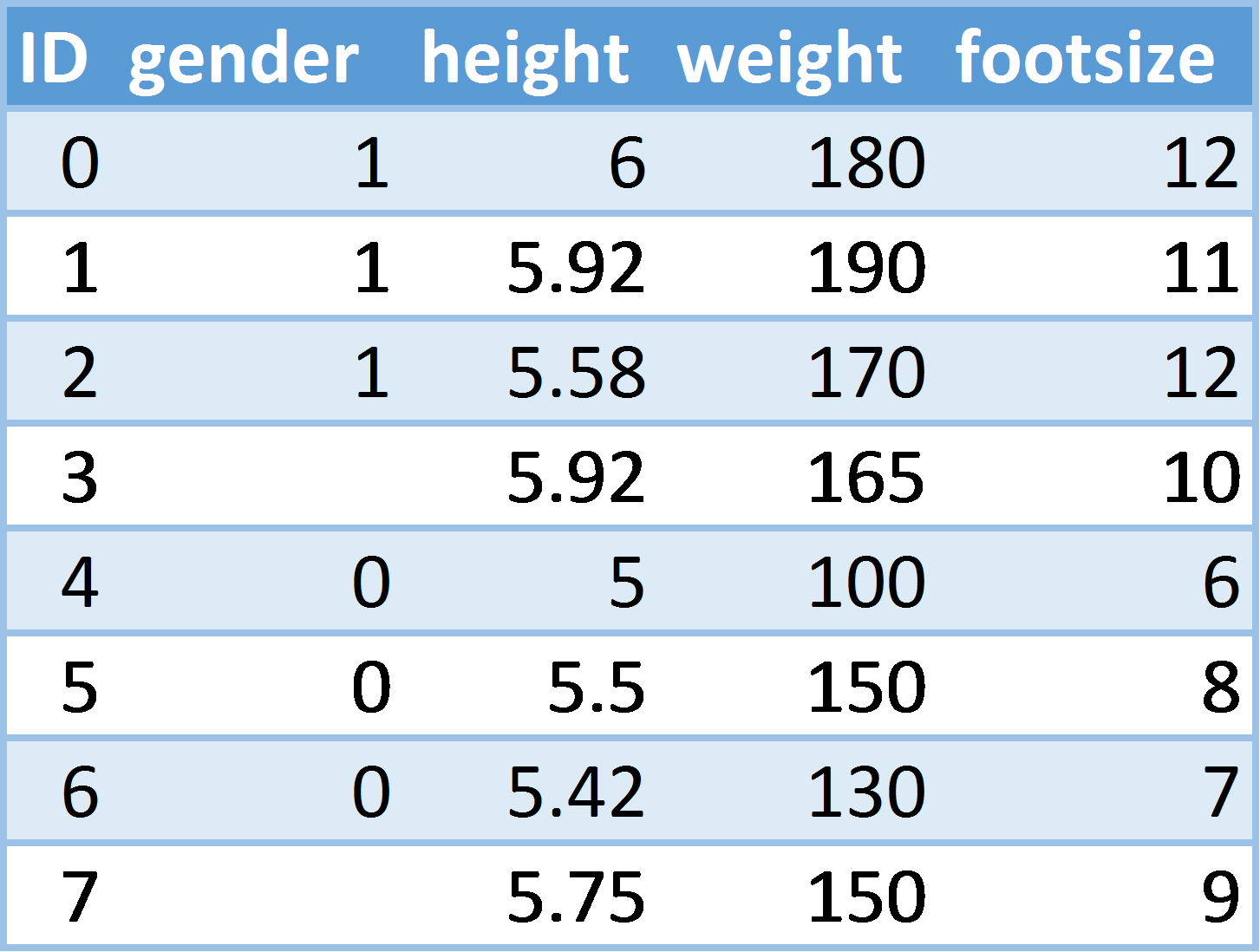}
\end{subfigure}
\hfil
\begin{subfigure}[b]{.61\linewidth}
\includegraphics[width=\linewidth,clip=true,trim=0 0 160pt 0]{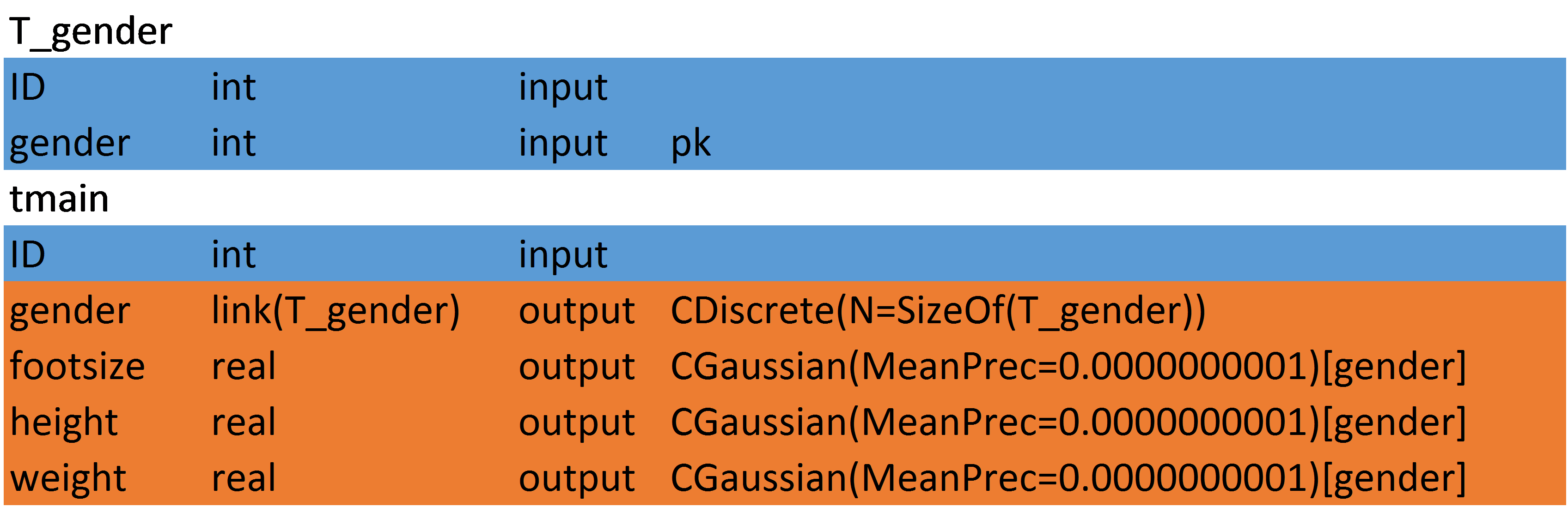}
\end{subfigure}
\caption{Naive Bayes result state after Code Listing~\ref{fNBPrim} runs on Figure~\ref{fNBs0}}
\label{fNBsf}
\end{figure}

\begin{code}
\begin{lstlisting}[linerange=naivebayes_derived]
namespace ModelWizard

module WizOpCompound =
    open Helper
    open OpMonad
    open WizOpSimple.SimpleHelper
    open WizardSchema
    open WizOpSimple
    open WizOpSimple

    type TypeDes = 
        | Nominal
        | Ordinal 
        | Real
        | Int
    
    

    /// Creates a Nominal table if it does not yet exist.  No effect if the column is already Nominal.
    let TypeNominal ttgt ctgt = 
        OPM {
        let! (Schema(schema) as schema0) = GetSchema
        let nominalTableName = "T_"+ctgt
        let table = getSchemaTableByName schema0 ttgt
        let column = getColumnByName table ctgt
        match column.ctype with
        | T_Link(link) when link = nominalTableName -> ()
        | _ ->  // check if table of unique values already exists
            match schema |> List.tryFind (fun decl -> decl.name = nominalTableName) with
            | None ->   let! _ = CreateTableUniques ttgt [ctgt]
                        ()
            | Some _ -> ()  // opLink will reorder tables if necessary
            do! Link nominalTableName [ctgt,ctgt] ttgt ctgt
        }
    (*
    //LATEX{ TypeNominal
    /// Create a new table of col's unique values and link col to it
    val TypeNominal: TableName -> col:ColumnName -> ValidOp<unit>
    //LATEX} TypeNominal
    *)


    /// THRESHOLD = 5% unique values.  Less than that means there is so much repetition we consider the column Nominal, even if it is all Int.
    [<Literal>]
    let THRESHOLD_NUMUNIQUE_NOMINAL = 0.05

    // Could return the new column type I suppose.
    let TypeInferColumn ttgt ctgt =
        OPM {
            let! schema,wdto = GetState
            let column = 
                let table = getSchemaTableByName schema ttgt
                getColumnByName table ctgt
            let coldata = 
                let tdto = getTDTOByName wdto ttgt
                let colidx = getDataColumnIndex tdto ctgt
                tdto.data.[*,colidx]
            // if there are a small number of unique values in the data set and they are all integer type, 
            // then call them Nominal
            let numUnique = coldata |> Seq.distinct |> Seq.length
            match coldata  |> Seq.filter ((<>) null) |> Seq.length with
            | 0 -> ()                                           // no data; do nothing
            | numNotNull ->
                match column.ctype with
                | T_Link tn when tn = "T_"+column.name -> ()    // already a Nominal column
                | _ ->
                    match (float numUnique) / (float numNotNull) with
                    | percentUnique when percentUnique < THRESHOLD_NUMUNIQUE_NOMINAL ->
                        do! TypeNominal ttgt ctgt
                    | _percentUnique ->
                        match column.ctype with
                        | T_Bool ->                             // change all Bool columns to Nominal
                            do! TypeNominal ttgt ctgt
                        | _ -> 
                            try do! TypeInt ttgt ctgt           // try typing the column as an int
                            with |_ ->
                                try do! TypeReal ttgt ctgt      // try typing the column as a real
                                with |_ ->
                                    do! TypeNominal ttgt ctgt   // when all else fails, type it Nominal   
        }
    (*
    //LATEX{ TypeInfer
    /// Examine col's data to infer its type
    val TypeInfer: TableName -> col:ColumnName -> ValidOp<unit>
    //LATEX} TypeInfer
    *)

    /// Infer the type of each Input column in the table
    let TypeInferTable ttgt =
        OPM {
            let! schema = GetSchema
            let table = getSchemaTableByName schema ttgt
            for col in  table.columns 
                        |> List.filter (function | {markup=Input} -> true | _ -> false) do
                do! TypeInferColumn ttgt col.name
        }
    (*
    //LATEX{ TypeInferTable
    /// TypeInfer each column in the table
    val TypeInferTable: TableName -> ValidOp<unit>
    //LATEX} TypeInferTable
    *)
    
    type ModelPrimitive =
        | Gaussian
        | Discrete
    
    let Model ttgt ctgt = 
        OPM {
            let! schema = GetSchema
            let tgtTable = getSchemaTableByName schema ttgt
            let col = getColumnByName tgtTable ctgt
            match col with
            | {ctype=T_Real }
            | {ctype=T_Int } ->     
                do! ModelGaussian ttgt ctgt
                //return Gaussian
            | {ctype=T_Upto _}
            | {ctype=T_Link _} 
            | {ctype=T_Bool }  ->   
                do! ModelDiscrete ttgt ctgt
                //return Discrete
            | col -> 
                doFail <| sprintf  "Cannot change to Discrete: column %A" (col) 
                //return Discrete
        }
        (*
        //LATEX{ Model
        /// Place an appropriate default model on a column based on its type
        val Model: TableName -> col:ColumnName -> ValidOp<unit>
        //LATEX} Model
        *)


    /// 1) Type the classifier column as Nominal and Model it.
    /// 2) For each ``free'' column in the table, Model it (if Input) and Index it by the classifier column.
    let NaiveBayes ttgt classname =
        OPM {
        do! TypeNominal ttgt classname
        do! Model ttgt classname
        let! schema = GetSchema
        let tgtTable = getSchemaTableByName schema ttgt
        let classcol = getColumnByName tgtTable classname
        let rngColList =    // collect columns in the range of Naive Bayes
            tgtTable.columns |> List.filter (fun col ->
                match (col,classcol) with
                | a,b when a.name = b.name -> false     // skip over classcol
                | {markup=BoundLink(_,_)},_           
                | {markup=Hyper _},_ -> false           // do not include BoundLinks or Hypers in the range
                | {name=name},_ when name=sID -> false  // do not include the ID column in the range
                | IndependentOf ->   // only include columns that have no preexisting relationship with classcol
                    true // assert reordering is possible
                | _ -> false
            )
        // for each column in rngColList, model it and index by classifier column
        for col in rngColList do
            if col.markup = Input
            then do! Model ttgt col.name
            do! Index ttgt col.name classname []
        }

    module mockup =
        //LATEX{ naivebayes_derived
        let NaiveBayes tn classname = OPM {
            do! TypeNominal tn classname
            do! ModelDiscrete tn classname
            let! schema,_ = GetState
            let table = getSchemaTableByName schema tn
            let classcol=getColumnByName table classname
            let rngColList=  // collect feature columns
                table.columns |> List.filter (fun col ->
                    match (col,classcol) with
                    | a,b when a.name = b.name -> false
                    | {name=n},_ when n="ID"   -> false
                    | IndependentOf -> true
                    | _             -> false)
            // Model & Index features by classcol
            for col in rngColList do
                if col.markup = Input
                then do! Model tn col.name
                do! Index tn col.name classname []}
        //LATEX} naivebayes_derived
        (*
        //LATEX{ NaiveBayes
        /// Construct Naive Bayes model taking 'free' columns of tn as features.
        val NaiveBayes: tn:TableName -> classname:ColumnName -> ValidOp<unit>
        //LATEX} NaiveBayes
        *)


    (*
    //LATEX{ ExactInfer
    /// Find, capture & return columns exactly determined by the column list
    val ExactInfer:TableName->domCols:ColumnName list->ValidOp<ColumnName list>
    //LATEX} ExactInfer
    *)
    /// Infer and return the range columns exactly determined by the domain columns.
    // Will call CreateTableUniques if necessary.
    let ExactInfer (ttgt:TableName) (cdomlist:ColumnName list) :ValidOp<ColumnName list> =
        let cdomarr = (Array.ofList cdomlist)
        OPM {
            let! Schema(schemain) as schema,data = GetState
            let tgtTable = getSchemaTableByName schema ttgt
            let tdtoTarget = getTDTOByName data ttgt
            match WizOp.getBestFD_Restricted_WithDomain tgtTable tdtoTarget cdomarr with
                | None -> return []
                | Some (domset, rngset, _fdmap) -> 
                    let domNamesArr = tdtoTarget.colnames |> Array.subarray domset 
                    let domNames = domNamesArr |> List.ofArray
                    System.Diagnostics.Debug.Assert(domNamesArr |> Array.forall (fun dn -> Array.exists ((=)dn) cdomarr))
                    let rngNamesArr = tdtoTarget.colnames |> Array.subarray rngset 
                    let rngNames = rngNamesArr |> List.ofArray

                    // check to see if domNames is singleton. 
                    //      Yes ==> check to see if we have a table for it. No => Type it Nominal. Yes => NoOp
                    //      No  ==> CreateTableUniques on the concatenation if it doesn't already exist.
                    let! domlinkname =
                        OPM {
                        if Array.length domNamesArr = 1
                        then
                            let dn = domNames.[0]
                            match schemain |> List.tryFind (fun decl -> decl.name = dn) with
                            | None -> 
                                do! TypeNominal ttgt dn
                                return dn
                            | Some _ -> return dn
                        else
                            // check to see if we already have the combined table. If not create it.
                            let! combotablename = 
                                OPM {
                                let defaultname = "T_" + (smashIntoString "_" domNamesArr)
                                match schemain |> List.exists (fun decl -> decl.name = defaultname) with
                                | false -> 
                                    let! tn = CreateTableUniques ttgt domNames
                                    return tn
                                | true -> return defaultname
                                }
                            // And make a new column for the new table
                            let! newlinkname = NewColumn ttgt (combotablename.[2..]) (ModelWizard.WizOpSimple.LinkColumn combotablename)
                            // And link the new table to the original
                            let domMap = domNames |> List.map (fun n -> n,n)
                            do! Link combotablename domMap ttgt newlinkname
                            return newlinkname
                        }

                    // Now call out each Exact.
                    for rn in rngNamesArr do
                        do! Exact ttgt domlinkname rn
                    return rngNames
        }

    /// Automatic table normalization. Be careful what you wish for.
    let ExactInferTable (ttgt:TableName) :ValidOp<unit> =
        failwith "nyi"



    /// Get a list of the table names that ttgt has columns linking to them.
    let GetNontrivialLinkedTableNames ttgt =
        OPM {
            let! schema,_ = GetState
            let tgtTable = getSchemaTableByName schema ttgt
            return tgtTable.columns |> List.collect (fun col -> 
                match col.ctype with 
                | T_Link tn -> 
                    let rt = getSchemaTableByName schema tn
                    if rt.columns |> List.exists (fun c -> c.name <> sID && not c.isPrimaryKey)
                    then [col.name,tn] 
                    else []
                | _ -> [])
        }

//    /// Get a list of columns that are link type
//    let GetLinkColumns ttgt =
//        OPM {
//            let! schema,_ = GetState
//            let tgtTable = getSchemaTableByName schema ttgt
//            return tgtTable.columns |> List.filter (fun col -> match col.ctype with |T_Link tn -> true | _ -> false)
//        }
    
    [<Literal>]
    let sNumberClusters = "NC" :ColumnName
    [<Literal>]
    let sCluster = "cluster" :ColumnName

    let GetConcreteColumns ttgt = OPM {
        let! schema,_ = GetState
        let tgtTable = getSchemaTableByName schema ttgt
        let freeCols =  
            tgtTable.columns 
            |> List.filter (
                function
                | {name=n} when n="ID" || n=sCluster   -> false
                | {isPrimaryKey=true} -> false
                //| {markup=Input}   -> false
                | {markup=Hyper _}
                | {markup=Param _}
                | {markup=BoundLink _} -> false
                | {ctype=T_Link tn} -> 
                    let rt = getSchemaTableByName schema tn
                    rt.columns |> List.exists (fun c -> c.name <> sID && not c.isPrimaryKey) |> not
                | _                -> true )
            //|> List.map (fun c -> c.name)
//            for c in freeCols do
//                if c.markup = Input
//                then do! Model ttgt c
        return freeCols
    }

    /// Uses NewColumn and Index to apply the Inferno pattern recursively.
    /// Returns a list of column links from this table; plus a function to change the # of clusters for each linked cluster table.
    /// Base case (leaf table): cluster the table with 4 clusters.
    /// Recursive case (interior table): cluster determined by product of all tables linked to. Index free variables by cluster.
    let rec Inferno (ttgt:TableName) :ValidOp<(ColumnName list * (int -> ValidOp<unit>)) list> =
        OPM {
            /// first-order links
            let! links = GetNontrivialLinkedTableNames ttgt
            let! linkLists = OPM {
                match links with
                | [] -> // base case
#if USE_HYPER
                    let! colNC = NewColumn ttgt sNumberClusters (WizOpSimple.NewColType.HyperIntColumn 4)
                    let! clusterColName =  NewColumn ttgt sCluster (WizOpSimple.NewColType.UptoRefSumColumn [colNC,[]])
                    return [[clusterColName]]
#else
                    let! clusterColName = NewColumn ttgt sCluster (NewColType.UptoColumn 4)
                    let modFun k = TypeUpto k ttgt clusterColName
                    return [[clusterColName], modFun]
#endif
                | _ ->  // recursive case
                    // P2: handle naming nicer
                    let linkLists = ref []
                    for (c,t) in links do
                        let! ll = Inferno t
                        // preprend the column linking to the cluster linkList of t
                        linkLists := (List.map (fun (m,vop) -> c::m,vop) ll) @ !linkLists
                    // get the tables we link to; allow duplicate links to same table; might have semantic differences, e.g. goodGuy to People, badGuy to People, use both for clustering
                    //let remoteNCs = links |> List.map (fun (c,_) -> c, [sNumberClusters])
                    // index colK by the cluster of each table we link to
                    //let remoteKs =  links |> List.map (fun (c,_) -> c, [sCluster])
                    //let! colNC = NewColumn ttgt sNumberClusters (WizOpSimple.NewColType.HyperRefSumColumn remoteNCs)
                    //let! colK = NewColumn ttgt sCluster (WizOpSimple.NewColType.UptoRefSumColumn [colNC,[]])
//                    for (lc,list) in remoteKs do
//                        do! Index ttgt colK lc list
                    return !linkLists
            }
            // gather free columns to index by colK
            let! rngColList=  GetConcreteColumns ttgt
            for c in rngColList do
                if c.markup = Input
                then do! Model ttgt c.name
                for ll in linkLists do
                    match ll with
                    | llhead::llbody,_ ->
                        do! Index ttgt c.name llhead llbody
                    | [],_ -> failwith "impossible"
            return linkLists
        }
        (*
        //LATEX{ InfernoOp
        /// Construct Inferno clustering model, recursing on linked tables.
        /// Return a list of column links and functions to modify # of clusters
        val Inferno:TableName->ValidOp<(ColumnName list*(int->ValidOp<unit>)) list>
        //LATEX} InfernoOp
        *)

    module private ToShow = 
        //LATEX{ InfernoCode
        let rec Inferno ttgt:ValidOp<(ColumnName list*(int->ValidOp<unit>))list>=
            OPM {
            let! links = GetNontrivialLinkedTableNames ttgt       // (A)
            let! linkLists = OPM {
                match links with
                | [] -> // base case
                    let! clusterColName = 
                            NewColumn ttgt "cluster" (NewColType.UptoColumn 4)
                    let modFun k = TypeUpto k ttgt clusterColName // (B)
                    return [[clusterColName], modFun]
                | _ ->  // recursive case
                    let linkLists = ref []                        // (C)
                    for (c,t) in links do
                        let! ll = Inferno t
                        linkLists := List.map (fun (m,vop) -> c::m,vop) ll 
                                     @ !linkLists                 // (D)
                    return !linkLists}
            let! rngColList=  GetConcreteColumns ttgt             // (E)
            for c in rngColList do
                if c.markup = Input then do! Model ttgt c.name
                for ll in linkLists do
                    match ll with
                    | llhead::llbody,_ ->
                        do! Index ttgt c.name llhead llbody
                    | [],_ -> failwith "impossible"
            return linkLists}
        //LATEX} InfernoCode


    let sweepNumberClusters clusterFun scoreFun isHighBest = OPM {
        let mem = ref []
        for i = 1 to 6 do
            do! clusterFun i
            let! score = scoreFun
            printfn "#Clusters= %d, Score= %f" i score
            mem := (i,score) :: !mem
        printfn "ALL: %A" !mem
        let bestK,score =
            if isHighBest
            then List.maxBy snd !mem
            else List.minBy snd !mem
        do! clusterFun <| bestK
        return bestK,score,!mem
        }
    
    (*
    //LATEX{ SweepNumberClusters
    /// Find best number of clusters using the modify-#-of-cluster functions
    /// & scoring function, e.g. ScoreLogEvidence (higher scores better) or 
    /// CrossValidate (lower scores better). Returns plottable score data.
    val SweepNumberClusters: clusterFun:(int->ValidOp<unit>) -> scoreFun:
    ValidOp<float> ->isHighBest:bool -> ValidOp<int*float* (int*float) list>
    //LATEX} SweepNumberClusters
    *)

    let private counter = ref 1

    open FSharp.Charting

    let saveRes trackBest filename =
        let tosave = (fun (a,_,c,d) -> a,c,d)
        use stream = new System.IO.FileStream(filename, System.IO.FileMode.Create)
        let bn = new System.Runtime.Serialization.Formatters.Binary.BinaryFormatter()
        let tmp = List.map (tosave) trackBest
        bn.Serialize(stream,tmp)

    let loadRes filename =
        use stream = new System.IO.FileStream(filename, System.IO.FileMode.Open)
        let bn = new System.Runtime.Serialization.Formatters.Binary.BinaryFormatter()
        let tmp = bn.Deserialize(stream) :?> ((string list * int * (int*float) list) list)
        List.map (fun (a,c,d) -> a, (fun _ -> NoOp), c, d) tmp

    let plotRes trackBest methodName =
        let title=
            "Optimal at "+(
                List.map (fun (cnl,_clusterFun,curBest,sweep) ->
                    sprintf "%s=%d" (List.head cnl) curBest
                ) trackBest
                |> List.reduce (fun a b -> a+", "+b)
            )
        let chart =
            List.map (fun (cnl,_clusterFun,curBest,sweep) ->
                Chart.Line( (sweep |> List.sortBy fst), Name=(List.head cnl))
            ) trackBest
            |> Chart.Combine
            |> Chart.WithXAxis(Title="#clusters")
            |> Chart.WithYAxis(Title=methodName)
            |> Chart.WithTitle(Text=title)
            |> Chart.WithLegend(true)
        chart.ShowChart() |> ignore
        chart.SaveChartAs(methodName+".png",ChartTypes.ChartImageFormat.Png)
        
        

    let printRes trackBest = 
        printfn "RESULT after %d iterations:" !counter
        counter := !counter + 1
        List.iter (fun (cnl,_clusterFun,curBest,sweep) ->
            printfn "%A: best=%d sweep=%A" cnl curBest sweep
        ) trackBest
        List.iter (fun (cnl,_clusterFun,curBest,sweep) ->
            printfn "%A\t%d" (List.head cnl) curBest
        ) trackBest
        printfn "~\t%s" (trackBest |> List.head |> (fun (_,_,_,swe)->swe) |> List.map (fst >> sprintf "%d") |> List.sort |> List.reduce (fun a b -> a+"\t"+b) )
        List.iter (fun (cnl,_clusterFun,curBest,sweep) ->
            printfn "%A\t%s" (List.head cnl) (sweep |> List.sortBy fst |> List.map (snd >> sprintf "%f") |> List.reduce (fun a b -> a+"\t"+b) )
        ) trackBest
    
    

    let rec sweepAllIterative trackBest scoreFun isHighBest = OPM {
        let changed = ref false
        let refNew = ref []
        for (cnl,clusterFun,curBest,sweep) in trackBest do
            printfn "Now Sweeping # of clusters on: %A" cnl
            let! (bestK,score,all) = sweepNumberClusters clusterFun scoreFun isHighBest
            printfn "Best # of clusters on %A is %d with score %f" cnl bestK score
            if curBest <> bestK then changed := true
            refNew := (cnl,clusterFun,bestK,all) :: !refNew
        refNew := List.rev !refNew
        printRes !refNew
        if !changed
        then 
            let! res = sweepAllIterative (!refNew) scoreFun isHighBest
            return res
        else return (!refNew)
        }

    let SaveDoAllSweep infRes scoreFun isHighBest methodName = OPM {
        let trackBest = List.map (fun (x,y) -> x,y,4,[]) infRes
        let! res = sweepAllIterative trackBest scoreFun isHighBest
        saveRes res (methodName+".bin")
        plotRes res methodName
        |> ignore
        }
\end{lstlisting}
\caption{NaiveBayes derived operation}
\label{lNaiveBayesDerived}
\end{code}

Figure~\ref{fNBsf} shows the final Naive Bayes model after running the rest of Code Listing~\ref{fNBPrim}. 
\texttt{ModelGaussian} is similar to \texttt{ModelDiscrete} in that it allows inference of mean and precision for a Gaussian distribution generating a column.

Naive Bayes is a common pattern in machine learning,
and as \cite{lucenaprobabilistic} concurs, probabilistic programming languages naturally express Naive Bayes.
The derived operation in Code Listing~\ref{lNaiveBayesDerived}
provides one such expression to facilitate its construction on future datasets. 
It offers 
abstraction in that we no longer need worry about its component operations or F\# code and can treat it as a black box. Alternatively, we gain understanding into how Naive Bayes works by looking at its code. The derived operation simplifies Code Listing~\ref{fNBPrim} to the one-liner 
\texttt{do! NaiveBayes "tmain" "gender"}.

We anticipate users use derived operations in two ways. On a basic level, users may record a list of operations on a dataset, initially taken ad-hoc, for replay on a similar dataset. 
After recognizing that a list of operations is an instance of a more common modeling pattern, 
as in the case of Naive Bayes, 
a user can abstract and generalize the derived operation to apply more widely. 
These operations include both transformations on the dataset, like the conversion of string gender values to links in a new table, and construction of a Tabular model.

\section{Hybrid Model: Internet Plant Sales}
\label{sInternetPlantSales}
We now turn to a real world example: modeling Internet plant sales. We use more advanced pre-processing, capturing functional dependencies, in combination with the previous section's Naive Bayes model.

\begin{figure}
\centering
\includegraphics[width=0.90\linewidth]{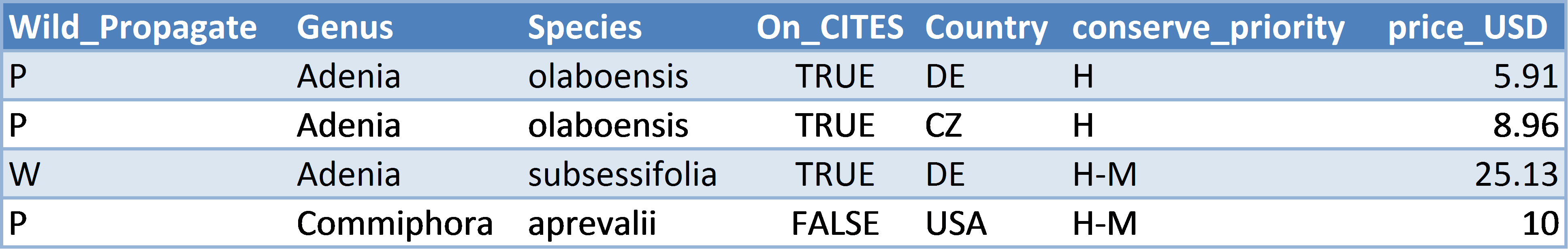}
\caption{Internet Plant Sales data}
\label{fPlantsInput}
\end{figure}

Figure~\ref{fPlantsInput} shows a few rows from a dataset, acquired from staff at the Royal Botanic Gardens in Kew UK who are interested in predicting Wild\_Propagate: whether a sale is of a wild plant or a propagate plant grown in a greenhouse. 
Predicting whether a plant is wild or propagate is a key step in identifying illegal plant sales. 
Perhaps more important than prediction, it is important to understand how strongly other columns impact Wild\_Propagate, 
including sale price, country of sale, plant conservation priority, and
status on the CITES list of endangered plant species \cite{CITES}.

In database literature \cite{date1990introduction}, a functional dependence (FD) between domain columns and a range column hold when 
domain column values uniquely determine range column value. 
FDs may hold for continuous domains 
(as in the relationship $y=x^2$ where $x$ functionally determines $y$), 
but in this thesis we restrict FDs to only use discrete domains.
We sometimes refer to an FD as an ``exact'' dependence, 
because the domain of an FD \textit{exact}ly determines the value of its range.
The Internet plant sales dataset has two FDs: that Genus and Species determine conservation priority and On\_CITES status. 

\begin{figure}
\centering
\includegraphics[width=0.90\linewidth,clip=true,trim=0 0 250pt 0]{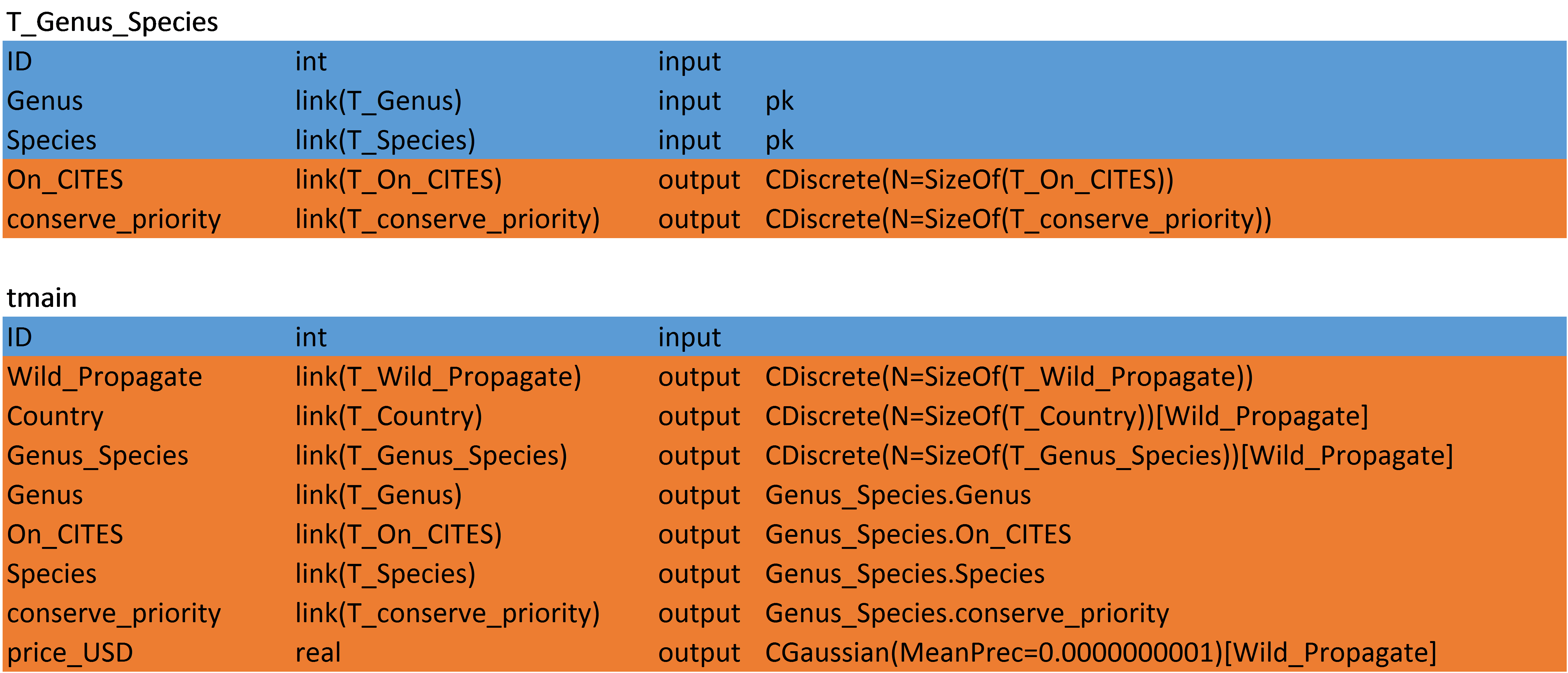}
\caption[EFD-augmented Naive Bayes model classifying Wild\_Propagate]{EFD-augmented Naive Bayes model classifying Wild\_Propagate \newline Tables of unique values for nominal variables omitted.}
\label{fPlantsModel}
\end{figure}

Code Listing~\ref{fPlantsScript} presents a script we use to create a hybrid model, 
shown in Figure~\ref{fPlantsModel}, for the Internet plant sales data.  
We infer the types of all columns in the dataset via \texttt{TypeInferTable},
create a separate table Genus\_Species explicitly enforcing 
two discovered functional dependencies via \texttt{ExactInfer} (described below), 
add a standard \texttt{NaiveBayes} on the remainder of tmain,
and complete the model by modeling the determined columns in Genus\_Species.
Ongoing research efforts build on this model 
to investigate how machine learning may aid CITES treaty enforcement
via plant sales inspections.



\begin{code}
\lstinputlisting[linerange=EFDex]{DemoSimple.fs}
\caption{ModelWizard script to construct plant sales model}
\label{fPlantsScript}
\end{code}

\subsection{ExactInfer and Exact: Functional Dependencies}
\label{sExact}
\texttt{ExactInfer} implements an algorithm to discover the best
functional dependency given a set of domain columns.  
``Best'' in this case refers to the functional dependence with largest range.
We could for example, only find in the Internet plant sales data that 
Genus and Species functionally determine On\_CITES status, 
but it would lead to a missed opportunity to find that Genus and Species 
functionally determine both On\_CITES and conserve\_priority.

We implement the derived operation \texttt{ExactInfer} in three phases.
\begin{enumerate}
\item Find all columns functionally determined by the given domain columns.
Do not consider columns that have dependencies.
\item If there is more than one range column, create a new table 
with domain columns as primary keys by calling \texttt{CreateTableUniques}.
\item For each range column, perform an \texttt{Exact}.
\end{enumerate}

\texttt{Exact} is a primitive operation that performs the steps below 
to take advantage of a functional dependency.
Translations of each step to the Internet plant sales 
\texttt{Exact tmain Genus\_Species On\_CITES} are given in brackets.

Given a main table, a range column in that main table, and a link-type domain column linking to a domain table [main table tmain, range column On\_CITES, and domain column Genus\_Species in tmain linking to the domain table T\_Genus\_Species (created by \texttt{ExactInfer})],
\begin{enumerate}
\item Create a new column in the domain table with the same name, type and model as the range column in the main table. 
[Create On\_CITES column in T\_Genus\_Species.]
\item Set the data of the new column to the values functionally determined by the primary key values in the same row. We implement this by building a map from domain column values to the unique range column value determined by the domain.
[Set the data of On\_CITES in T\_Genus\_Species to values determined by the FD in tmain from Genus\_Species to On\_CITES.]
\item Delete the data of the range column in the main table wherever the domain table link is not missing. 
There is no actual loss of data here because one may recover the data 
by joining the main table to the domain table.
Instead, think of the data deletion as a form of data compression.
[Delete the data of On\_CITES in tmain in rows where Genus\_Species is not missing.]
\item Modify the model of the range column in the main table to a reference to the range column in the domain table through the domain table link.
[Change the model of On\_CITES in tmain to Genus\_Species.On\_CITES.]
\end{enumerate}

Successively applying \texttt{ExactInfer} on a table's smallest possible 
domain set forms a basis for table normalization to Boyce-Codd Normal Form 
(BCNF).
Table normalization is the process of decomposing a table into smaller tables to eliminate redundancies.
Normalization to BCNF requires that the only FDs present in a table are either
trivial FDs where the range is a subset of the domain,
or FDs where the domain is a superkey (a superset of a candidate key: a set of columns that uniquely determine the FD's range).
We meet these requirements.

One could imagine extending the \texttt{ExactInfer} infrastructure 
to discover all the FDs within a table 
and not just the ones with a particular given domain,
similar to automated table normalization work in \cite{bahmani2008automatic}.
However, fully automated table normalization tends 
to discover spurious FDs, that is, 
FDs present in a dataset but not true in general.
Nevertheless, it is possible to improve \texttt{ExactInfer}'s 
FD-discovery algorithms by following recent work in \cite{heise2013scalable} 
and older work in its references.

\subsection{``Pre-processing'' during Modeling}
FD-handling is typically a pre-processing step, part of table normalization.
In fact, a user that knows about the FDs in a dataset will almost surely account for them before beginning the modeling process at the point where he architects dataset table layout.
What if the user does not know about a dataset FD, perhaps when there are hundreds of columns?
This is the ideal scenario for ModelWizard to help an analyst discover and take advantage of a dataset fact during the modeling process.
We leave it to future studies to determine whether the additional flexibility from conducting pre-processing at the same time as modeling will help in practice.

It is worth considering how the model would work in the case that we do not 
use a separate table T\_Genus\_Species to capture functional dependencies.
In this case, we would have a single table, tmain, with On\_CITES and conserve\_priority indexed by Genus and Species.  
Indexing the range columns by Genus and Species in the main table 
instead of a separate table models a FD in a looser sense, 
because we no longer restrict the range columns to a unique value 
given domain column values.

In the limit of an infinite dataset, the single-table model 
is equivalent to multi-table one because the CDiscrete distributions on On\_CITES and conserve\_priority collapse to point mass distributions for each unique (Genus, Species) pair.
Fixing a (Genus, Species) pair, the collapse occurs after running Bayesian updating on infinite datapoints that are all the same.
In the case of finite data, 
the single-table model incurs some predictive performance penalty.
This penalty grows small quickly if prior distributions are not too strong and dataset size is at least moderately large.

Both the single-table and multi-table model perform better 
than not capturing a FD at all, i.e., 
indexing On\_CITES and conserve\_priority by Wild\_Propagate.
This is because using tightly dependent columns---the range and domain columns---too grossly violates the Naive Bayes assumption that features are independent given Wild\_Propagate.

A prime future improvement to Figure~\ref{fPlantsModel}'s model is capturing the ordinal nature of column conserve\_priority.  
A common way to do this is to create a table of real-valued thresholds for each value of 
conserve\_priority (H, H-M, M, ...) and model the chosen conserve\_priority as a real-valued variable discretized by those thresholds.
An ``ordinal column template'' may make a nice addition to ModelWizard's primitive operations,
and it serves as another activity typically performed during preprocessing 
that aids users during modeling.

We imagine many other kinds of pre-processing activities 
from related work may enhance interactive model construction.  
The Bellman system presents algorithms 
to discover possible foreign key dependencies, 
columns formed from a concatenation or format change from other columns,
different ways two tables may be joined through intermediate tables,
and more,
by comparing summary structures computed for database tables
\cite{dasu2002mining}.
BayesDB offers capabilities to do similar preprocessing steps 
in the presence of more general data error
through concepts like ``stochastic primary key and foreign key,''
which can be taken as primary keys containing duplicates due to noise
and foreign keys containing changed or non-matching values due to noise
\cite{baxter2014bayesdb}.
Integrating probabilistic models of internal database structure 
in the style of BayesDB
with models of column value distributions 
in the style of Tabular
is an open avenue for future work.

\section{User-Movie-Rating Recommendation}
\label{sApplicationInferno}
For another real world application, we turn to movie recommendation 
as popularized by the Netflix challenge \cite{bennett2007netflix}.
We show how to build a general clustering model using a ModelWizard script and evaluate its predictive performance.
We specifically compare against Singh and Graepel's application of InfernoDB
to movie recommendations, as detailed in their first example 
that runs on synthetic user-movie-ratings data \cite{SinghGraepelInferno12, singh2013automated}.

\begin{figure}
\centering
\includegraphics[height=5em]{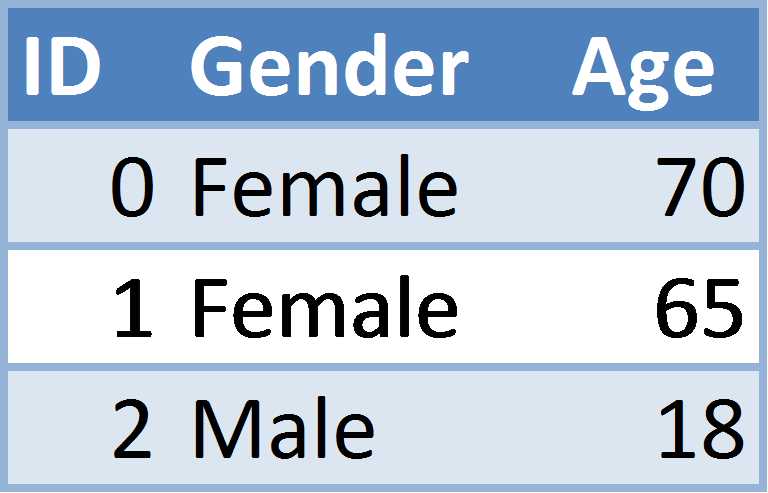} $\quad$
\includegraphics[height=5em]{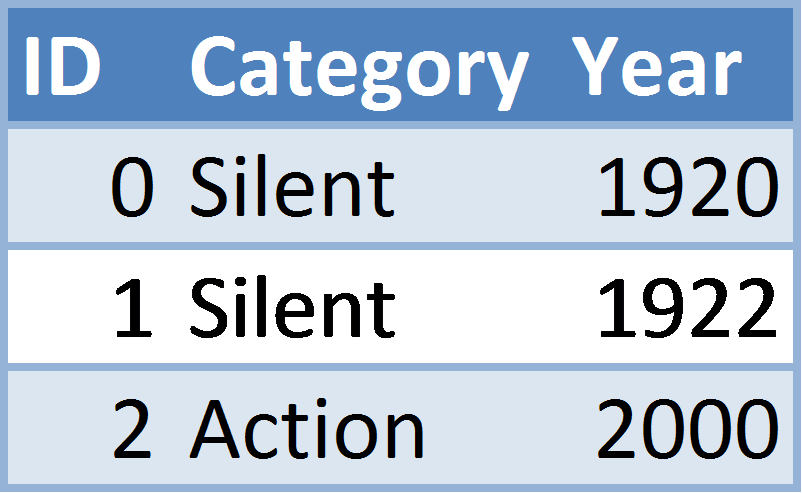} $\quad$
\includegraphics[height=5em]{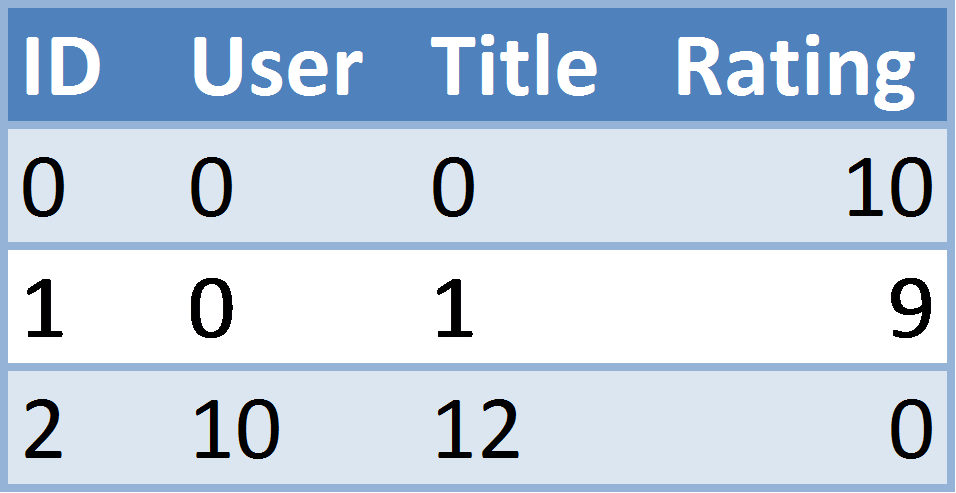}
\caption{Synthetic movie recommendation data snippet, from
\cite{SinghGraepelInferno12, singh2013automated}}
\label{fInfernoData}
\end{figure}

Our data is identical to Singh and Graepel's synthetic dataset, consisting of 
a table of users with attributes gender and age, 
a table of movies with attributes category and year, and 
a table of ratings with a rating from 0 to 10 and links to a user and movie.
The dataset is small, having two movie genres, 20 users, 30 movies and 25 ratings.
Figure~\ref{fInfernoData} illustrates a few rows.

\subsection{InfernoDB: Schema-Derived Clustering Models}
Singh and Graepel introduce the concept of automatically 
generating machine learning models based on the schema of a database,
including types of columns in tables and foreign key
relationships between tables. 

A column's type determines its base distribution: 
Real-valued columns have a Gaussian distribution 
and discrete (link) columns have a Discrete distribution. 
We replace the Bernoulli distribution used by Singh and Graepel 
with equivalent Discrete\footnote{Recall that we use Discrete as a synonym for a Categorical distribution.} distributions of dimension two.

Each table has a number of clusters. 
Tables with no foreign links (``leaf tables'') have a fixed number of clusters, chosen by hand since the number of clusters is a crucial hyperparameter.
Tables with foreign links (``body tables'') have a number of clusters equal to the product of linked tables' numbers of clusters.
Thus, a table with foreign links has clusters equivalent 
to the cross product of the clusters of tables it links to.

Each row has a distribution over the clusters in its table.
The distribution over clusters determines the parameters 
of each column's base distribution. 
The resulting parameters are a weighted average over each cluster's parameters, 
with weights determined by a Discrete distribution over clusters. 
Values for each column are drawn from their base distribution 
with these resulting parameters.
For example, in a table with an age column that has 4 clusters, 
the age for a row in that table is drawn from a Gaussian distribution 
with parameters determined by that row's distribution 
over the four clusters by taking a weighted average.

Each cluster's parameters are drawn from an uninformative prior distribution.
These priors are Gaussian ($\mu = 0$, $\sigma =10^8$) for the mean of a Gaussian, 
Gamma ($k = 1$,$\theta = 10$) for the precision of a Gaussian, and 
Dirichlet $\alpha = 1$ for the probability vector of a Discrete.
We chose these hyperparameters to match those used by Singh and Graepel.

Presented another way, Singh and Graepel generate a clustering model
formed inductively on tables:\footnote{We assume table links are acyclic; 
there must exist some table with no outgoing foreign keys.}

\begin{enumerate}
\item Table $T$ base case: no foreign links: create a hidden cluster variable that determines parameters of the distribution behind each column of $T$. 
The distribution behind a column depends on its type.
Fix the number of clusters heuristically.
\item Table $T$ recursive case: $T$ has foreign links: create a hidden cluster variable determined by the cross product of clusters of rows in tables that $T$ links to.
A particular row's distribution over clusters is determined by the distribution over clusters in each linked row in foreign tables.
\end{enumerate}

This model uses \textit{soft} clustering in that we allow a row to take 
a non-point-mass distribution over clusters. 
Calculations involving a row's cluster take weighted averages
over this distribution.  This is also true of foreign links in the case of missing
values for user and movie; we infer distributions over possible users and movies.
This process is called \textit{link prediction}. Its use case is when we have rows in our 
ratings table with a rating and movie, for example, and we wish to calculate 
which users are most likely to have given the movie that rating.
A challenge to link prediction is high dimension distributions when there are many users or movies.

\subsection{Inferno Operation with Cluster Setting}
Our \texttt{Inferno} implementation in ModelWizard has one major difference from 
Singh and Graepel's specification: 
it returns a map\footnote{The map is in the form of an association list, with elements of the form (key, value).} 
from ``leaf'' tables to a function that returns a \texttt{ValidOp} 
that, when later called, modifies the number of clusters in the corresponding leaf table.
The extra returned functions allow the user to change the number of clusters 
after running the \texttt{Inferno} operation. 
Leaf tables have four clusters by default.

\begin{code}
\lstinputlisting[linerange=InfernoCode]{WizOpCompound.fs}
\caption{Inferno ModelWizard Operation}
\label{fInfernoCode}
\end{code}

Code Listing~\ref{fInfernoCode} lists the \texttt{Inferno} operation.
Here are a few notes of explanation, matching Code Listing~\ref{fInfernoCode}'s commented annotations:
\begin{enumerate}[label={(\Alph*)}]
\item[(A)] \texttt{GetNontrivialLinkedTableNames ttgt} returns a list of table names that \texttt{ttgt} links to. We exclude ``trivial'' links to tables for nominal variables with no non-primary-key columns.

\item[(B)] \texttt{modFun} is a function to change the number of clusters set one line above.

\item[(C)] We construct \texttt{linkLists} by mutable assignment due to an artifact of \texttt{OpMonad}. This has the same effect as a fold with the additional ability to execute \texttt{ValidOp}s in the body of the fold.

\item[(D)] prepends the current table's linked ColumnNames to the list of ColumnNames in linked tables, building a correct list of ColumnNames across links to leaf tables.

\item[(E)] \texttt{GetConcreteColumns ttgt} returns the columns that will be clustered by \texttt{ttgt}'s new cluster column. These are all the columns in the table that are not link columns to a ``nontrivial'' table (see (A)) and are not the ID or cluster column. We \texttt{Model} these columns if they are input.
\end{enumerate}

\subsection{Hyperparameter Sweeps over Number of Clusters}
We use the resulting cluster-number-change functions from \texttt{Inferno} 
to find the optimal numbers of clusters 
by hyperparameter sweep using cross-validation.
We fashion our algorithm as iterative coordinate descent as follows:

\noindent Let $userk$ and $titlek$ be numbers of clusters for the user and title table, respectively. 
\begin{enumerate}
\item Initialize $userk, titlek := 4$ (from \texttt{Inferno}).
\item \label{iStep2} Hold $titlek$ fixed and calculate the best scoring $userk$
between 1 and 6 by running cross-validation 6 times and comparing scores.
\item Hold $userk$ fixed and calculate the best scoring $titlek$
between 1 and 6 by running cross-validation 6 times and comparing scores.
\item If $userk$ or $titlek$ changed, goto step~\ref{iStep2} else stop.
\end{enumerate}
The algorithm generalizes to any number of leaf tables.
We may also imagine a variant that does not fix the minimum or maximum number of clusters but instead searches for local score minima.

We use 5-fold cross-validation on the rating column 
to determine the best scoring numbers of clusters. 
To do this, we select 20\% of the rows in table tmain 
to hold out their ratings value. 
We train our model to score on the remaining data 
and predict the values for the unknown 20\% of rows. 
We use root mean square error (RMSE) 
to calculate error on predicted ratings versus their held out truths.
Repeat for each randomly allocated partition of the dataset into 5 divisions.
The best scoring model is the one with lowest mean cross-validation error.

\begin{figure}
\centering
\includegraphics[width=0.7\linewidth]{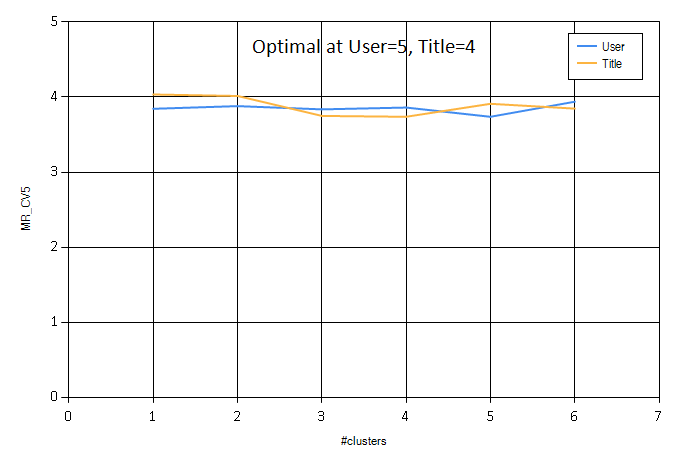}
\caption{Cross-validation mean scores for locally varying numbers of clusters}
\label{fInfernoCrossValidation}
\end{figure}

\begin{figure}
\centering
\includegraphics[width=0.90\linewidth,clip=true,trim=0 0 200pt 0]{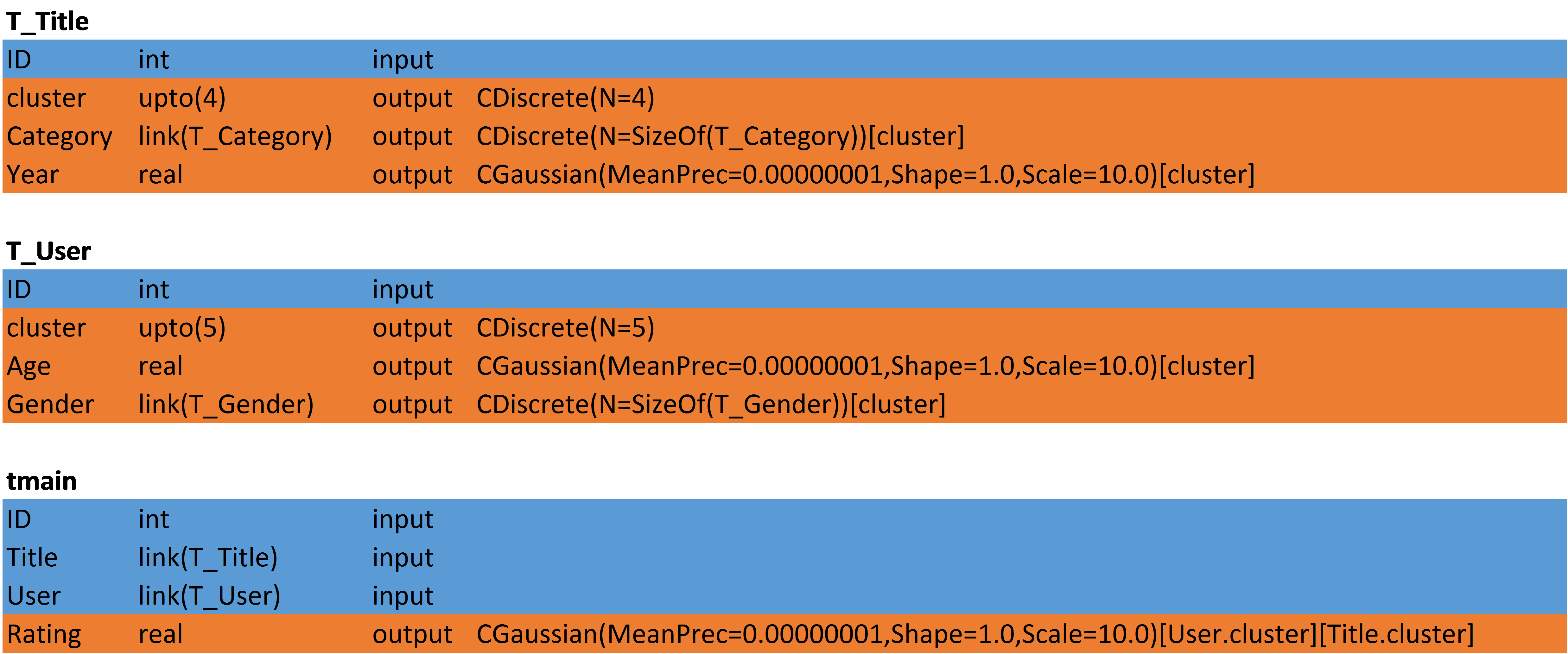}
\caption{Tabular model after Inferno and sweeping numbers of clusters}
\label{fInfernoModel}
\end{figure}

Figure~\ref{fInfernoCrossValidation} 
shows results of our sweep on Figure~\ref{fInfernoData}'s dataset,
and Figure~\ref{fInfernoModel} shows the resulting Tabular model (omitting trivial nominal tables of unique values).
We graph cross-validation errors of other numbers of clusters as different from the calculated optimal number of clusters: 5 for users and 4 for movies.
The $userk$ line holds $titlek$ constant at its calculated optimal number of clusters and vice versa.

The numbers of clusters are optimal for predicting ratings.  
They are not necessarily optimal for predicting age, gender, category or year
because we do not include prediction error from these columns in our RMSE plots.
To include additional column prediction error in our calculation,
we would have to define weights representing how much we care 
about one column's error relative to others, similar to a utility function.
Defining column weights becomes arbitrary, and so we arbitrarily chose 
weight one on rating and zero everywhere else.

\subsection{Performance on Missing Data}
To evaluate how well our ModelWizard implementation captures 
the concept behind InfernoDB, we create graphs 
of the same form as Figure 3 in \cite{singh2013automated}.
The graphs show model performance on predicting different columns with different numbers of randomly selected held out values. 
Unlike the previous section's procedure to find the optimal number of clusters,
we hold out values from all non-link columns: age and gender of a user, category and year of a movie, and rating. 
There are 125 data values to hold out.

\begin{figure}
\centering
\includegraphics[width=0.65\linewidth,clip=true,trim=15pt 10pt 35pt 10pt]{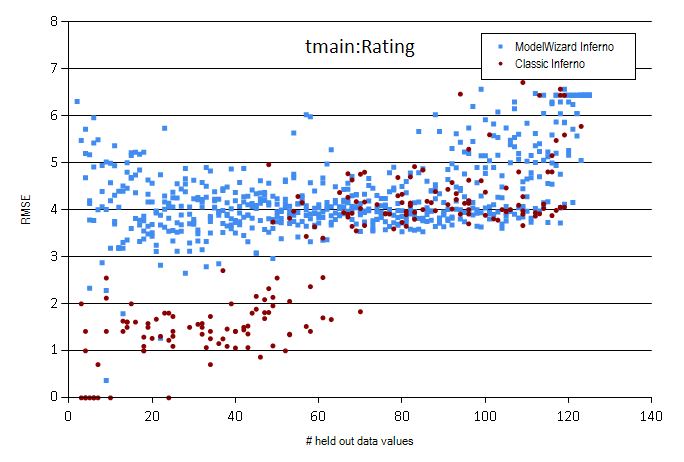}

\includegraphics[width=0.65\linewidth,clip=true,trim=15pt 10pt 35pt 10pt]{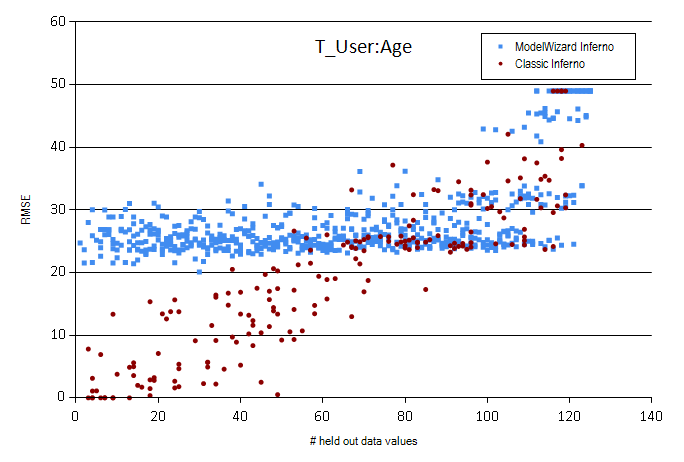}

\includegraphics[width=0.65\linewidth,clip=true,trim=15pt 10pt 35pt 10pt]{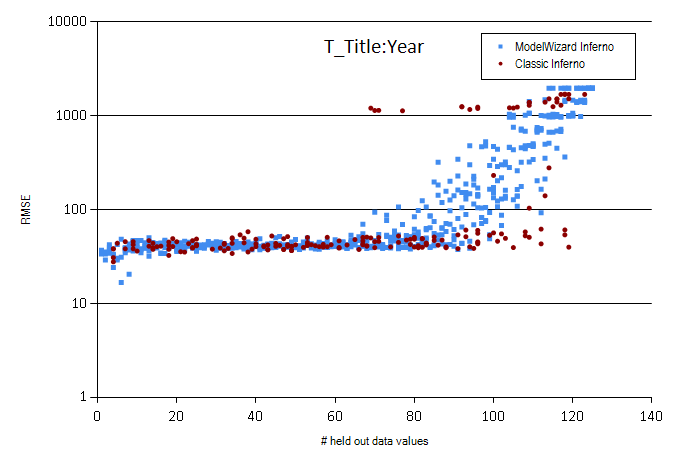}
\caption{\texttt{Inferno} model performance comparison varying missing cell count}
\label{fInfernoMissingGraph}
\end{figure}

Figure~\ref{fInfernoMissingGraph} shows the performance of the model 
derived from \texttt{Inferno} using the calculated optimal number of clusters
(that is, optimal for predicting ratings).
For each number of held out data points on the x-axis,
we plot five independent rounds of predicting that many held out values.
Each round may have different held out values, and for each we calculate 
RMSE.
Viewing several independent rounds per x-axis value highlights RMSE variance depending on how entries fall into training and test sets.
The original InfernoDB set similarly plots multiple rounds of predicting 
a given number of held out values, except that the number of rounds is not necessarily five; the classic InfernoDB test did not fix the number of rounds.

The ModelWizard version of InfernoDB has roughly equivalent performance as the original version at medium to high numbers of missing data cells. 
For low numbers of missing data cells, 
the ModelWizard version appears to perform worse.
We are unsure why this is so.
The cause could be structural, indicating that there is a major difference 
between the two versions of the InfernoDB algorithm 
such as using different numbers of clusters, 
or inconsequential, perhaps from miscopied data sources or a similar reason.

\subsection{Alternatives}
Matrix factorization is a popular alternative to cluster models for recommendation systems.
In matrix factorization, each user and movie has a hidden vector of $d$ dimensions,
where each dimension can be thought of as a trait. Compatibility between users and movies
is determined by the inner product of their trait vectors, resulting in a real-valued number
which is shifted to the average rating.
Traits of the same sign (in one dimension of the vector) contribute toward a higher rating and opposite-signed traits contribute toward a lower rating.
Other models in the literature worth consideration are in the category of collaborative filtering; see \cite{breese1998empirical} for an overview.


Another alternative is to consider rating as an ordinal variable instead of a real-valued one.
As in Section~\ref{sInternetPlantSales}'s Internet plant sales application, 
this involves ordering ratings and inferring 
their latent threshold values that delineate and effectively discretize ratings.

\begin{figure}
\centering
\includegraphics[height=5em]{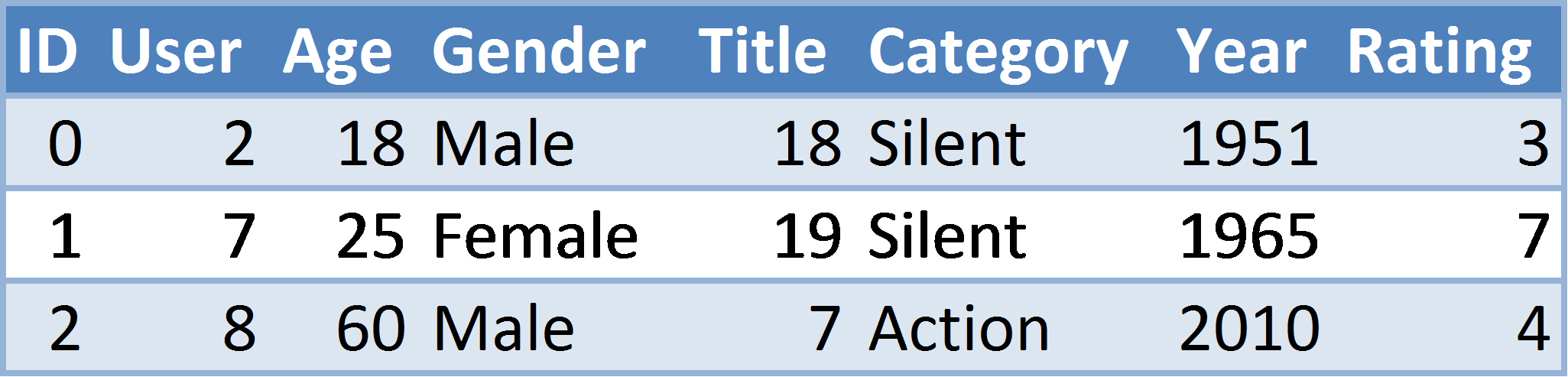}
\caption{Denormalized version of Figure~\ref{fInfernoData}'s movie data}
\label{fInfernoDataDenorm}
\end{figure}

As an alternative to Figure~\ref{fInfernoData}'s 
multi-table user-movie-ratings schema, 
the Inferno model also works for denormalized data 
for which we discover relationships in the dataset.
For example, suppose we receive data in the form of Figure~\ref{fInfernoDataDenorm}.
We may use \texttt{ExactInfer} operations as in Section~\ref{sInternetPlantSales}
to deduce a table schema that captures exact functional dependencies and then run \texttt{Inferno} on the resulting schema.  
The result is identical, and it shows how we may explore the space of dataset schemas while we explore the space of machine learning models in ModelWizard.

\chapter{Discussion}
\label{cDiscussion}

\section{Addressing Big Data's Challenges}
\label{sBigData}

A challenge to model construction is modern \textit{big data}, often characterized by four components \cite{EmbleyBigData2013}:
\begin{itemize}
\item Volume: large dataset size
\item Velocity: high rate of data acquisition, decreasing data value with time
\item Veracity: data uncertainty
\item Variety: multiple data sources, types and dimensions
\end{itemize}

Volume was not a consideration in ModelWizard's initial design, 
as hinted from small to moderate dataset sizes in the Applications chapter.
Luckily, there are many ways we imagine extending ModelWizard to better handle large datasets.
A good starting point is adding additional data representations such as sparse data storage.

Operating on data stored in a remote database instead of in-memory data could considerably increase scalability to large data.
The Excel workbook data model already has the ability 
to accept remote database connections, which ModelWizard can tap into
as it already does for data stored in spreadsheets as outlined in Section~\ref{sAdvancedOps}.
We may leverage Excel's remote database support by rewriting operations
to build remote queries and updates, perhaps using SQL language, instead of transforming data in memory.
Constraints on data map to database column constraints, 
and user-defined data transformation operations 
map to stored procedures executed by a SQL engine at the database.
NoSQL and NewSQL databases \cite{grolinger2013data}
are also in scope, though each additional database supported will require a separate set of drivers that operations interface with to access data---
unless recent projects such as SQL++ \cite{ong2014sql++},
D4M \cite{kepnerAssoc2015}, or the larger BigDawg proposal \cite{dugganbigdawg}
succeed in providing a common language interface to many types of databases.

The shift in thinking is to construct a function that will check and modify data, 
throwing an exception for data that does not fit an operation 
rather than eagerly modify data.
Viewing operations as database transactions additionally 
enables parallel execution of nonconflicting operations---that is, 
operations that do not modify the same part of Schema or Data as another operation that reads or modifies that part.

Sampling, lazy evaluation and background evaluation are further promising techniques 
that would help ModelWizard handle larger datasets.
Instead of running data-transforming operations on a dataset right away, 
we could run operations on samples from a dataset to see a preview of how the operation will affect the entire dataset.
The same principle holds for inference operations; one can gauge how well a model performs by running on a sample.
After obtaining preliminary results, 
ModelWizard can start a background process to run the operation on larger and larger samples, up to the full dataset, asynchronously returning progressively more accurate estimations of an operation's effect.
This allows us to retain interactivity using quick estimations on samples,
without losing correctness and accuracy when the user does not mind waiting.

ModelWizard script reuse will help in the case of many datasets with similar columns.  
A user may create a ModelWizard script that operates on one dataset and share that script with colleagues that have similar datasets, 
who can then apply the script with little to no change.
It is easy to imagine libraries of ModelWizard scripts developed and downloaded for data scientists within and across organizations.

The first component of velocity, high rate of data acquisition, 
is outside ModelWizard's scope.  
ModelWizard is designed to operate on fixed datasets with unknown structure rather than streaming data. 
A compromise is using ModelWizard to build a well-performing machine learning model for a fixed segment of streaming data on a development server, 
and then extract the machine learning model 
and implement it on a production server for streaming analysis.
C\# code for Tabular models built from ModelWizard is available through the Tabular GUI.
Thus to reemphasize: ModelWizard is a tool for model exploration and construction, not for top-performance model deployment.

The second component of velocity, decreasing data value with time, is directly modelable.
We can include the time a data value is recorded as itself another column in the dataset.
For example, take the problem of object tracking with dataset columns $position$ and $time$. 
If our goal is to infer the object's position at time $t$, 
position observations close to time $t$ are valuable whereas position observations farther before or after time $t$ yield decreasingly less information.
We may account for time dependence in modeling by including time difference between observations in the variance of position difference between observations.

Veracity, the phenomena of data uncertainty, 
is also directly modelable.
Including data accuracy and precision information as additional columns
enables us to use those columns in probabilistic modeling.
For example, we could imagine extending Section~\ref{sAppleFreefallExample}'s apple freefall example by including the imperfect mechanism by which we record apple elevations.
Such an addition will not change the mean, most likely value of a predicted elevation (assuming the same mechanism records all data points), 
but it will return higher, more realistic variances.
It also allows us to more strongly weigh sample rows that we have more confidence in, such as rows from using a higher precision measuring device.

ModelWizard handles data variety fairly well, especially heterogeneous data types.
ModelWizard's machine learning models operate on nominal, string-valued data, numeric data, and in the future, ordinal data.
Multiple data sources are no problem as ModelWizard represents each with a separate table, from which we may create cross-table links.

High dimensional data is not a strong suit for ModelWizard.
Imagine importing raw image processing data:
a single megapixel image in truecolor 
(one row) will have a million columns, each with a 24-bit integer.
Large numbers of columns are not ideal for manual analysis, though one could write ModelWizard operations to process them systematically.
A better use case for interactive modeling with ModelWizard 
is after feature extraction and dimensionality reduction, when each column has an interpretable meaning to data analysts.

\section{Extending Model Safety to Model Inferability}
ModelWizard's ValidState, ValidOp and OpMonad constructs guarantee a base level of model safety: 
every model constructed by ModelWizard has a valid Tabular counterpart and will pass Tabular's type checking.
Is it possible to extend model safety to guarantee that models constructed by ModelWizard are \textit{inferable}, that is, can be run by the back-end inference engine to produce posterior distributions on parameters and missing values?
Such a strong guarantee would enable data scientists to freely conjure models using the suite of operations available (respecting types so as not to throw exceptions), without worrying whether the model they construct will run.

The short answer is with great difficulty.
Providing inferability guarantees 
is inherently specific to choice of inference engine.
A formal proof requires modeling the syntax and semantics of a probabilistic program to prove that a certain subset of programs are inferable by a particular inference engine, 
and then prove that ModelWizard operations will only generate programs (after Tabular compilation) within that inferable subset.
All probabilistic programs are theoretically inferable,
but different inference engine implementations will support some probabilistic program families more than others.

As a heuristic example derived from the author's experience, 
Infer.NET's expectation propagation algorithm 
performs more reliably for models with real-valued columns.
Infer.NET's variational message passing algorithm 
performs more reliably for models with discrete columns.
Some models containing both real-valued and discrete columns 
are inferable by neither algorithm, a problem that surfaced during model exploration for Section~\ref{sInternetPlantSales}'s Internet Plant Sales application.
A chief cause of non-inferability is unimplemented message passing factors 
due to open research problems in variational inference.

\section{Extending to Usable Interactivity}
\label{sExtendUsable}
Many researchers believe usability is a key factor in the extent of success for probabilistic programming tools \cite{GordonAgendaProbProg13}.
To truly make incremental model construction accessible to data scientists,
we imagine a GUI or IDE wrapping ModelWizard's state, sketched in Figure~\ref{fModelWizardGUI}.
Regular users will find model exploration easier using a GUI, and 
``power users'' who want to create custom operations may do so via script editing.

\begin{figure}
\centering
\begin{tikzpicture}[align=center,font=\fontsize{11.5}{16}\selectfont]
\draw  (-4,3.5) rectangle (4.5,2.5) node [pos=.5] (v1) {Menu of Operations};
\draw  (4.5,2.5) rectangle (2,-2.5) node [pos=.5] (v2) {Operation \\ History with \\ Inference \\ Performance \\ Scores};
\draw  (-4,2.5) rectangle (2,0) node [pos=.5] (v3) {Current Tabular Schema};
\draw  (-4,0) rectangle (2,-2.5) node [pos=.5] (v4) {Data Preview};
\end{tikzpicture}
\caption{ModelWizard GUI Sketch}
\label{fModelWizardGUI}
\end{figure}

The GUI would show the currently constructed State, 
with the ability to zoom in on Tabular Schema and a preview or sample of Data.
At the right, we envision a list of operations performed so far.
The GUI may tag operations with a model score based on given scoring criteria, 
such as cross-validation on a particular column of interest.
Model log evidence is a useful default.

The menu at the top of Figure~\ref{fModelWizardGUI} contains buttons that trigger operations, grouped by category such as ``Typing'' and ``Base Models.''  
Right clicking on columns in the Schema or Data preview will list operations relevant to that column's type.  
At a more advanced stage, we imagine suggesting operations based on common modeling paradigms learned from users modeling similar datasets.

A particularly powerful meta-feature is \textit{undo}: the ability to roll back an operation and quickly try another one.
One can implement undo by extending the OpMonad computation expression to include State monad functionality by tracking past states (or diffs between states) within the monad.
Even in the current ModelWizard implementation,
users could save intermediate states
via ordinary F\# let bindings, 
as demonstrated by state \texttt{s1} in Figure~\ref{fBayesNetSearch},
after which performing undo is simply reusing saved state.
Undo enables users can quickly try an operation, 
see whether it makes sense or is silly, 
and quickly proceed or rollback.

The idea for a ModelWizard GUI is inspired by Microsoft Power Query's data transformation script editor \cite{webb2014power} and Proof General's Coq theorem proving script editor \cite{aspinall2000proof}.
Both editors lower the barrier for new users,
raise situational awareness,
and increase productivity while writing scripts.
We wish the same for ModelWizard users tackling model exploration.

In addition to visualization and interactive editing, 
a growing theme in systems design is to integrate the ``whole machine 
learning pipeline'' from data mining and enrichment to data cleaning and modeling
into a common framework, as outlined in~\cite{Morton2014} to support the activities of
\textit{data enthusiasts} that do not have formal training in data science.
We view ModelWizard as a forward step in creating such an integrated analytics 
framework, paritcularly in the combination of preprocessing with modeling.
Much future work (or combination with existing tools) is needed 
to incorporate data enrichment and cleaning.

%
%

\section{Interactive Theorem Proving Analogy}
Model construction bears many similarities to theorem proving. 
Both benefit from interactivity as a happy medium between two extremes.

\textit{Manually} proving theorems on pencil and paper requires expert knowledge and training to navigate the space of all possible proofs and formalize one that proves the theorem.
The approach works as with manual model construction but at high labor cost, sometimes taking years of study in the proof's domain.

\textit{Automatically} proving theorems requires powerful algorithms. 
For example, the automated SMT solver Z3 \cite{de2008z3} succeeds for simple theorems but will fail for complex enough theorems due to the size of the space of all proofs. 
Automated model construction similarly fails in an unrestricted model space.

We aspire to use the design of interactive theorem provers as inspiration for interactive model construction. 
Imagine operations as model construction tactics, lower ones like ModelDiscrete akin to Coq's rewrite tactic \cite{bertot2004interactive} 
and higher ones like ExactInfer akin to Isabelle's sledgehammer tactic \cite{paulson2010three}.
Like tactics, operations may fail by throwing an exception and leaving state unchanged.

\section{Conclusion}


Probabilistic programs offer a universal space to create machine learning models for a dataset.  
ModelWizard presents a new take on constructing machine learning models,
focusing on iterative construction one-to-one with model exploration,
rather than traditional specification of a probabilistic program entirely at once.

Our primary contribution is planting the seeds for an interactive framework,
applicable to any language and inference engine,
to view data transformation and machine learning models 
in terms of composable building blocks we call operations, 
glued together by program code (F\# in our case).
In doing so we simultaneously realize realms of 
pre-processing, manual analysis, automated analysis and modeling itself.
We further realize these realms in naturally familiar program structure 
such as conditionals, iteration, and procedural abstraction,
along with opportunity for classic programming language boons
such as static analysis, syntactic safety and compilation optimization.

We proffer ModelWizard as a solution scheme to three scientific goals: 
prediction, retained from inference on constructed machine learning models;
understanding, built from model decomposition into understandable operations;
and generalization, found through application of operations to new problems and domains.
We postulate future work building on our framework's principles 
will naturally integrate with and augment data scientists' 
\textit{modus operandi} to discover our world's true structure.
\hfill $\blacksquare$ 

\newpage


\addcontentsline{toc}{chapter}{Appendices}

\renewcommand{\thesection}{\Alph{section}}
\setcounter{section}{0}

\section{OpMonad Translation} \label{AOpMonadQuote}
As a demonstration for how \texttt{let!}, \texttt{do!} and other monadic syntax
desugars into vanilla F\# code,
we present a simplified version of the derived operation \texttt{TypeInferTable} 
on the target table \texttt{tmain}, 
followed by its translation into 
an F\# quotation that evaluates to a \texttt{ValidOp<unit>} value.
The quotation includes calls to computation expression functions,
defined in Code Listing~\ref{fOpMonad} and bolded in the code below.

\lstset{keywordstyle=[2]{\bfseries},keywords=[2]{For,Delay,Bind,Return,Zero,Combine,ReturnFrom}}
\begin{lstlisting}
OPM {let! schema = GetSchema
	 let table = getSchemaTableByName schema "tmain"
	 for col in  table.columns do			  // infer each column's
		do! TypeInferColumn "tmain" col.name} // type in tmain
		
Quotations.Expr<ValidOp<unit>> =
Call (Some (Value (FSI_0008+OpMonad)), Delay,
  [Lambda (unitVar,
   Call (Some (Value (FSI_0008+OpMonad)), Bind,
	 [PropertyGet (None, GetSchema, []),
	  Lambda (_arg1,
		 Let (schema, _arg1,
			Let (table,
			  Call (None, getSchemaTableByName,
				[schema, PropertyGet (None, "tmain", [])]),
			  Call (Some (Value (FSI_0008+OpMonad)), For,
			    [Coerce (PropertyGet (Some (table),
			   						  columns, []), IEnumerable`1),
				 Lambda (_arg2,
					Let (col, _arg2,
					 Call (Some (Value (FSI_0008+OpMonad)),
					  Bind,[Call (None,
								 TypeInferColumn,
								 [PropertyGet (None, "tmain", []),
								  PropertyGet (Some (col), name, [])]),
							Lambda (_arg3,
							  Call(Some (Value (FSI_0008+OpMonad)),
							   Return, [Value (<null>)]))])))]))))]))])
\end{lstlisting}

\section{ModelWizard API of Primitive and Derived Operations} \label{AModelWizardAPI}
\texttt{P} stands for a Primitive operation; others are derived from primitives using \texttt{OpMonad}.
Excel-interfacing and ``helper'' operations are omitted.
\vspace{1em}

Core bridge from \texttt{State} to F\#:
\lstinputlisting[aboveskip=0pt,belowskip=0pt,linerange={GetState}]{OpMonad.fsi}

Column typing:
\lstinputlisting[aboveskip=0pt,belowskip=0pt,linerange={TypeReal}]{OpMonad.fsi}
\lstinputlisting[aboveskip=0pt,belowskip=0pt,linerange={TypeUpto}]{OpMonad.fsi}
\lstinputlisting[aboveskip=0pt,belowskip=0pt,linerange={NewColumn}]{OpMonad.fsi}

Table management:
\lstinputlisting[aboveskip=0pt,belowskip=0pt,linerange={CreateTableUniques}]{OpMonad.fsi}
\lstinputlisting[aboveskip=0pt,belowskip=0pt,linerange={Link}]{OpMonad.fsi}

Derived typing:
\lstinputlisting[aboveskip=0pt,belowskip=0pt,linerange={TypeNominal}]{WizOpCompound.fs}
\lstinputlisting[aboveskip=0pt,belowskip=0pt,linerange={TypeInfer}]{WizOpCompound.fs}
\lstinputlisting[aboveskip=0pt,belowskip=0pt,linerange={TypeInferTable}]{WizOpCompound.fs}

Base machine learning models:
\lstinputlisting[aboveskip=0pt,belowskip=0pt,linerange={ModelGaussianDiscrete}]{OpMonad.fsi}
\lstinputlisting[aboveskip=0pt,belowskip=0pt,linerange={Model}]{WizOpCompound.fs}

\clearpage
Model coupling:
\lstinputlisting[aboveskip=0pt,belowskip=0pt,linerange={Index}]{OpMonad.fsi}
\lstinputlisting[aboveskip=0pt,belowskip=0pt,linerange={RegOps}]{OpMonad.fsi}

Derived machine learning models:
\lstinputlisting[aboveskip=0pt,belowskip=0pt,linerange={NaiveBayes}]{WizOpCompound.fs}
\lstinputlisting[aboveskip=0pt,belowskip=0pt,linerange={InfernoOp}]{WizOpCompound.fs}

Exact functional dependencies:
\lstinputlisting[aboveskip=0pt,belowskip=0pt,linerange={Exact}]{OpMonad.fsi}
\lstinputlisting[aboveskip=0pt,belowskip=0pt,linerange={ExactInfer}]{WizOpCompound.fs}

Inference:
\lstinputlisting[aboveskip=0pt,belowskip=0pt,linerange={ScoreLogEvidence}]{CallTabular.fs}
\lstinputlisting[aboveskip=0pt,belowskip=0pt,linerange={CrossValidatekFoldRMSE}]{CrossValidate.fs}
\lstinputlisting[aboveskip=0pt,belowskip=0pt,linerange={SweepNumberClusters}]{WizOpCompound.fs}
\lstinputlisting[aboveskip=0pt,belowskip=0pt,linerange={MissingDataAnalysis}]{MultiCV.fs}


%
%
%
%

\bibliographystyle{alpha}
\newpage
\addcontentsline{toc}{chapter}{Bibliography}		
\bibliography{Dylan_MSThesis}

\end{document}